\documentclass[aps,pra,twocolumn,amsmath,amssymb,superscriptaddress,showpacs,
]{revtex4-2}
\usepackage{bbm}
\usepackage{graphics}
\usepackage{graphicx}
\usepackage{amsfonts}
\usepackage{amssymb}
\usepackage{ulem}
\usepackage{float}
\usepackage{bm}
\usepackage[urlcolor=blue,colorlinks=true,citecolor=blue,linkcolor=blue,pdfstartview={FitH},bookmarks=false]{hyperref}
\usepackage{CJK}

\begin{document}

\title{Characterizing quantum criticality and steered coherence in the  \emph{XY}-Gamma chain}

\author{Zhuan Zhao}
 \affiliation{College of Physics, Nanjing University of Aeronautics and Astronautics, Nanjing, 211106, China}
 \affiliation{MIIT Key Laboratory of Aerospace Information Materials and Physics, Nanjing University of Aeronautics and Astronautics, Nanjing 211106, China}

\author{Tian-Cheng Yi}
\affiliation{Beijing Computational Science Research Center, Beijing 100193, China}

 \author{Ming Xue}
 \affiliation{College of Physics, Nanjing University of Aeronautics and Astronautics, Nanjing, 211106, China}
 \affiliation{MIIT Key Laboratory of Aerospace Information Materials and Physics, Nanjing University of Aeronautics and Astronautics, Nanjing 211106, China}

\author{Wen-Long You}
 \email{wlyou@nuaa.edu.cn}
 \affiliation{College of Physics, Nanjing University of Aeronautics and Astronautics, Nanjing, 211106, China}
 \affiliation{MIIT Key Laboratory of Aerospace Information Materials and Physics, Nanjing University of Aeronautics and Astronautics, Nanjing 211106, China}

\begin{abstract}
In this paper, we show that an effective spin Hamiltonian with various types of couplings can be engineered using quantum simulators in atomic-molecular-optical laboratories, dubbed the \emph{XY}-Gamma model. We analytically solve the one-dimensional short-range interacting case with the Jordan-Wigner transformation and establish the phase diagram. In the gapless phase, an incommensurate spiral order is manifested by the vector-chiral correlations. Between distinct gapped phases, a logarithmic scaling behavior of local measures, including spin correlations and the steered quantum coherence, is identified for the quantum critical points, yielding a compelling value of the correlation-length critical exponent.
We derive explicit scaling forms of the excitation gap near the quantum critical points. The extracted critical exponents reveal the quantum phase transition on the boundary of
Tomonaga-Luttinger liquid belongs to Lifshitz universality class.
Our results may provide useful insights into the underlying mechanism in quantum criticality for state-of-the-art experiments of quantum simulation.
\end{abstract}

\date{\today}

\maketitle

\section{Introduction}
The exploration of 
quantum phenomena
is an outstanding challenge and has been one of the most active   arenas in condensed matter physics~\cite{sachdev2008quantum}.
Novel forms of phases
are enriched by 
ongoing discoveries in transition metal compounds~\cite{Fe,MnCO,zhang2021strain,nature2021},
such as spin-orbit-entangled electronic phases~\cite{muhlbauer09skyrmion,yu2010real}. In particular, the quantum spin liquid (QSL) has recently emerged as a new paradigm in correlated electron physics,
as it holds promise for the potential application of quantum computing and quantum information.
One avenue towards QSL is the focus on highly frustrated materials, exemplified by either the geometrical frustration or the exchange frustration~\cite{Kiatev2014,kiatev2014b,HK1,HK2}.
The triangular, kagome, and pyrochlore structures are categorized into the first type,
while the actively sought-after Kitaev magnets belong to the second type. In this context,
much attention has been recently devoted towards $4d$ and $5d$ transition metal compounds,
due to the interplay between spin-orbit couplings and electronic correlations~\cite{material2018,material2020,Ru2019}.
The playground to search for the QSL was recently extended to $3d$ transition metal compounds~\cite{xu20possible}.
Despite these endeavors, the Kitaev QSL state has not been conclusively identified among most 
candidate materials.
Certain long-range order (LRO) is unexpectedly found in a variety of
Mott insulators, indicating the existence of more conventional types of exchanges beyond a dominant bond-directional Kitaev interaction in these nonideal materials.

Since engineering robust QSL states remains challenging in spin-orbit-entangled candidate materials, an
alternative route for exploring novel phases of matter and forms of entanglement is the experimental implementation
with the help of other quantum systems in the laboratory. After Feynman's proposal in 1982~\cite{feynman1982computer}, the field of quantum simulation has been developing rapidly for decades and nowadays enables the investigation of quantum systems in a programmable fashion. Especially, recent 
advances of the
laser technology and the laser
manipulation of atomic gases have made it feasible to implement a wide class of analog quantum simulations in atomic-molecular-optical (AMO) laboratories. Quantum simulators have been realized on
a few platforms, 
e.g., ultracold atoms~\cite{trotzky2008time}, polar molecules~\cite{yanbo13nature},
trapped ions~\cite{roos14nature,monroe14nature},
photonic systems\,\cite{hung2016quantum,Douglas15quantum,Gonz15quantum},
and Rydberg atom arrays~\cite{ABrowaeys20np,lukin21nature256}, etc.
These systems can be finely tuned in a sufficiently precise way and observed in real time. The effective many-body Hamiltonian can be incorporated from recent developments of simulating quantum  magnetism and related quantum dynamics using atoms interacting with the same quantum modes, wherein the quantum channel can be the guided modes in the photonic crystal waveguides~\cite{hung2016quantum}, the photon of cavity modes~\cite{monika20heisenberg,chiocchetta2021cavity},
and the Rydberg dressing states~\cite{zoller15designing}.
With present architectures of quantum simulators, a generic  Hamiltonian consisting of flexible coupling graphs
can be freely realized, offering the opportunity to implement, simulate, and experimentally test fundamental paradigmatic model Hamiltonians.

This paper is organized as follows. In Sec.~\ref{sec:simulation}, we discuss the possible engineering of the
\emph{XY}-Gamma model using the AMO system facilitated by
coherent photon-mediated Raman transitions.
In our scenario, the independent control of  \emph{XX}-,\emph{ YY}-,\emph{ XY}- and \emph{YX}- terms can be achieved by a double $\Lambda$ scheme in neutral atoms.
Section~\ref{model} is devoted to exact solutions of the one-dimensional (1D) short-range interacting Hamiltonian using the Jordan-Wigner transformation. Two-site correlations and dimer correlations are then analyzed.
Next, the quantum steered coherence is investigated in Sec.~\ref{SQC}. A discussion and summary  follow in Sec.~\ref{summary}.

\section{Effective model with Photon-mediated atom-atom Interactions}\label{sec:simulation}
Recently, using photons to mediate controllable atom-atom interactions has been a well-established paradigm in AMO quantum simulation~\cite{zoller15designing,Mivehvar19toolbox,Mivehvar21review,chiocchetta2021cavity}.
The constituents of the engineered quantum spin models are not restricted to only Heisenberg-type
interactions~\cite{lev11frustration,Mivehvar17disorder,Esslinger18formation}.
The complex photon-mediated interaction graphs, including both their amplitudes and interaction ranges,
can be arbitrarily programed by state-of-the-art techniques involving more laser beams, posing envisioned synthetic quantum matter.
For instance, it is well known that the LRO is prohibited by the Mermin-Wagner theorem for short-range interacting models with a continuous symmetry in one spatial dimension~\cite{ren22longrange},
while the counterpart on a two-dimensional (2D) bipartite lattice is generally expected to host the long-range N\'{e}el order for any spin magnitude $S$,
although a rigorous proof of the existence of LRO in a 2D $S=1/2$ Heisenberg model is still lacking~\cite{you2009long,tang1989long,lin92long}.
The long-range \emph{XY} order is induced in an \emph{XXZ} chain by single-mode-cavity-mediated infinite-range interactions~\cite{li21long}. Instead, QSLs are stabilized in a 2D isotropic Heisenberg model by power-law decaying interactions in  multimode cavities~\cite{chiocchetta2021cavity}.

\begin{figure}
  \centering
  \includegraphics[width=0.56\columnwidth]{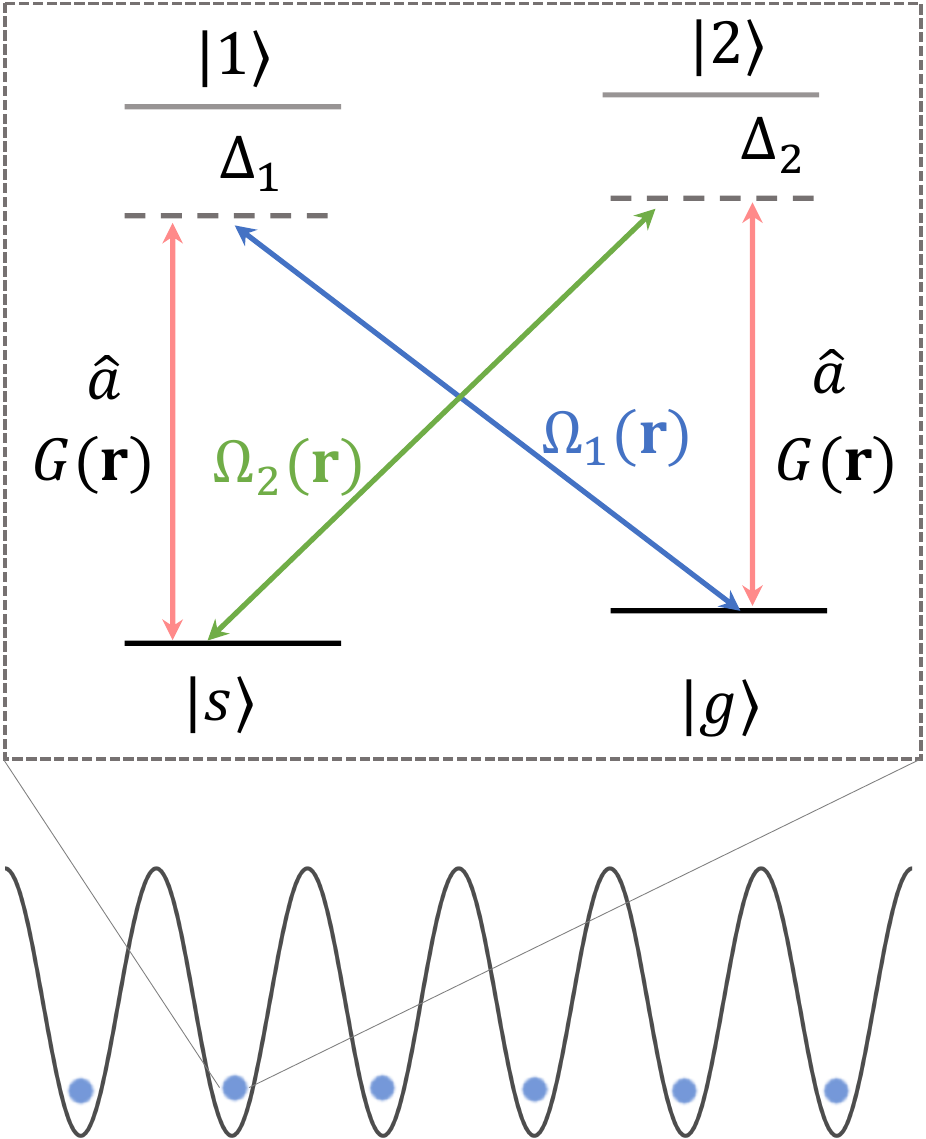}
  \caption{
Schematic setups. Atoms are trapped in 1D optical lattice with four-level energy diagram (inset).
Two pump lasers with Rabi frequencies $\Omega_1$ (blue) and $\Omega_2$ (green) induce $\sigma$-transitions between atomic ground and excited states $|g\rangle\leftrightarrow|1\rangle$ and
$|s\rangle\leftrightarrow|2\rangle$, respectively.
A 
cavity mode (red) 
induces $\pi$-transitions between $|s\rangle \leftrightarrow |1\rangle$ and  $|g\rangle\leftrightarrow |2\rangle$.}
  \label{fig:cavity_scheme}
\end{figure}

In the following,
we consider $N$ atoms that are trapped tightly in a 1D optical lattice,
as depicted in Fig.~\ref{fig:cavity_scheme}.
A double $\Lambda$ scheme of an atomic-level diagram is assumed,
where two 
internal atomic states $\{|s\rangle, |g\rangle\}$ in the ground-state manifold
represent the pseudospin-$1/2$ states with energy $\{0, \omega_g\}$ and two auxiliary excited states $\{|1\rangle, |2\rangle\}$ exist with energy
$\{\omega_1, \omega_2\}$, respectively.
A photon mode at frequency $\omega_\mathbf{k}$ with the field operator $\hat{a}_\mathbf{k}$
induces $\pi$-transitions between atomic ground and excited states $|s\rangle\leftrightarrow|1\rangle$ and $|g\rangle\leftrightarrow|2\rangle$,
where $\mathbf{k}$ denotes the index of bosonic modes.
$G_\mathbf{k}(\bf{r})$ is the corresponding spatially dependent coupling strength.
Two pump lasers, denoted as $(L1, L2)$ with Rabi strengths $(\Omega_1, \Omega_2)$
and frequencies $(\omega_{L1}, \omega_{L2})$,
are implemented to induce $\sigma$ transitions between atomic ground and excited states, i.e., $|g\rangle\leftrightarrow|1\rangle$ and
$|s\rangle\leftrightarrow|2\rangle$, respectively.

Under rotating wave approximation, the atom-light hybrid system is described by the Hamiltonian $\hat{H} =\hat{H}_0+\hat{H}_{AL}$,
where $\hat{H}_0$ is the free Hamiltonian consisting of photon fields $\hat{a}_\mathbf{k}$ and atomic Zeeman energy levels $\hat{\sigma}_j^{aa}$,
and $\hat{H}_{AL}$ is the atom-light interaction Hamiltonian ($\hbar=1$ throughout):
\begin{eqnarray}
  \hat{H}_{0}&=&\sum_{\mathbf{k}} \omega_{\mathbf{k}}\hat{a}^\dagger_\mathbf{k}\hat{a}_\mathbf{k}
  +\sum_{j}\sum_{\alpha}\omega_\alpha\hat{\sigma}^{\alpha\alpha}_j,\\
  \hat{H}_{\rm AL}&=&\sum_j \left[{\Omega_1(\mathbf{r}_j)}e^{-i\omega_{L1}t}\hat{\sigma}_j^{1g}
  +{\Omega_2(\mathbf{r}_j)}e^{-i\omega_{L2}t}\hat{\sigma}_j^{2s}+{\rm H.c.}\right]\nonumber\\
  &&+\sum_j\sum_{\mathbf{k}}\left[G_{\mathbf{k}}(\mathbf{r}_j)\hat{a}_{\mathbf{k}}(\hat{\sigma}_j^{1s}+\hat{\sigma}_j^{2g})+{\rm H.c.}\right].
\end{eqnarray}
Here the atomic transition operators are defined as $\hat{\sigma}_j^{ab}\equiv|a\rangle_j\langle b|$ with four atomic energy levels  $a,b=\{g,s,1,2\}$ and $j$ labels the site.
Two-atom interaction is synthesized by two Raman transitions, where the photon field $\hat{a}_\mathbf{k}$
provides the two-body correlation between the two atoms, as depicted in Fig.\,\ref{fig:cavity_scheme}.

Working in a rotating frame defined by
$\hat{U}$=$\exp\{i[\sum_{j}(\omega_{-}\hat{\sigma}_j^{gg}
    +\omega_{+}\hat{\sigma}^{11}_j+\omega_{L2}\hat{\sigma}^{22}_j)
   +\sum_\mathbf{k}\omega_{+}\hat{a}_\mathbf{k}^\dagger\hat{a}_\mathbf{k}]t\}$ with $\omega_{\pm}$=$(\omega_{L2}\pm\omega_{L1})/2$, the transformed Hamiltonian
$\tilde{H}=U \hat{H} U^\dagger+i(\partial_t U) U^\dagger$ reads
\begin{eqnarray}
\tilde{H}&=&-\sum_{\mathbf{k}} \Delta_\mathbf{k} \hat{a}^\dagger_\mathbf{k}\hat{a}_\mathbf{k}
+\sum_j \left[\delta \hat{\sigma}^{gg}_j  -\Delta_1\hat{\sigma}^{11}_j-\Delta_2\hat{\sigma}^{22}_j\right]\nonumber\\
&&+\sum_j\left[{\Omega_1(\mathbf{r}_j)}\hat{\sigma}_j^{1g}
+{\Omega_2(\mathbf{r}_j)}\hat{\sigma}_j^{2s}+{\rm H.c.}\right]\nonumber\\
&&+\sum_j\sum_{\mathbf{k}}
\left[G_{\mathbf{k}}(\mathbf{r}_j)\hat{a}_{\mathbf{k}}(\hat{\sigma}_j^{1s}+\hat{\sigma}_j^{2g})+{\rm H.c.}\right]\label{eq:Hrot},
\end{eqnarray}
where
$\Delta_{\mathbf{k}}\equiv(\omega_{L1}+\omega_{L2})/2-\omega_{\mathbf{k}}$,
$\Delta_1\equiv (\omega_{L1}+\omega_{L2})/2-\omega_1$,
$\Delta_2\equiv \omega_{L2}-\omega_2$ and
$\delta=\omega_g-(\omega_{L2}-\omega_{L1})/2$.
Supposing the frequencies of pump lasers and bosonic modes are all far detuned from the atomic transitions, i.e., $\Delta_1$ and $\Delta_2$ are much larger 
than the Rabi coupling coefficients
$|\Omega_1|, |\Omega_2|$, and $|G_\mathbf{k}|$,
we can safely eliminate the excited states $|1\rangle$ and $|2\rangle$ to obtain the effective
Hamiltonian in the ground-state manifold (see Appendix \ref{DerivationOfEffectiveHamiltonian}),
\begin{eqnarray}
  \hat{H}_{gs}&=&\sum_j\left[(\delta+V_g(\mathbf{r}_j))\hat{\sigma}_j^{gg}+
  V_s(\mathbf{r}_j)\hat{\sigma}_j^{ss}\right]\nonumber\\
&&  +\sum_j\left[\hat{\sigma}_j^{gs}\hat{\Omega}(\mathbf{r}_j)
  +\hat{\Omega}^\dagger(\mathbf{r}_j)\hat{\sigma}_j^{sg}\right],\label{eq:Hgs_a}
\end{eqnarray}
where $\hat{\Omega}(\mathbf{r})= \Omega_1^\ast \sum_\mathbf{k}G_\mathbf{k}(\mathbf{r})\hat{a}_{\mathbf{k}}/\Delta_1
+ \sum_\mathbf{k}G^\ast_\mathbf{k}(\mathbf{r})\hat{a}^\dagger_{\mathbf{k}}/\Omega_2 $  and the light shifts
for ground states $V_g(\mathbf{r})={|\Omega_1|^2}/{\Delta_1}+{\sum_{kk^\prime}G^\ast_\mathbf{k}G_{\mathbf{k}^\prime}\hat{a}^\dagger_{\mathbf{k}}\hat{a}_{\mathbf{k}^\prime}}/{\Delta_2}$,
$V_s(\mathbf{r})={|\Omega_2|^2}/{\Delta_2}+{\sum_{kk^\prime}G^\ast_\mathbf{k}G_{\mathbf{k}^\prime}\hat{a}^\dagger_{\mathbf{k}}\hat{a}_{\mathbf{k}^\prime}}/{\Delta_2}$.
In the adiabatic limit due to large detuing or large dissipation $\kappa$
of modes $\hat{a}_\mathbf{k}$,
the photon field can be approximated by its steady-state value,
\begin{eqnarray}\hat{a}_{\mathbf{k}}^{\rm ss}=
\sum_j\, G^\ast_{\mathbf{k}}(\mathbf{r}_j)
\left(\frac{\Omega_1}{\Delta_1}\hat{\sigma}^-_j+
\frac{\Omega_2}{\Delta_2}\hat{\sigma}^+_j\right)/{\tilde{\Delta}_\mathbf{k}}, \end{eqnarray}
with $\tilde{\Delta}_\mathbf{k}=\Delta_\mathbf{k}+i\kappa-
\sum_j \,|G_\mathbf{k}(\mathbf{r}_j)|^2\left(\hat{\sigma}_j^{ss}/{\Delta_1}
+\hat{\sigma}_j^{gg}/{\Delta_2}\right)$.
Note that here we have included the
cavity dissipation $\kappa$ phenomenologically.
By adiabatically eliminating the photon degree of freedom,
in terms of Pauli operators,  $\sigma^x_j=\hat{\sigma}_j^{sg}+\hat{\sigma}_j^{gs}, \sigma_j^{y}=i(\hat{\sigma}^{sg}_j-\hat{\sigma}^{gs}_j), \sigma^z_j=\hat{\sigma}_j^{gg}-\hat{\sigma}_j^{ss}$,
we obtain the effective spin 
Hamiltonian in a compact form,
\begin{eqnarray}
  \hat{H}_{\rm int}\!
  &=&\sum_{ij}\!(J_{ij}^x \sigma_i^x \sigma_j^x
  +J_{ij}^y\sigma_i^y \sigma_j^y)\!+\!J_{ij}^{\rm DM}(\sigma_i^x \sigma_j^y
  -\sigma_i^y\sigma_j^x)\nonumber\\
  &+&\sum_{ij} J_{ij}^{\rm SO}\left(\sigma_i^y \sigma_j^x
  + \sigma_i^x \sigma_j^y\right) + \sum_{j} h_j^z {\sigma}_j^z.\label{eq:Hamiltonian0}
\end{eqnarray}
Here $h_j^z=\delta/2+ V_g(\mathbf{r}_j)-V_s(\mathbf{r}_j)$, $J_{ij}^x=2$({Re}${[\Lambda_0]}$+{Re}${[\Lambda_1]}$),
$J_{ij}^y=$2({Re}${[\Lambda_0]}$$-${Re}${[\Lambda_1]}$),
$J_{ij}^{\rm DM}=$2{Im}$[\Lambda_0]$, and $J_{ij}^{\rm SO}=$-2{Im}$[\Lambda_1]$, where {Re} ({Im}) indicates the real (imaginary) part of a complex variable,  $\Lambda_{0,1}\equiv\Lambda_{0,1}(\mathbf{r}_i,\mathbf{r}_j)$ with
$\Lambda_0(\mathbf{r},\mathbf{r}^\prime)=\sum_\mathbf{k}
[\frac{\Omega_1^\ast(\mathbf{r})\Omega_1(\mathbf{r}^\prime)}{\Delta_1^2\tilde{\Delta}_\mathbf{k}}G_\mathbf{k}(\mathbf{r})G^\ast_{\mathbf{k}}(\mathbf{r}^\prime)
+\frac{\Omega_2(\mathbf{r})\Omega_2^\ast(\mathbf{r}^\prime)}{\Delta^2_2\tilde{\Delta}^\ast_\mathbf{k}}G_\mathbf{k}^\ast(\mathbf{r})G_{\mathbf{k}}(\mathbf{r}^\prime)]$
and $\Lambda_1(\mathbf{r},\mathbf{r}^\prime)=\sum_\mathbf{k}[
\frac{\Omega_1^\ast(\mathbf{r})\Omega_2(\mathbf{r}^\prime)}{\Delta_1\Delta_2\tilde{\Delta}_\mathbf{k}}G_\mathbf{k}(\mathbf{r})G^\ast_{\mathbf{k}}(\mathbf{r}^\prime)
+\frac{\Omega_2(\mathbf{r})\Omega_1^\ast(\mathbf{r}^\prime)}{\Delta_1\Delta_2\tilde{\Delta}^\ast_\mathbf{k}}G_\mathbf{k}^\ast(\mathbf{r})G_{\mathbf{k}}(\mathbf{r}^\prime)]$.
The detailed definitions of coefficients in Eq.\,(\ref{eq:Hamiltonian0}) are derived in Appendix\,\ref{DerivationOfEffectiveHamiltonian}.

The first bracketed term of Eq.~\eqref{eq:Hamiltonian0} corresponds to the conventional \emph{XY}-type interactions, and the term in the second brackets denotes the $z$ component of
Dzyaloshinskii-Moriya interactions (DMIs).
The cross-couplings in the third brackets between $x$ and $y$  spin components are referred to as symmetric off-diagonal $\Gamma$ interactions.
Exotic forms like DMIs
originated from spin-orbit couplings~\cite{dzyaloshinsky1958thermodynamic,DM1959,XYDM1960,DM1960b} and were firstly devised to account for the
weak ferromagnetism in antiferromagnetic crystals~\cite{Fe,MnCO,zhang2021strain,nature2021},
favoring chiral states such as spin spirals and skyrmions~\cite{muhlbauer09skyrmion,yu2010real}. Concurrently,
the importance of pervasive off-diagonal $\Gamma$ interactions can be traced back to the study of the
Kitaev-Heisenberg model~\cite{Kiatev2014,kiatev2014b,HK1,HK2}.
Further research suggests that the symmetric off-diagonal $\Gamma$ interactions should also be taken into account to explain the possible QSLs observed in experiments~\cite{Luo2021unveiling,Luo2021unusual,you2020Lifshitz,Alberto2020phase,Erik2021entanglement,Yang2020comprehensive}.

\section{Exact solution and correlations}\label{model}
Note that the cavity-mediated spin-spin interactions in Eq.~(\ref{eq:Hamiltonian0}) have infinite range
if only a single cavity mode is involved. The finite-range interactions are achieved by using a multimode
cavity. The photon modes can be
the guide modes in photonic crystal waveguides with quasimomentum $\mathbf{k}$\,\cite{,Douglas15quantum,Gonz15quantum,hung2016quantum},
or the near-degenerate transverse cavity  modes\,\cite{Lev18tunable,Guo21nature}. 
In particular, the multifrequency driving also enables finite-range interactions between intracavity atoms
in a single-mode cavity, which has been recently realized experimentally\,\cite{periwal2021programmable}.
Therefore,
an effective Hamiltonian with tunable interaction strength and interaction range can be constructed by using multimodes $\{\hat{a}_\mathbf{k}\}$.
In this case,
the interference of cavity modes may render the beyond-nearest-neighbor
couplings to be negligibly small.

In the following, we concentrate on the spin models
for an ensemble of spin-1/2 interacting particles
on a 1D lattice with nearest‐neighbor interactions only.   The spin Hamiltonian can be rewritten as
\begin{eqnarray}
\hat{H}&=&\sum_{j=1}^{N}J\left(\frac{1+\gamma}{2}\sigma_j^x\sigma_{j+1}^x+\frac{1-\gamma}{2}\sigma_j^y\sigma_{j+1}^y\right)\nonumber\\
&&+\sum_{j=1}^{N} \left[\Gamma\left(\sigma_j^x\sigma_{j+1}^y+\alpha\sigma_j^y\sigma_{j+1}^x\right)+h\sigma_j^z\right],
\label{eq:Hamiltonian1}
\end{eqnarray}
where the antiferromagnetic coupling  $J\equiv J_{j,j+1}^x+J_{j,j+1}^y$
between the nearest-neighbor atoms is set up as an energy unit for simplicity unless otherwise stated, i.e., $J=1$, $\gamma\equiv(J_{j,j+1}^x-J_{j,j+1}^y)/J$
serves as the anisotropy parameter, $\Gamma \equiv J_{j,j+1}^{\rm DM}+J_{j,j+1}^{\rm SO}$ characterizes the amplitude of off-diagonal exchange interactions,
$\alpha\equiv (J_{j,j+1}^{\rm SO}-J_{j,j+1}^{\rm DM})/ \Gamma$ denotes the relative coefficient of off-diagonal exchange couplings, and $h$ represents the strength of the uniform transverse field.
The $\Gamma$ term reduces to the DMI for $\alpha=-1$ and the symmetric off-diagonal exchange interaction for $\alpha=1$.
In what follows, we impose periodic boundary conditions (PBCs) with $\sigma_{N+1}^a\equiv\sigma_{1}^a$ ($a=x,y,z$).

The motivations of exploring the quantum criticality in the \emph{XY}-Gamma model are two fold. On one hand, the low-dimensional quantum magnets have been particularly of concern owing to their evident quantum aspects and substantial corrections to classical counterparts. The quantum criticalities have been explored in a few magnetic materials. The notable examples range from the spin-1/2 Ising ferromagnet LiHoF$_4$~\cite{bitko1996quantum}, 
SrCo$_2$V$_2$O$_8$~\cite{cui2019quantum},
Cs$_2$CoCI$_4$~\cite{kenzelmann2002order}, and CoNb$_2$O$_6$~\cite{coldea2010quantum} to  BaCo$_2$V$_2$O$_8$~\cite{PhysRevLett.123.027204}
as well as the spin-1 ferromagnetic Heisenberg chain NiNb$_2$O$_6$~\cite{chauhan20tunable}. To date, quantum phase transitions (QPTs) of analog models have been studied in different contexts, such as the \emph{XY} model with DMIs~\cite{XYDMYTC}.
The simultaneous appearance of off-diagonal exchange $\Gamma$ interactions
and \emph{XY}-type interaction in the presence of external fields, especially counteracting the disordered state,
is less systematically clear.
One the other hand, the merit of 
Eq.~(\ref{eq:Hamiltonian0}) resides in its exact solvability. The analytical results render the possibility to calculate accurately the experimental measurable quantities, in particular various dynamic ones, and thus serve as a benchmark for more sophisticate models. As we shall demonstrate, the extracted critical exponents
can be relevant to the experimental measurement of thermodynamic quantities and information measures.

\begin{figure}[ht]
\centering
\includegraphics[width=0.9\columnwidth]{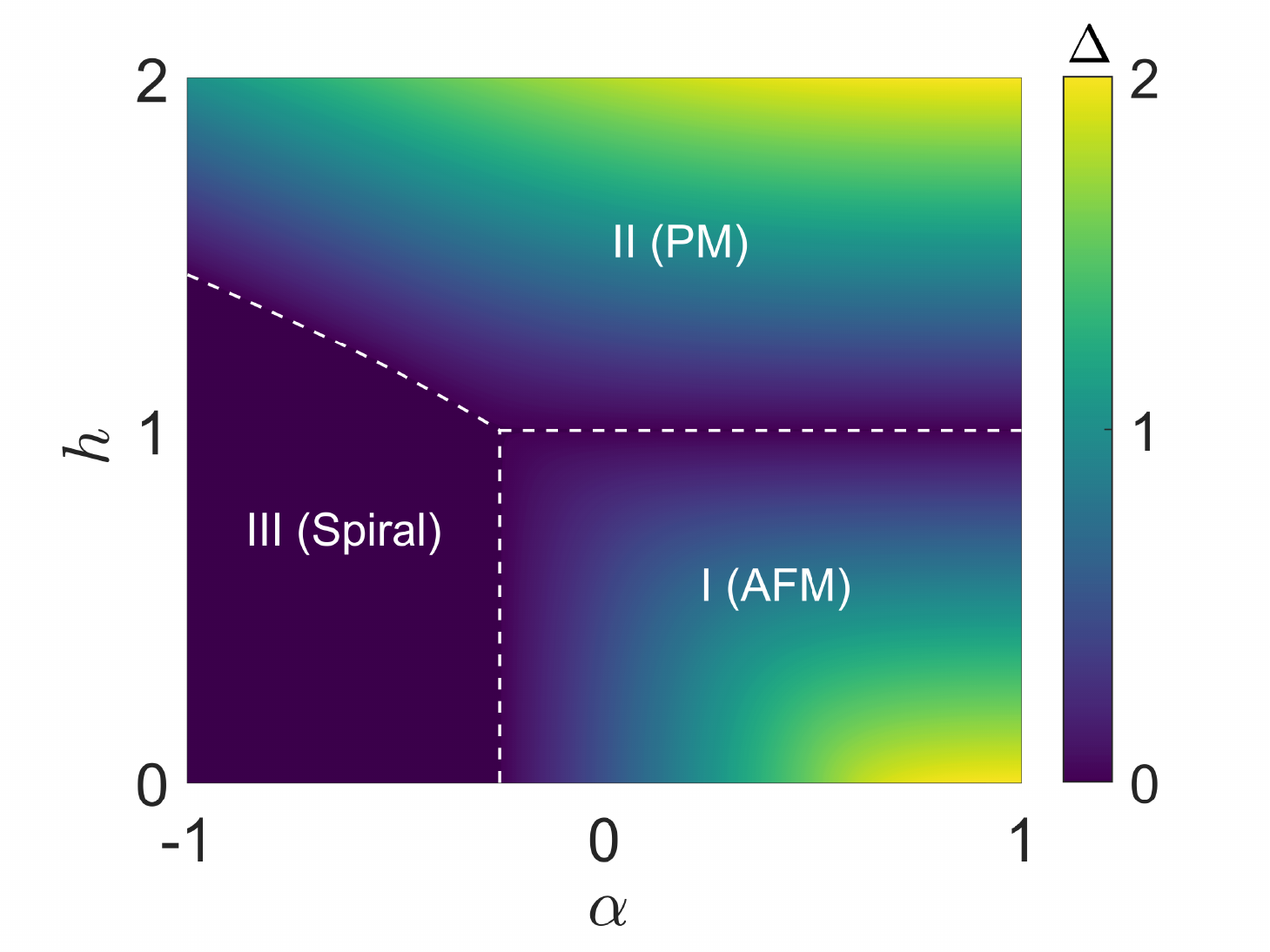}
\caption{The contour map of excitation gap with respect to $\alpha$ and $h$.
The white dashed lines correspond to the critical lines separating
 phases I (AFM), II (PM), and III (Spiral) of the \emph{XY}-Gamma model with $J=1.00$, $\Gamma=0.60, \gamma=0.60$.}
\label{phasediagram}
\end{figure}

The Hamiltonian in Eq.~(\ref{eq:Hamiltonian1}) is analytically solved by means of Jordan-Wigner, Fourier, and Bogoliubov transformations.
The detailed diagonalization procedure is shown in  Appendix \ref{energyspect1}.
Ultimately, the Hamiltonian can be brought into a diagonal form of a 
less fermion in the momentum space,
\begin{eqnarray}
\hat{H}&=& \sum_{k}{\varepsilon}_{k}({b}_{k}^{+}{b}_{k}-\frac{1}{2}),
\end{eqnarray}
where the energy spectrum of fermionic quasiparticles is
\begin{eqnarray}
{\varepsilon}_{k}&=&
2\sqrt{[\Gamma^2(\alpha+1)^2+\gamma^2]\sin^2{k}+(\cos{k}-h)^2} \nonumber\\ &&-2\Gamma(1-\alpha)\sin{k} .
\label{varepsilon}
\end{eqnarray}
With the excitation energy at hand, the energy gap $\Delta$=$\min_k {\varepsilon}_{k}$ can be
determined. As shown in Fig. \ref{phasediagram}, the ground-state phase diagram consists of three phases by varying $\alpha$ and $h$.  The horizontal segment $h_{c,1}=1$ for $\alpha > - {\gamma^2}/{4\Gamma^2}$ separates the gapped phase I and phase II, while the gapless phase III is encompassed by the separatrices
 $\alpha_{c,1}=- \gamma^2/(4\Gamma^2)$ for $h\le 1$
and $\alpha_{c,2}=(1-h^2-\gamma^2)/(4\Gamma^2)$ for $h>1$, or equivalently, $h_{c,2}=\sqrt{1-\gamma^2-4\Gamma^2\alpha}$  for $\alpha<- \gamma^2/(4\Gamma^2)$.

Typically, one can define spatial and temporal characteristic lengths that have diverging behavior as the control parameter $\lambda$ approaches the threshold value $\lambda_c$. This diverging property of characteristic lengths with the correlation-length critical exponent $\nu$  and the dynamical critical exponent $z$ enables one to define universality classes.   The critical behavior
is determined by those low-energy states near
the critical mode.
The dynamical exponent $z$ relates the scaling of energy 
to length scales,
which can be retrieved by the shape of the spectra near the gap closing mode $\Delta \sim (k-k_c)^{z}$. As $\lambda$ approaches $\lambda_{c}$, the gap vanishes as $\Delta\sim(\lambda-\lambda_{c})^{\nu z}$.  To this end, we expand
Eq.~(\ref{varepsilon}) at the critical line $h_{c,1}$ around the gap closing momentum $k_c=0$,
\begin{eqnarray}
   {\varepsilon}_{k} \simeq  2\sqrt{\gamma^2+\Gamma^2(1+\alpha)^2} \vert k\vert-2\Gamma(1-\alpha) k.
\end{eqnarray}
The relativistic spectra at the critical line $h_{c,1}$ imply that $z$ = 1.
The gap near $h_{c,1}$ is approximated as
\begin{equation}
\label{Delta}
\Delta \simeq
\,2\vert h-h_{c,1}\vert,
\end{equation}
and one then finds $\nu z=1$. In this case, the quantum critical point (QCP) between phase I and phase II belongs to 2D Ising universality class characterized by $\nu=1$, $z=1$. On the verge between gapped phase II and  gapless phase III, one can find the spectra vanish at an incommensurate momentum $k_c=\arccos(h_{c,2}^{-1})$,
 \begin{eqnarray}
\varepsilon_k \simeq\left[\frac{\Gamma(1-\alpha)\cos^2 k_c}{2\sin k_c}+\frac{(h_{c,2}^{-1}-h_{c,2})^2}{2\Gamma(1-\alpha)\sin k_c}\right](k-k_c)^2.\quad \quad\label{quadratic}
\end{eqnarray}
The above quadratic dispersion indicates the dynamical exponent $z = 2$.
While expanding the gap around the QCP from the upper threshold
one obtains the excitation as
\begin{equation}
\label{Delta2}
\Delta \simeq \frac{2(h_{c,2}-h_{c,2}^{-1})}{\Gamma(1-\alpha)\sin k_c} (h-h_{c,2}).
\end{equation}
The critical exponents $\nu=1/2$ and $z=2$ annotate that the QPT is in the so-called Lifshitz universality class~\cite{Rufo2019Multicritical}, which corresponds to the universality class of quantum criticality of free fermions.  %
In the case of  the I-III transition $\alpha_{c,1}$, the spectra are found to be quadratic in $k$ around the gap closing mode $k_c=\arccos h$ as
  \begin{eqnarray}
\varepsilon_k \simeq \frac{\sqrt{1-h^2}}{\Gamma(1-\alpha)} (k-k_c)^2, \label{quadratic2}
\end{eqnarray}
which yields $z$=2. Similarly, the gap around the critical point $\alpha_{c,1}$ from above obeys a power-law relation as
\begin{equation}
\label{Delta3}
\Delta \simeq \frac{4 \Gamma\sin k_c}{1-\alpha_{c,1}} (\alpha-\alpha_{c,1}).
\end{equation}
The scaling form 
in Eq.~(\ref{Delta3}) reveals that the transition also belongs to the Lifshitz universality class with $z$ = 2 and $\nu$ = 1/2.

\begin{figure}[ht]
  \includegraphics[width=0.85\columnwidth]{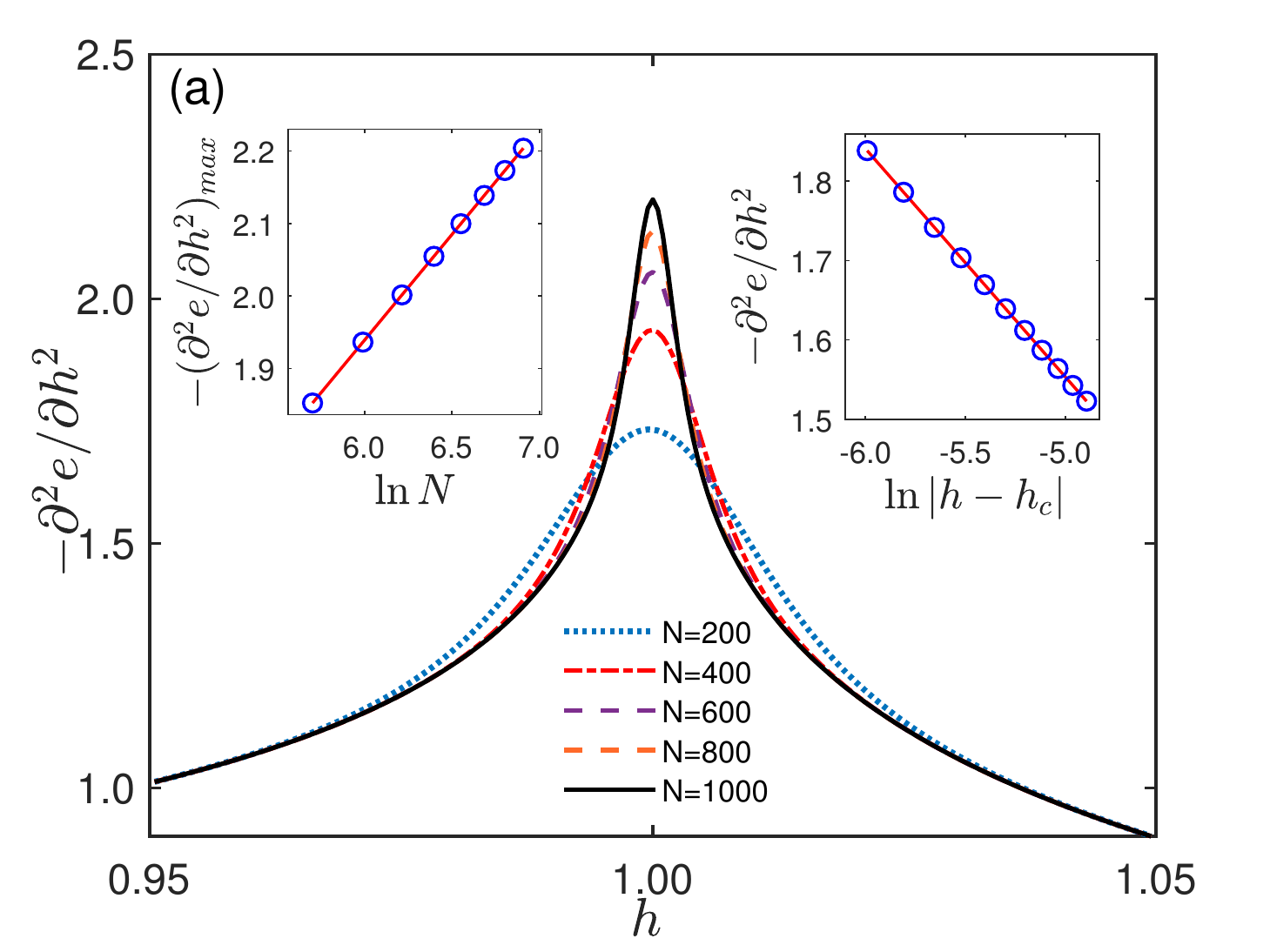}
  \includegraphics[width=0.85\columnwidth]{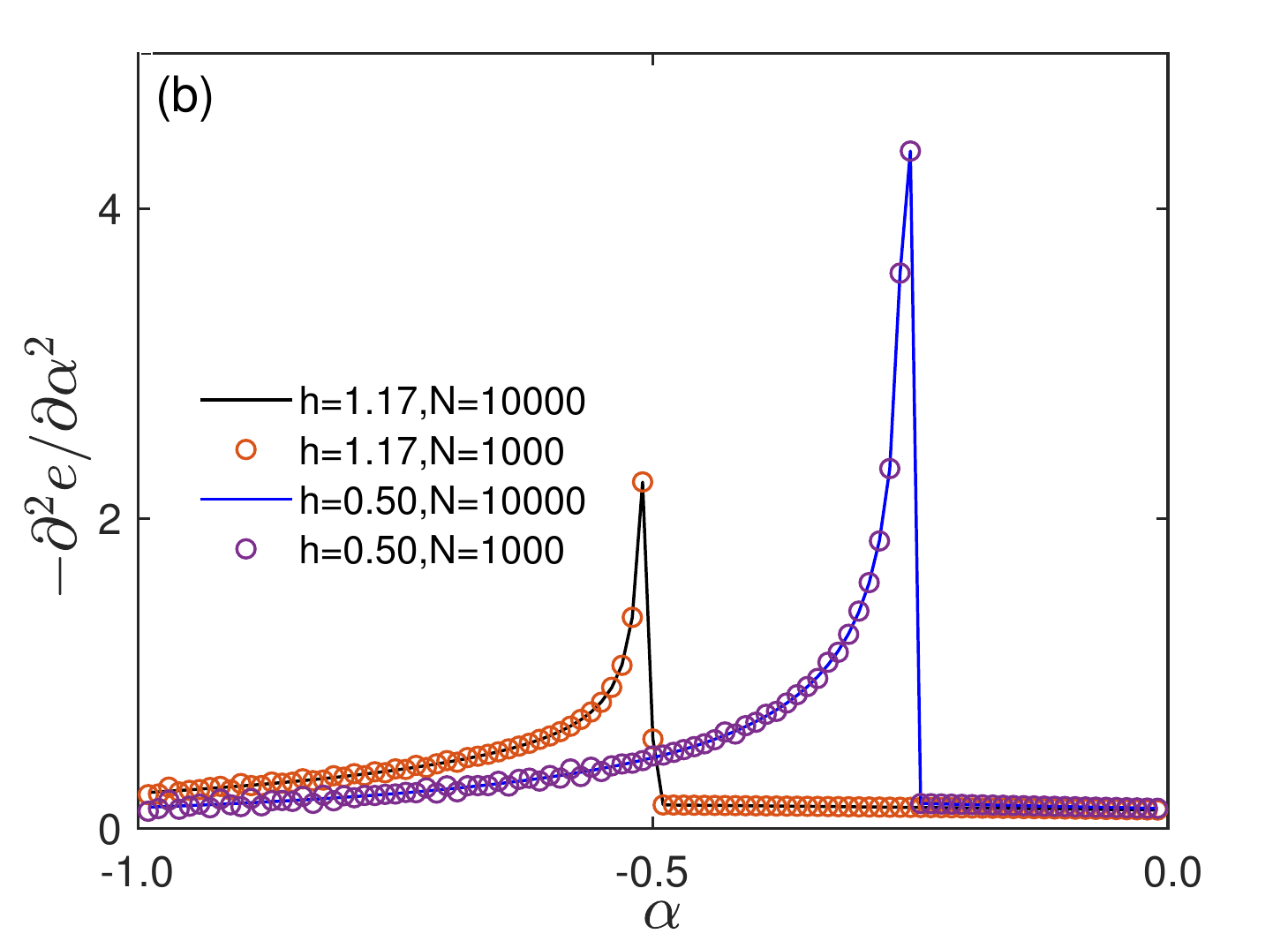}
  \caption{(a)  ${-\partial^2e/\partial h^2}$ as a function of $h$ with $\alpha=0.50$  for various system sizes.
  The left inset shows the scaling behavior between the maximum value of ${-\partial^2e/\partial h^2}$
  and the system size $N$. The right inset shows $-{\partial^2e/\partial h^2}$ in the vicinity of the critical point.  The symbols denote the numerical results, and the solid lines correspond to the linear fittings.
  (b) ${-\partial^2e/\partial \alpha^2}$ vs $\alpha$ with $h=0.50$ and $h=1.17$.
  Other parameters: $J=1.00$, $\gamma=0.60$, $\Gamma=0.60$.}
  \label{energyderivative}
\end{figure}

We also calculate the second derivative of the ground-state
energy density $e_0=-\sum_k \vert {\varepsilon}_{k} \vert/(2N)$ in
Fig.~\ref{energyderivative}, which showcases extreme values around critical points.
With increase of the system sizes, the peaks of $-\partial^2e_0/\partial h^2$ for $\alpha=0.50$ become
more pronounced. To be concrete, a logarithmic singularity across the QPT between phase I and phase II is identified as
\begin{eqnarray}
\left(-\frac{\partial^2{e_0}}{\partial{h^2}}\right)_{\rm max}&=&a_{\rm E} \ln N+c_1.
\end{eqnarray}
Meanwhile, in the vicinity of the critical point in the thermodynamic limit, one finds
 \begin{eqnarray}
\left(-\frac{\partial^2{e_0}}{\partial{h^2}}\right)&=& b_{\rm E}\ln |h-h_c|+c_2.
\end{eqnarray}
The numerical fittings in Fig.~\ref{energyderivative}(a) yield
$a_{\rm E}=0.2871\pm0.0058$, $c_1=0.1878$, $b_{\rm E}=-0.2887\pm0.0009$, and $c_2=0.1162$.
According to the logarithmic scaling ansatz~\cite{scalling2006},
the ratio $|a_E/b_E|$ equals the correlation-length exponent $\nu$ $\simeq 1$,
confirming that the QPT from phase I to  phase II coincides with a second-order transition.
The retrieved specific heat exponent $\alpha$ = $2-(d + z)\nu$=0 validates the scaling relation for the logarithmic scaling in $d=1$ dimension.
In contrast, one observes in Fig.~\ref{energyderivative}(b)
that $-\partial^2e_0/\partial \alpha^2$ exhibits a size-independent discontinuity at the critical points for $h=0.50$ and $h=1.17$,
which is a common feature of the transition between the gapless phase and the gapped phase~\cite{GammaLZA}.

\begin{figure}[ht]
\includegraphics[width=0.85\columnwidth]{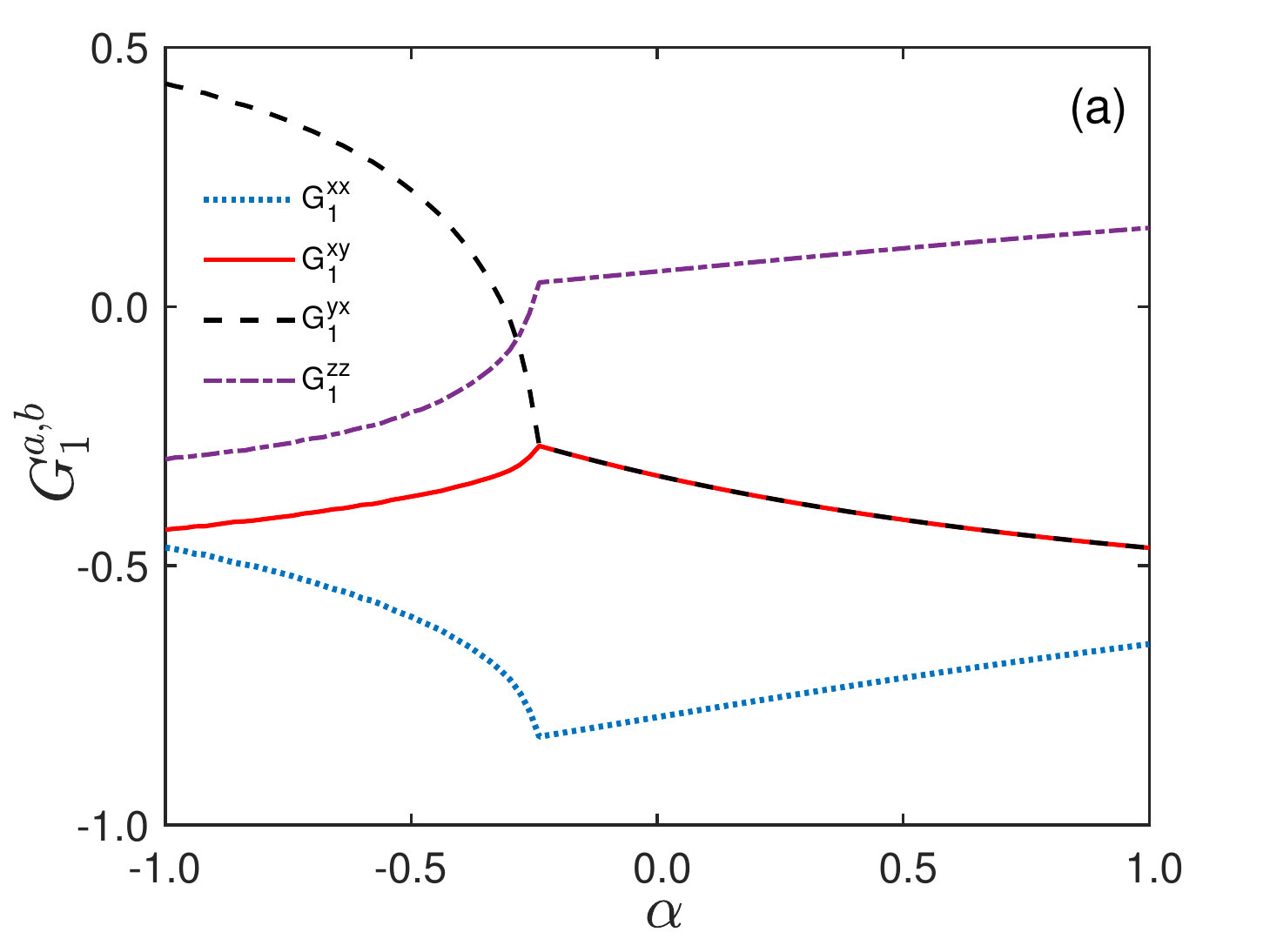}
 \includegraphics[width=0.85\columnwidth]{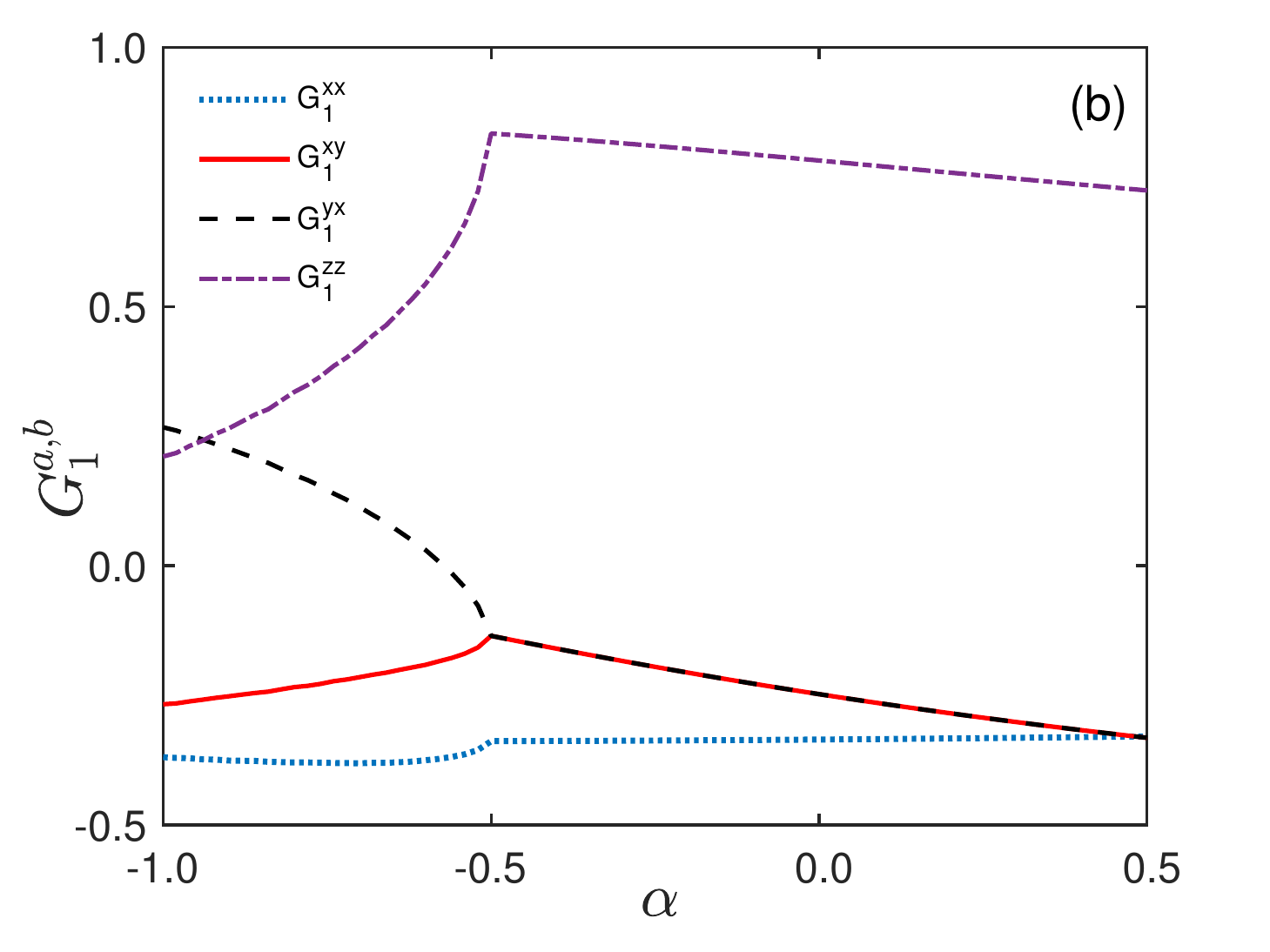}
\caption{ The nearest-neighbor correlation functions
with respect to $\alpha$ for (a) $h=0.50$ and  (b) $h=1.17$.  Other parameters: $N$=2000, $J=1.00$, $\Gamma=0.60$, $\gamma=0.60$.}
\label{G1}
\end{figure}

\begin{figure}[ht]
\includegraphics[width=0.9\columnwidth]{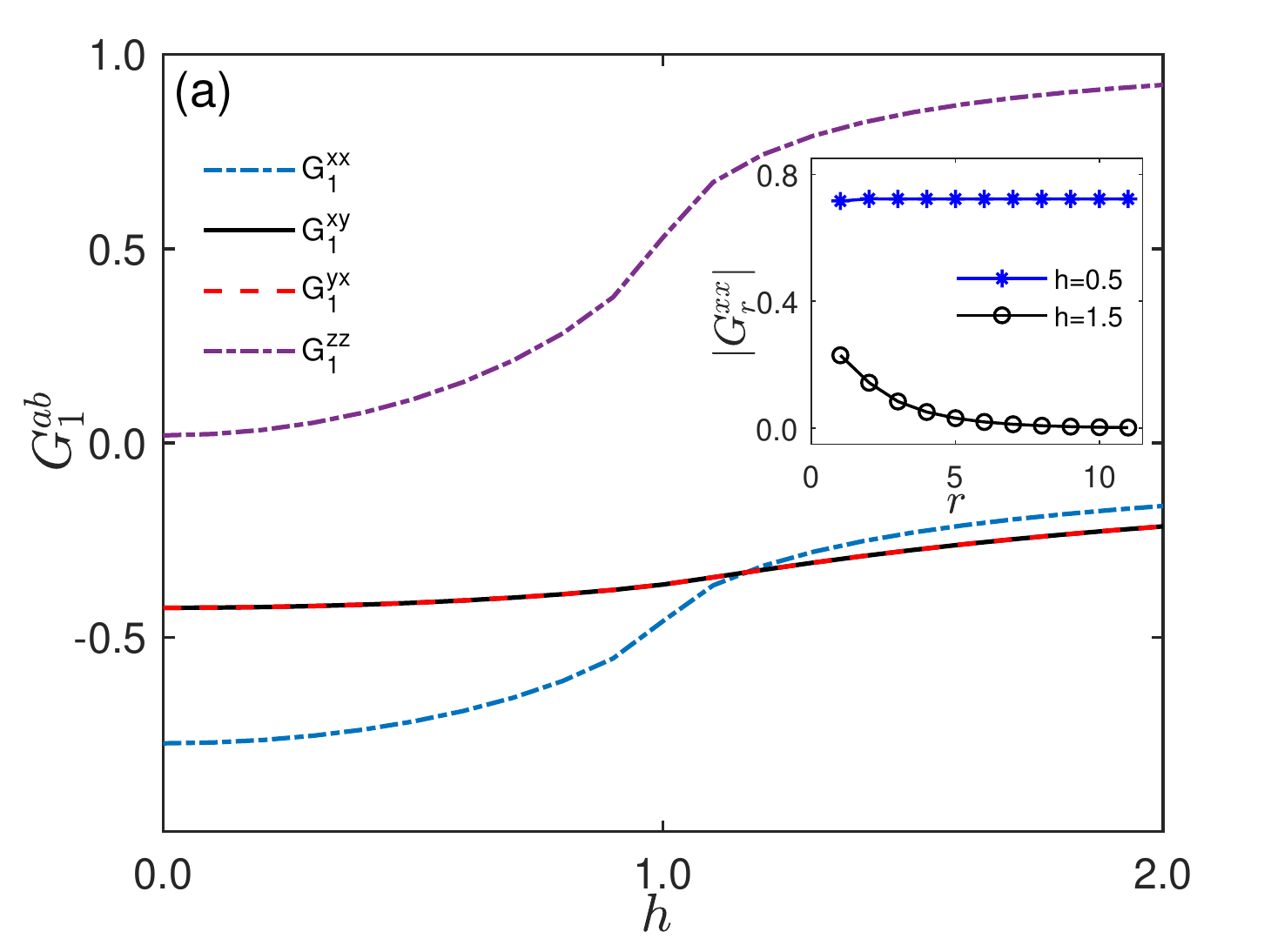}
 \includegraphics[width=0.9\columnwidth]{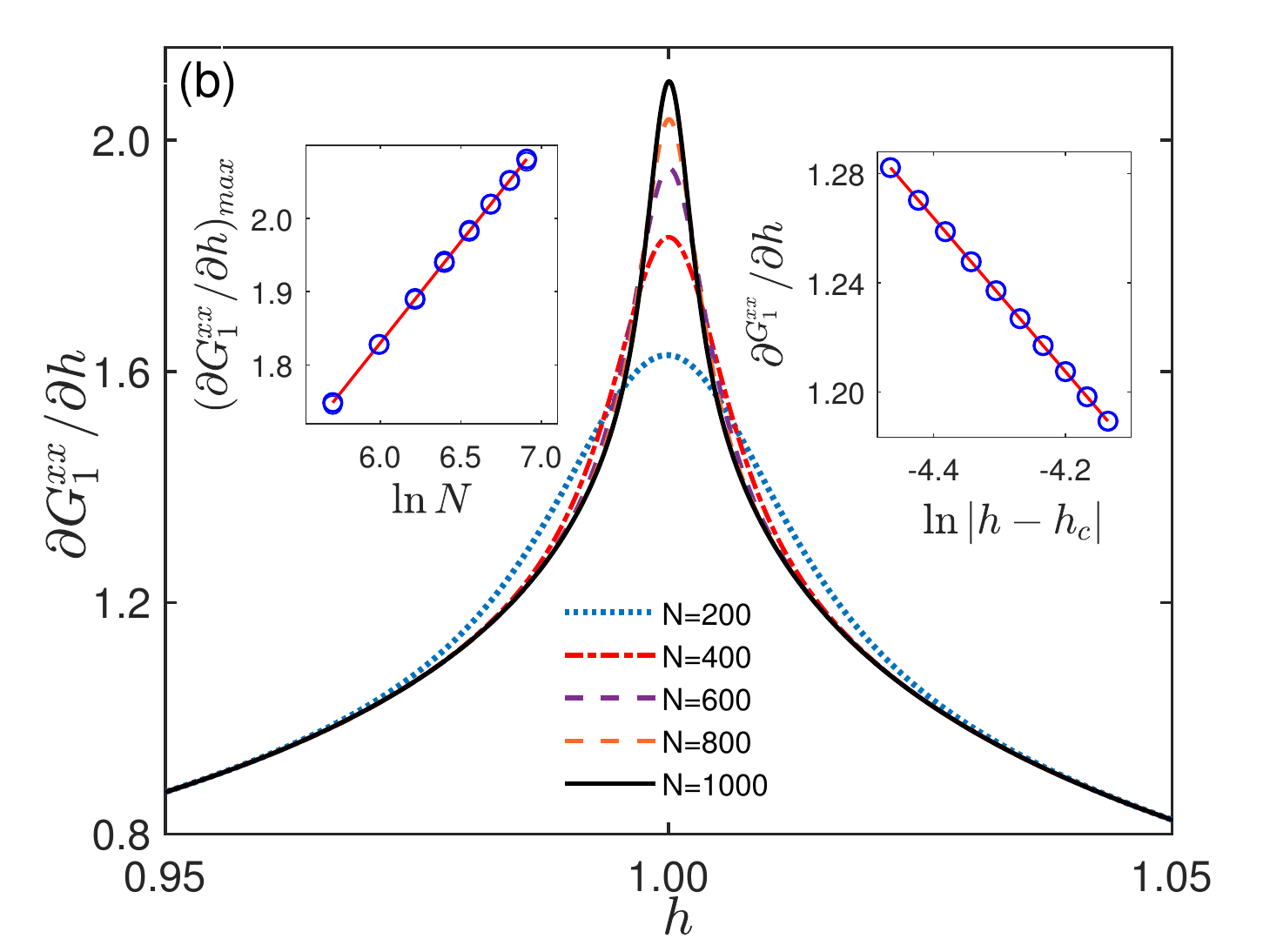}
\caption{(a) The correlation functions $G_{1}^{xx}$, $G_{1}^{zz}$, $G_{1}^{xy}$, $G_{1}^{yx}$ as a function of $h$ for $\alpha=0.50$. The inset shows $\vert G_r^{zz}\vert$ vs $r$ for $h=0.50$ and $1.50$ with $N=2000$.
 (b) The first-order derivative of $G_{1}^{xx}$ with $N=200$, $400$, $600$, $800$, $1000$. The left inset shows the scaling behavior between the maximum value of ${\partial G_{1}^{xx}/\partial h}$ and the system size $N$. The right inset shows $\partial G_{1}^{xx}/\partial h$ in the vicinity of the critical point. The symbols denote the numerical results, and the solid lines correspond to the linear fittings. Other parameters: $J=1.00$, $\gamma=0.60$, $\Gamma=0.60$.
}
\label{Gxx}
\end{figure}

The nature of the
ground state can be gained from the two-qubit correlation functions \begin{eqnarray}
G_{i,j}^{a,b}=\langle\sigma_i^a\sigma_j^b\rangle-\langle\sigma_i^a \rangle \langle\sigma_j^b\rangle
\end{eqnarray}
with $a$, $b$=$x$, $y$, $z$.
In fact, $G_{i,j}^{a,b}$ can be abbreviated as $G_{r}^{a,b}$ with $r=i-j$ due to the translational invariance of 
Eq.~(\ref{eq:Hamiltonian1}). A generic correlation ${\langle { \sigma}_{i}^{a}{ \sigma}_{j}^{b}\rangle}$
can be expressed as a Pfaffian form in terms of Wick's theorem~\cite{Pfaffian}, which is the determinant of the $2r\times 2r$ dimensional antisymmetric matrix.
The detailed calculation is exhibited in Appendix \ref{correlation function1}. One observes that the nearest-neighbor correlation functions display kinks across QCPs between the gapless phase and gapped phases. With increasing $\alpha$ for $h=0.50$ in Fig.~\ref{G1}(a), the dominant nearest-neighbor correlation changes from a positive value of $G_1^{yx}$ to a negative value of $G_{1}^{xx}$, implying a QPT from the gapless spiral phase to the gapped antiferromagnetic (AFM) phase.  Instead, the ruling correlations $G_{1}^{zz}$ in Fig.~\ref{G1}(b) for $h=1.17$ suggest that phase II belongs to the paramagnetic (PM) phase.  Similar trends of correlations for $\alpha=0.50$ are displayed in Fig.~\ref{Gxx}. The dominating nearest-neighbor correlations change from a negative value of $G_{1}^{xx}$ to a positive value of $G_{1}^{zz}$ across $h_{c,1}$. Therefore, the first-order derivative of $G_{1}^{xx}$ presents a pronounced peak at $h_{c,1}$ in Fig.~\ref{Gxx}(b).  We can further find the first-order derivative of $G_{1}^{xx}$ also follows a logarithmic divergence across the second-order QPT as
 \begin{eqnarray}
&&\left(\frac{\partial G^{xx}}{\partial{h}}\right)_{\rm max}= a_{\rm G} \ln{N}+c_3, \\
&&\left(\frac{\partial G^{xx}}{\partial{h}}\right)= b_{\rm G}\ln{|h-h_c|}+c_4,
\end{eqnarray}
where $a_{\rm G}=0.2789\pm0.0045$, $b_{\rm G}=-0.2811\pm0.0001$, $c_3=0.1758, $ and $c_4=0.0265$.
 In this case, one can speculate that $\nu\approx|a_{\rm G}/b_{\rm G}|$=$0.9922\pm0.0209$, which is quite close to the value retrieved from the second derivative of the ground-state energy density.

\begin{figure}[ht]
  \centering
  \includegraphics[width=0.92\columnwidth]{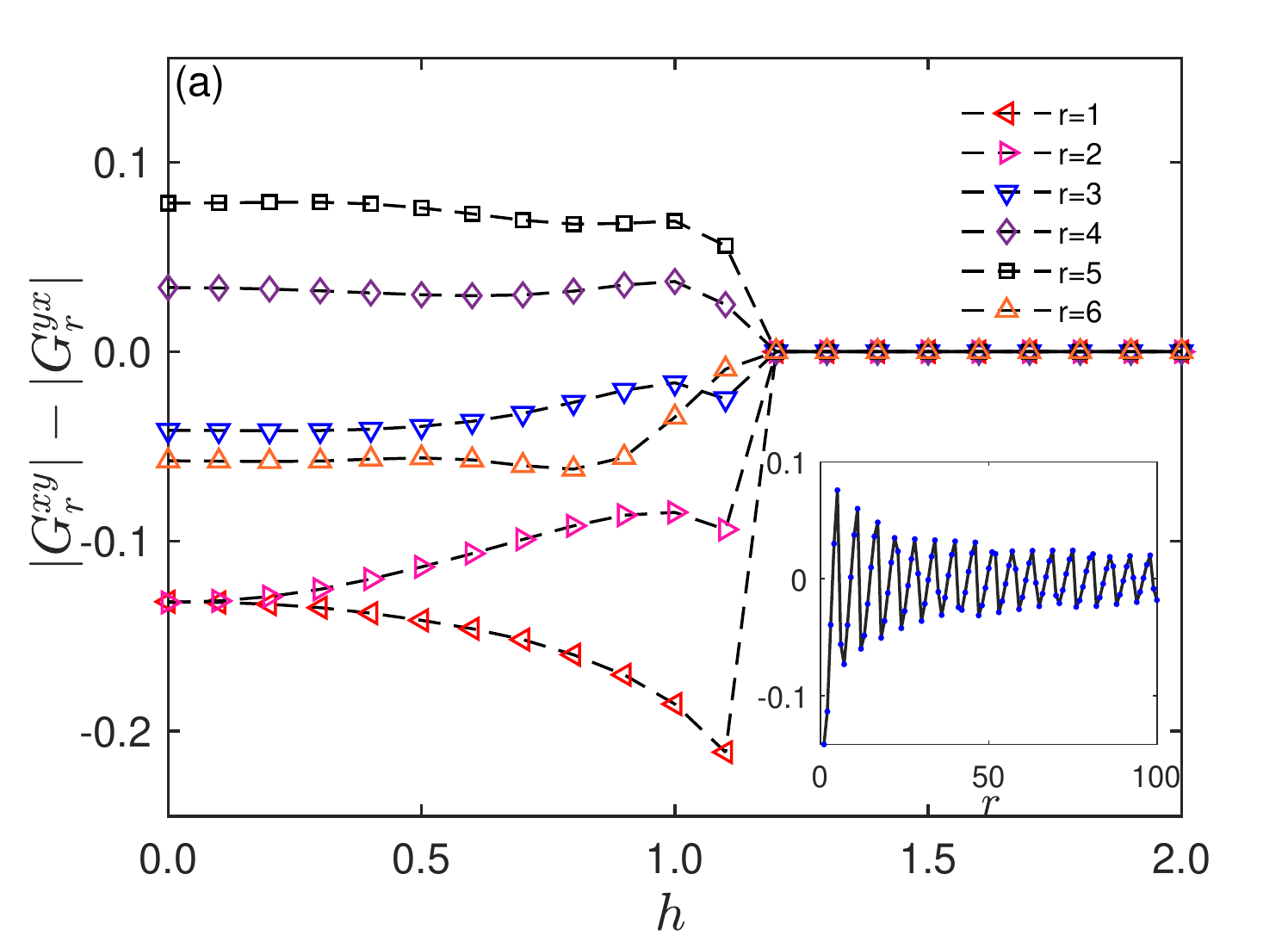}
  \includegraphics[width=0.92\columnwidth]{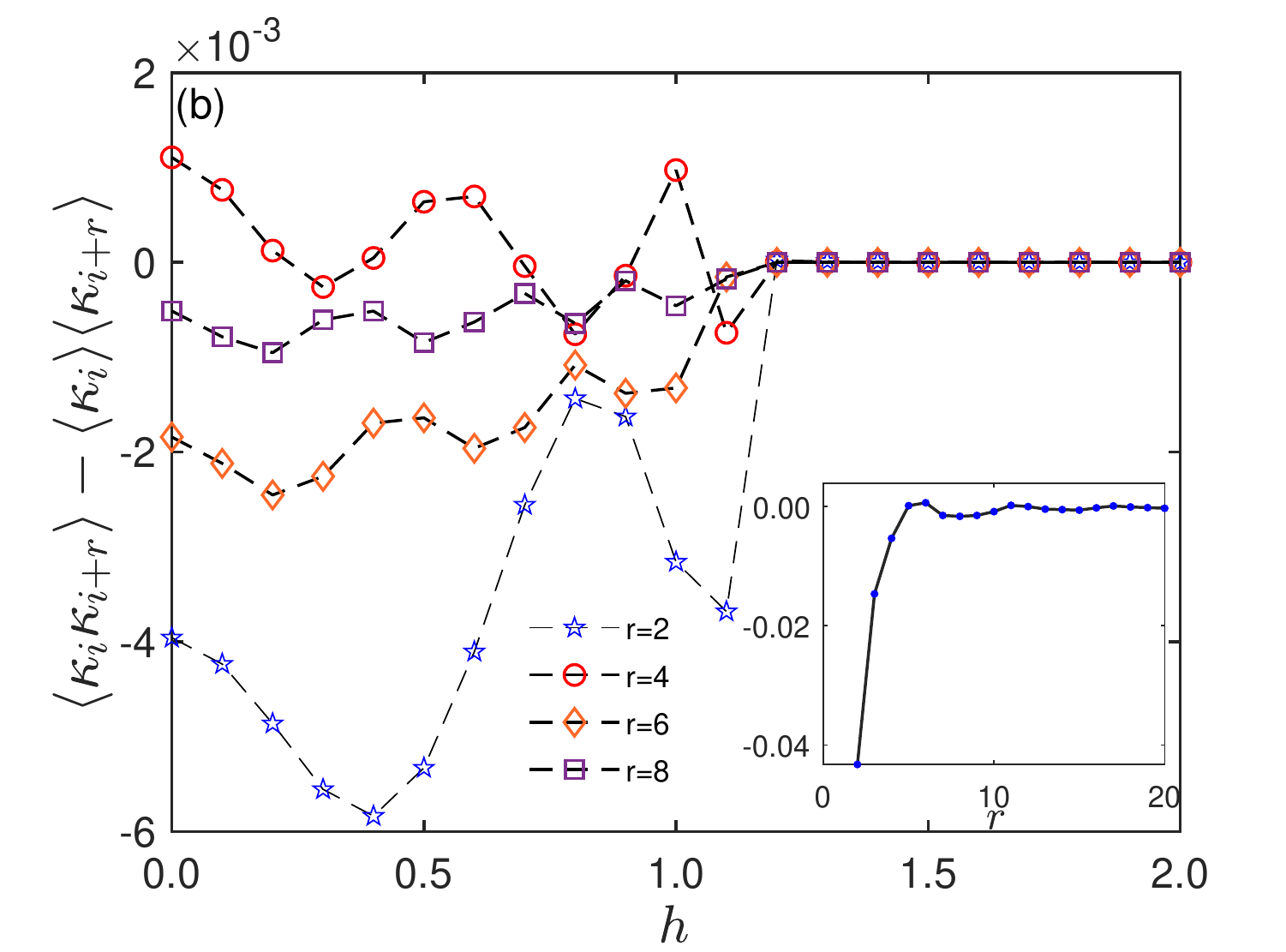}
  \caption{(a) The correlation function $|G_{i,i+r}^{xy}|-|G_{i,i+r}^{yx}|$ with respect to $h$ for different $r$. The inset shows $|G_{i,i+r}^{xy}|-|G_{i,i+r}^{yx}|$ vs $r$ for $h=0.50$. (b) The dimer correlation $\langle \kappa_{i} \kappa_{i+r}\rangle-\langle \kappa_{i} \rangle \langle \kappa_{i+r}\rangle $ with respect to $h$ for different $r$. The inset displays the dimer correlation vs $r$ for $h=0.50$. Other parameters: $N$=2000, $J=1.00$, $\alpha=-0.50$,$\gamma=0.60$, $\Gamma=0.60$.}
\label{Gr}
\end{figure}

It is well known that the AFM phase hosts N\'{e}el LRO, while the LRO is absent in the PM phase, as is unraveled in the inset of Fig. \ref{Gxx}(a). One can further notice that the amplitudes of $G_{1}^{xy}$ and $G_{1}^{yx}$ coincide in the gapped phases, while become unbalanced in the gapless phase. Hence, such a feature suggests that  $\vert G_{1}^{xy}\vert-\vert G_{1}^{yx}\vert$ is a well-defined order parameter to identify the spiral phase for the \emph{XY}-Gamma model. A natural question arises whether there is LRO in the gapless spiral phase. To probe this question, we numerically calculate the vector-chiral correlations $|G_{r}^{xy}|-|G_{r}^{yx}|$  for different distances $r$ in Fig.~\ref{Gr}(a), in which the absolute value is taken in order to remove the indeterminate sign of the numerical Pfaffian calculation. One can find that
$|G_{r}^{xy}|-|G_{r}^{yx}|$ is always zero in the gapped phase as a consequence of $G_r^{xy}=G_r^{yx}$,  while it remains finite in the gapless spiral phase. Upon increasing the distance $r$, the correlation presents an oscillating decline as $r^{-1/2}$ shown in Fig.~\ref{Gr}(a)~\cite{XYDMYTC},
suggesting the existence of a quasi-long-range order of an incommensurate spiral order. Next, to delve more deeply into the spiral order, we consider four-qubit correlations as exemplified
by the dimer correlation
\begin{eqnarray}
D_{j,j+r}=\langle\kappa_j \kappa_{j+r} \rangle- \langle\kappa_j\rangle \langle \kappa_{j+r} \rangle,
\end{eqnarray}
where the $z$-component vector chiral order parameter is defined by~\cite{mcculloch08vector,ueda2014vector}
\begin{eqnarray}
\kappa_j=\left(\vec{\sigma}_j \times \vec{\sigma}_{j+1}\right)^z.
\end{eqnarray}
We thus calculate the dimer correlation $D_{j,j+r}$ as a function of $h$ for $\alpha=0.50$ in Fig.~\ref{Gr}(b).
One observes that the dimer correlations oscillate in the spiral phase and tend to decay with increasing the distance $r$ between the dimers. 
We further find that dimer correlations persist only for a few sites, and thus the four-qubit correlations decay more rapidly than the two-qubit counterparts.

\section{STEERED QUANTUM COHERENCE}\label{SQC}
\begin{figure}[h]
  \centering
  \includegraphics[width=0.9\columnwidth]{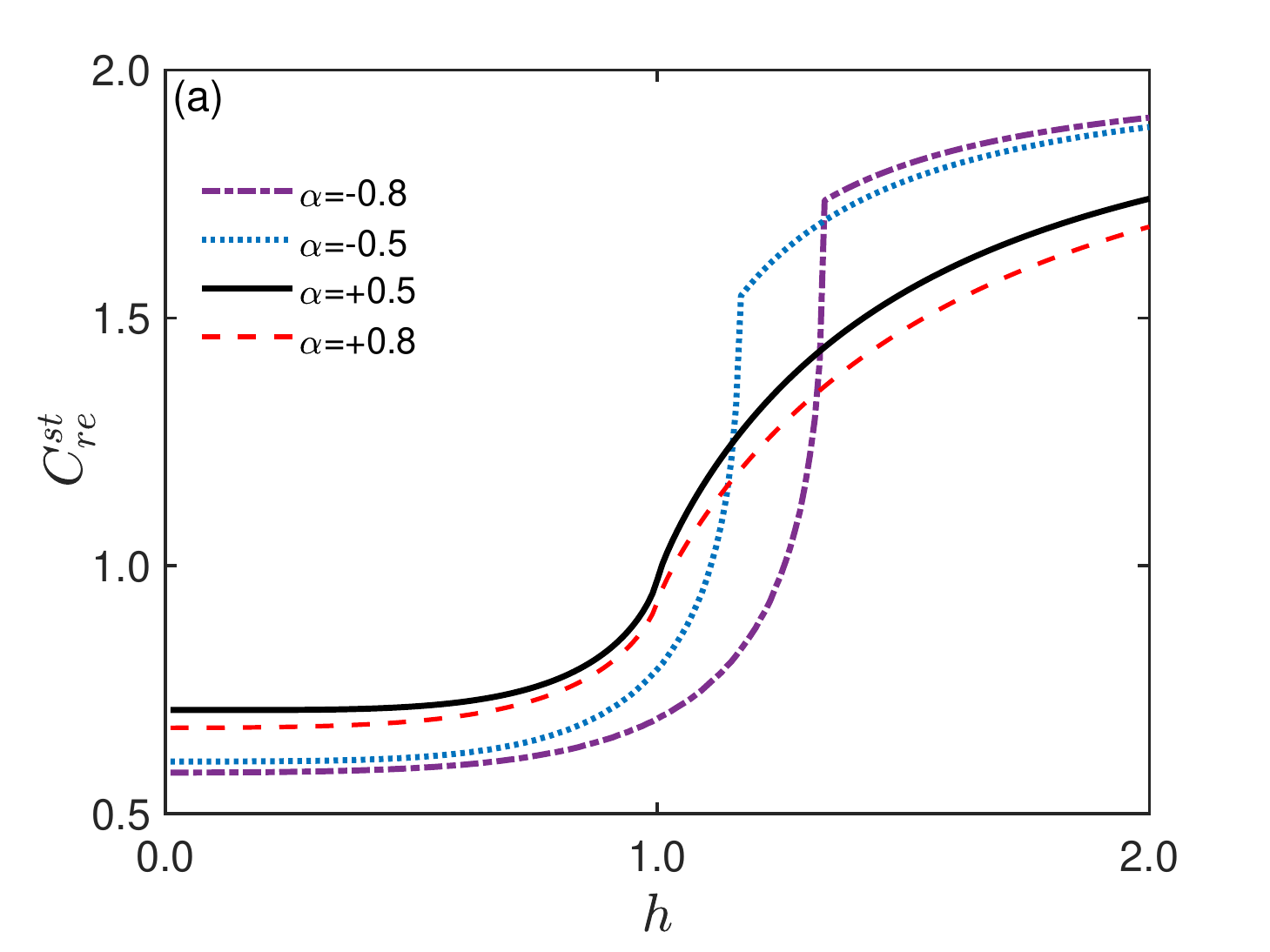}\\
  \includegraphics[width=0.9\columnwidth]{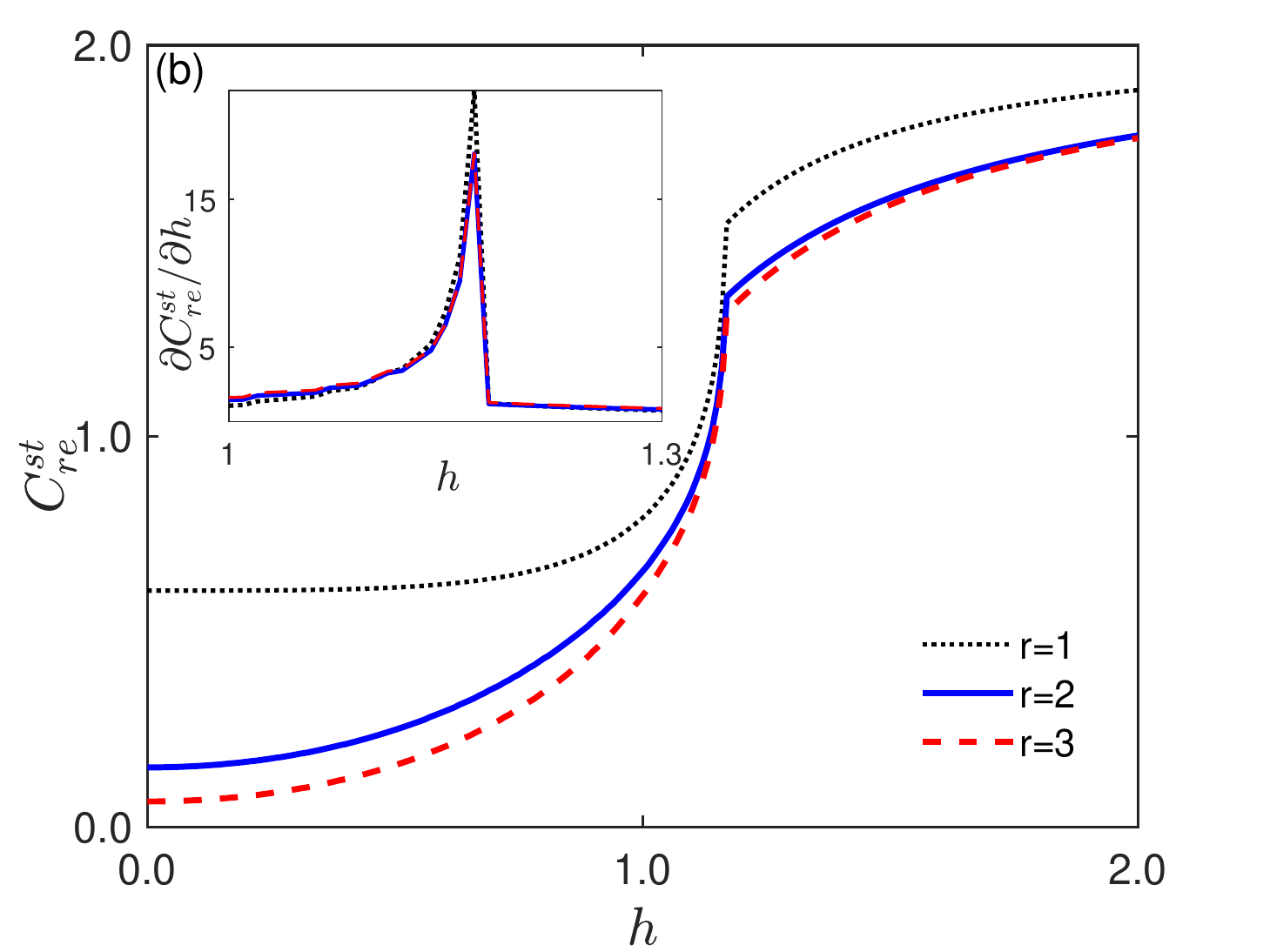}
  \caption{(a) The SQC $C_{re}^{st}$ vs the transverse field $h$. The dash-dotted, dotted, solid and dashed lines correspond to $\alpha=-0.80$, $-0.50$, $0.50$, and $0.80$, respectively.(b) $C_{re}^{st}$ vs $h$ for $\alpha=-0.50$. The dotted, solid  and dashed lines correspond to different distances between the two qubits $r=1$, $2$, and $3$. Inset shows the first derivative of $C_{re}^{st}$ with respect to $h$.  Other parameters: $N$=2000, $J=1.00$, $\gamma=0.60, \Gamma=0.60$.}
  \label{SQC1}
\end{figure}

\begin{figure}[h]
\includegraphics[width=0.9\columnwidth]{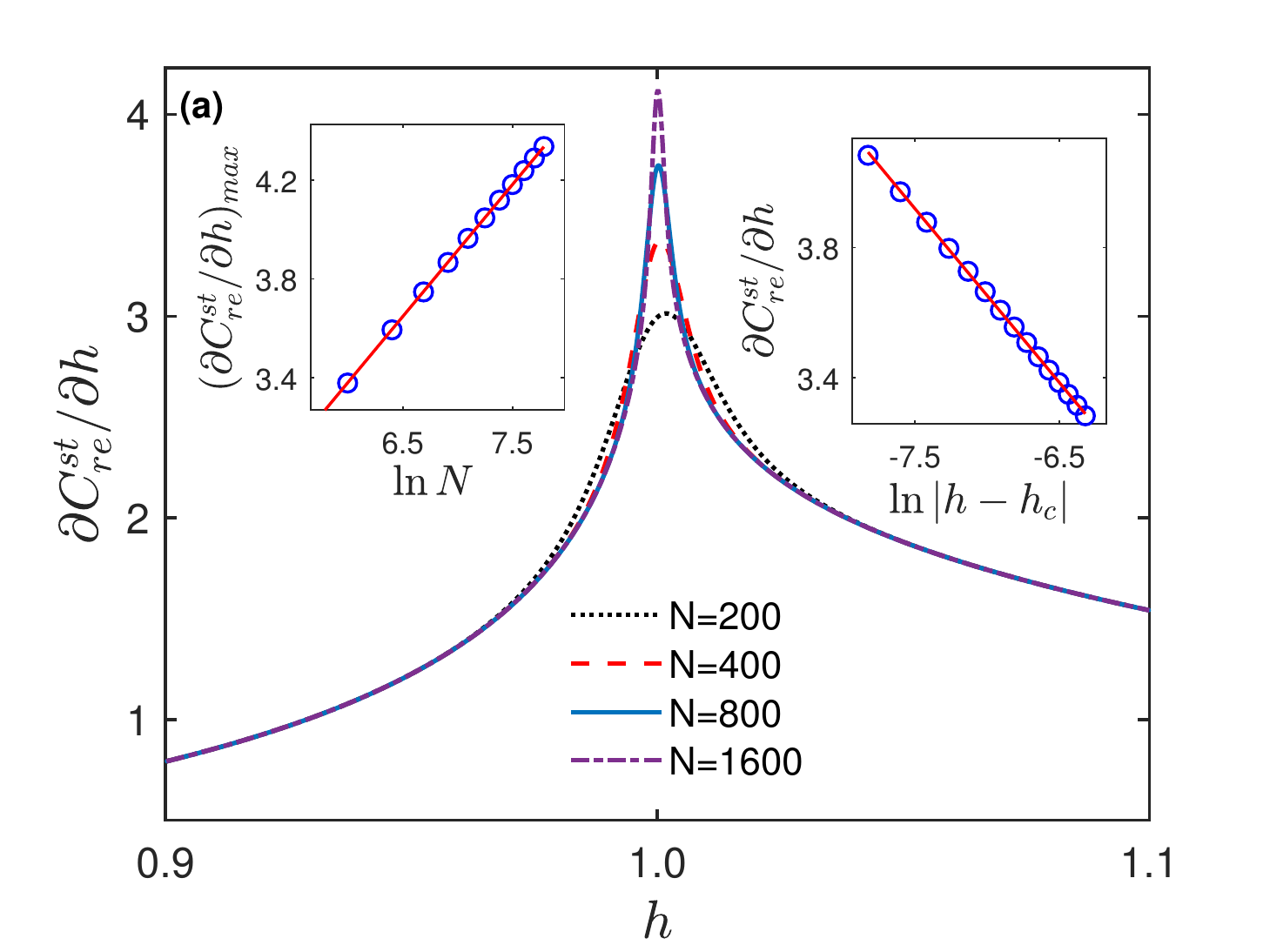}
 \includegraphics[width=0.9\columnwidth]{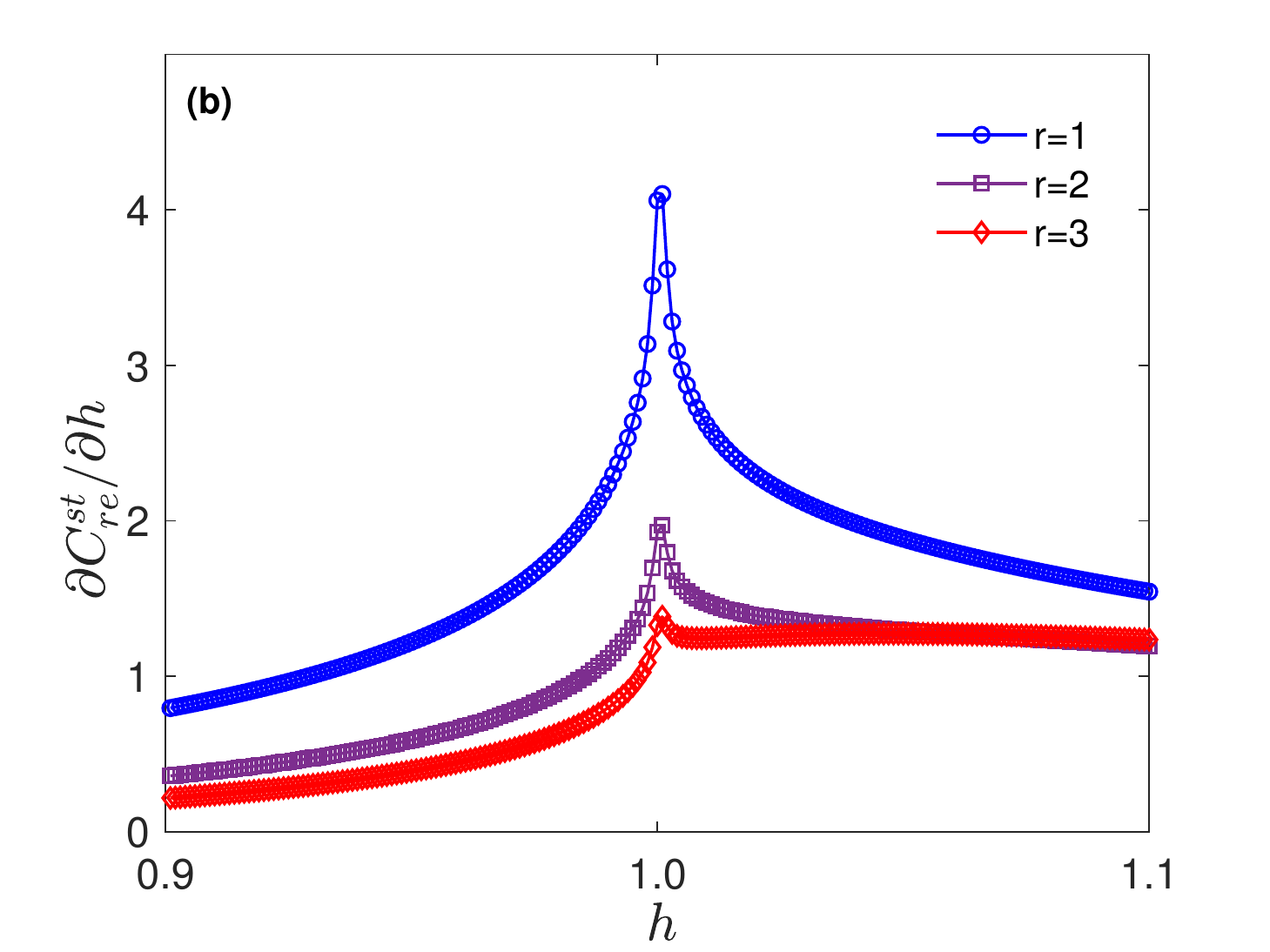}
\caption{(a)$C_{re}^{st}$ vs $h$ for $\alpha=0.50$.  The left inset shows the scaling behavior between the maximum value of ${\partial C_{re}^{st}/\partial h}$ and the system size $N$, and the right inset shows $\partial C_{re}^{st}/\partial h$ in the vicinity of the critical point. (b) The first derivative of $C_{re}^{st}$ with respect to $h$ for $\alpha=0.50$ with $r=1$, $2$, and $r=3$. Other parameters: $N$=2000, $J=1.00$, $\gamma=0.60, \Gamma=0.60$.}
\label{SQCr}
\end{figure}

In recent years, a few approaches inherited from quantum information have been employed to characterize the QPTs, such as quantum entanglement~\cite{vidal2003entanglement,osterloh2002scaling,gu2003entanglement},
quantum discord~\cite{ollivier2001quantum,modi2012classical},
quantum coherence~\cite{chen2016coherence}, and
fidelity susceptibility~\cite{you2007fidelity}.
These information measures have played the role of universal order parameters
and deepened understanding of the
quantumness of correlations at quantum criticality.
Quantum coherence, a landmark manifestation of quantum superposition, has been widely
recognized as a common necessary condition for both entanglement and other types of quantum correlations.
Given that, the precise control over the photon-mediated  interactions between atoms means that  any pair of constituent particles could be ideally coaxed into any desired quantum-mechanical superposition state. In particular, the neutral atoms couple weakly to
the environment, allowing relatively long coherence times. Hence, such an array of atoms can function as a versatile model
for the study of quantum coherence, which may ultimately help
alleviate
the adverse effects of decoherence in quantum computation and quantum information processing~\cite{Bloch2008quantum}.
Based on a rigorous framework to quantify coherence~\cite{Baumgratz14quantify},
a few measures have been proposed, including the relative entropy,  $l_1$ norm of coherence,
Wigner-Yanase-Dyson skew information, and Jensen-Shannon divergence.  These coherence-based indicators have been applied to identify QPTs in many-body systems~\cite{you17emergent,you18quantum,yi19criticality}.
However, their feasibility strongly depends on a careful choice of the specific basis in advance, so these conventional quantum coherence measurements may extract useless information if the reference basis is inappropriate.
To overcome the irrationality, the steered quantum coherence (SQC) was proposed recently~\cite{M.L.Hu2018,M.L.Hu2020,M.L.Hu2021}. The SQC is based on the mutually unbiased bases and shows its figure of merit in characterizing quantum criticality.

For a bipartite state $\rho_{AB}$ shared by  Alice and Bob, the SQC was defined by Alice's local measurements and classical communication between Alice and Bob.
To be explicit, Alice carries out one of some preagreed measurements ${\sigma^\mu}$ $({\mu=x,y,z})$ on qubit $A$ and informs Bob of  the chosen observable $\sigma^\mu$
and the outcome $a\in\{0,1\}$.  Bob's system then collapses to the ensemble states \{${p_{\mu,a},\rho_{B|\Pi_\mu^a}}$\}
with $p_{\mu,a}={\rm tr}(\Pi_\mu^a\rho_{AB})$ being the probability of Alice's outcome $a$,
and $\rho_{B|\Pi_\mu^a}={\rm tr}_A(\Pi_\mu^a\rho_{AB})/p_{\mu,a}$ being Bob's conditional state.
Bob can measure the coherence of the ensemble \{${p_{\mu,a},\rho_{B|\Pi_\mu^a}}$\} with respect to
the eigenbasis of either one of the remaining two Pauli operators $\sigma^\nu$ ($ \nu\neq\mu$).
After  all Alice's  possible measurements 
\{${\Pi_\mu^a}$\}$_{\mu=x,y,z}$ with equal probability,
the SQC of qubit $B$ can be defined as the following averaged quantum coherence:
\begin{eqnarray}
  C_{re}^{st}(\rho_{AB})\!&=&\!\frac{1}{2}\Sigma_{{\mu\neq\nu,a}}  p_{\mu,a}C_{re}^{\sigma^\nu}(\rho_{B|\Pi_\mu^a}),  \quad
\end{eqnarray}
where
\begin{eqnarray}
  C_{re}^{\sigma_\nu}(\rho) &=& S(\rho_d)-S(\rho).
\label{Cre1}
\end{eqnarray}
Here the relative entropy of coherence is used due to its clear physical meaning~\cite{you17emergent}, where
 $S(\rho)=-{\rm tr}(\rho \log_2\rho)$ stands for the von Neumann entropy of $\rho$ and $\rho_d$ is obtained from $\rho$ by removing all its off-diagonal entries.

Regarding $\mathbb{Z}_2$ and translation symmetries of Eq. (\ref{eq:Hamiltonian1}), in the bases spanned by the two-qubit product
states of eigenstate of $\sigma^z$, i.e., $\{\vert 0\rangle_i\otimes\vert 0\rangle_j$,
$\vert 0\rangle_i\otimes\vert 1\rangle_j$,
$\vert 1\rangle_i\otimes\vert 0\rangle_j$,
$\vert 1\rangle_i\otimes\vert 1\rangle_j\}$, where $\vert 0\rangle$ ($\vert 1\rangle$) denotes a spin-up (-down) state, the reduced density matrix $\rho_{ij}$ of two qubits $i$ and $j$ can be cast into an \emph{X}-state form,
\begin{equation}
\rho_{ij}=\left(\begin{array}{cccc}
  {u}^+&0&0&{z}_1\\
   0&{\omega}^+&{z}_2&0\\
    0&{z}_2^*&{\omega}^-&0\\
     {z}_1^*&0&0&{u}^-
  \end{array}\right),
  \label{jz}
\end{equation}
with
\begin{eqnarray}
&&{u}^{\pm}=\frac{1}{4}(1\pm2\langle\sigma_{i}^z\rangle+\langle\sigma_{i}^z\sigma_{j}^z\rangle), \\
&&{z}_1=\frac{1}{4}(\langle\sigma_{i}^x\sigma_{j}^x\rangle-\langle\sigma_{i}^y\sigma_{j}^y\rangle-i\langle\sigma_{i}^x\sigma_{j}^y\rangle-i\langle\sigma_{i}^y\sigma_{j}^x\rangle), \\
&&{z}_2=\frac{1}{4}(\langle\sigma_{i}^x\sigma_{j}^x\rangle+\langle\sigma_{i}^y\sigma_{j}^y\rangle+i\langle\sigma_{i}^x\sigma_{j}^y\rangle-i\langle\sigma_{i}^y\sigma_{j}^x\rangle), \\
&&{\omega}^{\pm}=\frac{1}{4}(1
-\langle\sigma_{i}^z\sigma_{j}^z\rangle).
\label{ys}
\end{eqnarray}

The SQC of two-qubit states as a function of $h$ for different $\alpha$ and  $r$ is plotted in Fig.~\ref{SQC1}.  $C_{re}^{st}$ shows a monotonic increase with respect to $h$, which
tends towards the maximum value 2.00, in contrast to a monotonically decreasing behavior of the relative entropy~\cite{XYDMYTC}. With increasing $h$, the SQC of two adjacent spins shows a smooth transition from the AFM phase to the PM phase, while $C_{re}^{st}$ exhibits a salient point across the transition from the spiral phase to the PM phase [cf. Fig.~\ref{SQC1}(a)].
For two qubits farther than the nearest neighbor, $C_{re}^{st}$ decreases with increasing $r$, but the positions of salient points are unchanged.
As is shown in Fig.~\ref{SQC1}(b), the nonanalyticity of the ground state at QCPs can be pinpointed by the discontinuity of the first-order derivative of the SQC. One can find that the coherence susceptibility, i.e., $\chi_{re}^{st}$ $\equiv$ $\partial C_{re}^{st}/\partial h$~\cite{chen2016coherence},  almost superposes onto each other around the QCPs for different $r$. Similarly, $\chi_{re}^{st}$ presents a pronounced peak at $h_{c,1}=1$ for $\alpha=0.50$ in Fig.~\ref{SQCr}(a) and the peaks become sharper and sharper as the system size increases, and it is expected to diverge in the thermodynamic limit. The singularity of $\chi_{re}^{st}$ is manifested in
the logarithmic scaling as
 \begin{eqnarray}
&&\left(\chi_{re}^{st}\right)_{\rm max}= a_{\rm C} \ln{N}+c_5, \label{Cst5}\\
&& \chi_{re}^{st}= b_{\rm C}\ln{|h-h_c|}+c_6.\label{Cst6} \quad
\end{eqnarray}
One further finds in Fig.~\ref{SQCr}(b) that the coherence susceptibilities $\chi_{re}^{st}$ for different $r$ also obey the logarithmic scaling.
The fitting parameters are listed in Table \ref{Fittingparameter}, and the extracted values of $\nu$ agree well with each other, although the deterioration of the precision with increasing $r$ can be easily noticed.

\begin{table}[bh!]
\caption{Fitting parameters $\{$$a_{\rm C}$, $b_{\rm C}$$\}$ of the slopes in logarithmic scaling of coherence susceptibility through relations Eqs.~(\ref{Cst5}) and (\ref{Cst6}) with $\nu=|a_{\rm C}/b_{\rm C}|$. Other parameters: $J=1.00$, $\gamma=0.60, \Gamma=0.60$, $\alpha= 0.50$.}
\label{Fittingparameter}
\begin{ruledtabular}
\begin{tabular}{ c c c c }
	$r$	&$a_{\rm C}$&$b_{\rm C}$&$\nu$ \\
\hline\\
1 &0.5084$\pm$0.0087&-0.5134$\pm$0.0091&0.9903$\pm$0.0355\\
2 &0.2065$\pm$0.0010&-0.2125$\pm$0.0049&0.9718$\pm$0.0272 \\
3 &0.0889$\pm$0.0027&-0.0933$\pm$0.0052& 0.9528$\pm$0.0777\\
\end{tabular}
\end{ruledtabular}
\end{table}

\section{Conclusion and Discussion}\label{summary}
In this work, we show that an elaborate scheme in atom-cavity systems can engineer an 
effective Hamiltonian composed of various types of couplings, including \emph{XY}, Dzyaloshinskii-Moriya, and symmetric off-diagonal $\Gamma$ interactions.
We explore the quantum criticality in the so-called \emph{XY}-Gamma model, in which only nearest-neighbor interactions
between the particles are allowed. The 
intricate interplay of diverse controlled exchange interactions between atoms in the presence of external fields
counteracts a rich variety of quantum phases at equilibrium.
The Hamiltonian can be rigorously solved through Jordan-Wigner and Bogoliubov transformation.
The exact solutions endow us with precise knowledge of ground-state properties.
For generic values of the parameters,
the phase diagram consists of the antiferromagnetic (AFM) phase, the paramagnetic (PM) phase,  and the gapless spiral phase. The second derivative of the ground-state energy diverges logarithmically across the AFM-PM transition, while it displays
a discontinuity at the critical point between the gapless spiral phase and gapped phases. Similar characteristics can be confirmed by the nearest-neighbor correlations and steered quantum coherence (SQC).
Moreover, we show that the gapless phase is characterized by a quasi-long-range order of an incommensurate spin spiral.
In a sense, the vector-chiral operator for two qubits is proven to act as a suitable order parameter for discerning the Tomonaga-Luttinger liquid. In contrast, the dimer order correlations vanish rapidly with the distance between the dimers.
The findings reveal that the incommensurate phase transitions away from the Tomonaga-Luttinger-liquid (TLL) phase all belong to the second-order transitions.
As a hallmark of critical phenomena in a continuous quantum phase transitions, critical points with the same set of critical exponents are categorized into a universality class, among which the dynamical exponent $z$ and the correlation-length exponent $\nu$ are most crucial.
For the quantum phase transitions between the AFM phase and the PM phase, one finds the correlation-length exponent can be obtained from several scaling forms including the gap and the correlation function as well as the SQC. The critical exponents $\nu = 1$ and $z = 1$ clearly indicate the transition  belong to two-dimensional Ising universality class. As regards the transitions from the TLL phase driven by either the off-diagonal exchange coupling ratio $\alpha$ or the magnetic field $h$, we obtain explicit forms for the energy gap near the critical points yielding
 $z$=2 and  $\nu$=1/2, signaling the critical points on the boundaries of the gapless phase are in the Lifshitz universality class. Thus, an experimental measurement of the correlation-length exponent $\nu$ and the dynamical exponent $z$ becomes tractable. The critical exponents $z$ and $\nu$ can be extracted from specific heat exponent $\alpha$ according to the hyperscaling relation $\nu+\nu z$=$2-\alpha$ or through
the Kibble–Zurek exponent $\mu=\nu/(1+\nu z)$~\cite{PhysRevB.99.094203,chepiga2021kibble}.  The finite-size effect theory developed for the Tomonaga-Luttinger liquid and the associated Lifshitz universality class speaks to the experimental verification for a finite number of atoms.

To conclude, the reported results may serve to test other approximate techniques used to study more realistic models.  This provides an interesting platform to understand the validity of characterizing tools in identifying unconventional transitions. The emergent phenomena in quantum many-body systems [Eq.~(\ref{eq:Hamiltonian0})] with cavity-induced long-range interactions await  further study. In particular, it becomes possible to realize nonequilibrium many-body phenomena in a controlled way~\cite{PhysRevLett.123.027204}, which are inaccessible for conventional solid-state materials. From both an experimental and a theoretical point of view, quantum simulations in the AMO systems offer outstanding possibilities for measuring  quantum coherence encoded in the many-body systems. Thus, the nature of the ground state and the coherence dynamics of the many-body systems can be unveiled, providing a hallmark of the TLL spin dynamics in the one-dimensional AFM chain. However, controlling atoms and coherence distillations with single-site resolution in an optical lattice remains a huge challenge.

\section*{ACKNOWLEDGMENTS}
The authors appreciate insightful discussions with Z.-A. Liu, X. Li, and T. Lv. W.-L. Y. is supported by the National Natural Science Foundation of China (NSFC) under Grant No.~12174194,
the startup fund (Grant No.~1008-YAH20006) of Nanjing University of Aeronautics and Astronautics (NUAA) , Top-notch Academic Programs Project of Jiangsu Higher Education Institutions (TAPP), and stable supports for basic institute research (Grant No.~190101). M. X. acknowledges support by the startup fund (Grant No.~1008-YAT21004) of NUAA
and the Open Research Fund Program of the State Key Laboratory of Low-Dimensional Quantum Physics (Grant No.~KF202111).
\bibliography{mybib0608}

\begin{thebibliography}{81}%
\makeatletter
\providecommand \@ifxundefined [1]{%
 \@ifx{#1\undefined}
}%
\providecommand \@ifnum [1]{%
 \ifnum #1\expandafter \@firstoftwo
 \else \expandafter \@secondoftwo
 \fi
}%
\providecommand \@ifx [1]{%
 \ifx #1\expandafter \@firstoftwo
 \else \expandafter \@secondoftwo
 \fi
}%
\providecommand \natexlab [1]{#1}%
\providecommand \enquote  [1]{``#1''}%
\providecommand \bibnamefont  [1]{#1}%
\providecommand \bibfnamefont [1]{#1}%
\providecommand \citenamefont [1]{#1}%
\providecommand \href@noop [0]{\@secondoftwo}%
\providecommand \href [0]{\begingroup \@sanitize@url \@href}%
\providecommand \@href[1]{\@@startlink{#1}\@@href}%
\providecommand \@@href[1]{\endgroup#1\@@endlink}%
\providecommand \@sanitize@url [0]{\catcode `\\12\catcode `\$12\catcode
  `\&12\catcode `\#12\catcode `\^12\catcode `\_12\catcode `\%12\relax}%
\providecommand \@@startlink[1]{}%
\providecommand \@@endlink[0]{}%
\providecommand \url  [0]{\begingroup\@sanitize@url \@url }%
\providecommand \@url [1]{\endgroup\@href {#1}{\urlprefix }}%
\providecommand \urlprefix  [0]{URL }%
\providecommand \Eprint [0]{\href }%
\providecommand \doibase [0]{https://doi.org/}%
\providecommand \selectlanguage [0]{\@gobble}%
\providecommand \bibinfo  [0]{\@secondoftwo}%
\providecommand \bibfield  [0]{\@secondoftwo}%
\providecommand \translation [1]{[#1]}%
\providecommand \BibitemOpen [0]{}%
\providecommand \bibitemStop [0]{}%
\providecommand \bibitemNoStop [0]{.\EOS\space}%
\providecommand \EOS [0]{\spacefactor3000\relax}%
\providecommand \BibitemShut  [1]{\csname bibitem#1\endcsname}%
\let\auto@bib@innerbib\@empty
\bibitem [{\citenamefont {Sachdev}(2008)}]{sachdev2008quantum}%
  \BibitemOpen
  \bibfield  {author} {\bibinfo {author} {\bibfnamefont {S.}~\bibnamefont
  {Sachdev}},\ }\bibfield  {title} {\bibinfo {title} {Quantum magnetism and
  criticality},\ }\href {https://www.nature.com/articles/nphys894} {\bibfield
  {journal} {\bibinfo  {journal} {Nature Physics}\ }\textbf {\bibinfo {volume}
  {4}},\ \bibinfo {pages} {173} (\bibinfo {year} {2008})}\BibitemShut {NoStop}%
\bibitem [{\citenamefont {Mazurenko}\ and\ \citenamefont
  {Anisimov}(2005)}]{Fe}%
  \BibitemOpen
  \bibfield  {author} {\bibinfo {author} {\bibfnamefont {V.~V.}\ \bibnamefont
  {Mazurenko}}\ and\ \bibinfo {author} {\bibfnamefont {V.~I.}\ \bibnamefont
  {Anisimov}},\ }\bibfield  {title} {\bibinfo {title} {Weak ferromagnetism in
  antiferromagnets:
  $\ensuremath{\alpha}\text{\ensuremath{-}}\mathrm{Fe}_{2}\mathrm{O}_{3}$ and
  $\mathrm{La}_{2}\mathrm{Cu}\mathrm{O}_{4}$},\ }\href
  {https://doi.org/10.1103/PhysRevB.71.184434} {\bibfield  {journal} {\bibinfo
  {journal} {Phys. Rev. B}\ }\textbf {\bibinfo {volume} {71}},\ \bibinfo
  {pages} {184434} (\bibinfo {year} {2005})}\BibitemShut {NoStop}%
\bibitem [{\citenamefont {Fink}\ and\ \citenamefont {Shaltiel}(1963)}]{MnCO}%
  \BibitemOpen
  \bibfield  {author} {\bibinfo {author} {\bibfnamefont {H.~J.}\ \bibnamefont
  {Fink}}\ and\ \bibinfo {author} {\bibfnamefont {D.}~\bibnamefont
  {Shaltiel}},\ }\bibfield  {title} {\bibinfo {title} {High-frequency resonance
  of a weak ferromagnet: {MnC${\mathrm{O}}_{3}$}},\ }\href
  {https://doi.org/10.1103/PhysRev.130.627} {\bibfield  {journal} {\bibinfo
  {journal} {Phys. Rev.}\ }\textbf {\bibinfo {volume} {130}},\ \bibinfo {pages}
  {627} (\bibinfo {year} {1963})}\BibitemShut {NoStop}%
\bibitem [{\citenamefont {Zhang}\ \emph {et~al.}(2021)\citenamefont {Zhang},
  \citenamefont {Liu}, \citenamefont {Dong}, \citenamefont {Wu}, \citenamefont
  {Zhang}, \citenamefont {Wang}, \citenamefont {Lu}, \citenamefont
  {R\"uckriegel}, \citenamefont {Wang}, \citenamefont {Duine}, \citenamefont
  {Yu}, \citenamefont {Luo}, \citenamefont {Shen},\ and\ \citenamefont
  {Zhang}}]{zhang2021strain}%
  \BibitemOpen
  \bibfield  {author} {\bibinfo {author} {\bibfnamefont {Y.}~\bibnamefont
  {Zhang}}, \bibinfo {author} {\bibfnamefont {J.}~\bibnamefont {Liu}}, \bibinfo
  {author} {\bibfnamefont {Y.}~\bibnamefont {Dong}}, \bibinfo {author}
  {\bibfnamefont {S.}~\bibnamefont {Wu}}, \bibinfo {author} {\bibfnamefont
  {J.}~\bibnamefont {Zhang}}, \bibinfo {author} {\bibfnamefont
  {J.}~\bibnamefont {Wang}}, \bibinfo {author} {\bibfnamefont {J.}~\bibnamefont
  {Lu}}, \bibinfo {author} {\bibfnamefont {A.}~\bibnamefont {R\"uckriegel}},
  \bibinfo {author} {\bibfnamefont {H.}~\bibnamefont {Wang}}, \bibinfo {author}
  {\bibfnamefont {R.}~\bibnamefont {Duine}}, \bibinfo {author} {\bibfnamefont
  {H.}~\bibnamefont {Yu}}, \bibinfo {author} {\bibfnamefont {Z.}~\bibnamefont
  {Luo}}, \bibinfo {author} {\bibfnamefont {K.}~\bibnamefont {Shen}},\ and\
  \bibinfo {author} {\bibfnamefont {J.}~\bibnamefont {Zhang}},\ }\bibfield
  {title} {\bibinfo {title} {Strain-driven dzyaloshinskii-moriya interaction
  for room-temperature magnetic skyrmions},\ }\href
  {https://doi.org/10.1103/PhysRevLett.127.117204} {\bibfield  {journal}
  {\bibinfo  {journal} {Phys. Rev. Lett.}\ }\textbf {\bibinfo {volume} {127}},\
  \bibinfo {pages} {117204} (\bibinfo {year} {2021})}\BibitemShut {NoStop}%
\bibitem [{\citenamefont {Liu}\ \emph {et~al.}(2021{\natexlab{a}})\citenamefont
  {Liu}, \citenamefont {Song}, \citenamefont {Wu}, \citenamefont {Yang},
  \citenamefont {Yang}, \citenamefont {Lu}, \citenamefont {Tian}, \citenamefont
  {Sun}, \citenamefont {Lu}, \citenamefont {Wang} \emph {et~al.}}]{nature2021}%
  \BibitemOpen
  \bibfield  {author} {\bibinfo {author} {\bibfnamefont {X.}~\bibnamefont
  {Liu}}, \bibinfo {author} {\bibfnamefont {W.}~\bibnamefont {Song}}, \bibinfo
  {author} {\bibfnamefont {M.}~\bibnamefont {Wu}}, \bibinfo {author}
  {\bibfnamefont {Y.}~\bibnamefont {Yang}}, \bibinfo {author} {\bibfnamefont
  {Y.}~\bibnamefont {Yang}}, \bibinfo {author} {\bibfnamefont {P.}~\bibnamefont
  {Lu}}, \bibinfo {author} {\bibfnamefont {Y.}~\bibnamefont {Tian}}, \bibinfo
  {author} {\bibfnamefont {Y.}~\bibnamefont {Sun}}, \bibinfo {author}
  {\bibfnamefont {J.}~\bibnamefont {Lu}}, \bibinfo {author} {\bibfnamefont
  {J.}~\bibnamefont {Wang}}, \emph {et~al.},\ }\bibfield  {title} {\bibinfo
  {title} {Magnetoelectric phase transition driven by interfacial-engineered
  {D}zyaloshinskii-{M}oriya interaction},\ }\href
  {https://doi.org/10.1038/s41467-021-25759-1} {\bibfield  {journal} {\bibinfo
  {journal} {Nature communications}\ }\textbf {\bibinfo {volume} {12}},\
  \bibinfo {pages} {5453} (\bibinfo {year} {2021}{\natexlab{a}})}\BibitemShut
  {NoStop}%
\bibitem [{\citenamefont {Mu\"uhlbauer}\ \emph {et~al.}(2009)\citenamefont
  {Mu\"uhlbauer}, \citenamefont {Binz}, \citenamefont {Jonietz}, \citenamefont
  {Pfleiderer}, \citenamefont {Rosch}, \citenamefont {Neubauer}, \citenamefont
  {Georgii},\ and\ \citenamefont {Bo\"oni}}]{muhlbauer09skyrmion}%
  \BibitemOpen
  \bibfield  {author} {\bibinfo {author} {\bibfnamefont {S.}~\bibnamefont
  {Mu\"uhlbauer}}, \bibinfo {author} {\bibfnamefont {B.}~\bibnamefont {Binz}},
  \bibinfo {author} {\bibfnamefont {F.}~\bibnamefont {Jonietz}}, \bibinfo
  {author} {\bibfnamefont {C.}~\bibnamefont {Pfleiderer}}, \bibinfo {author}
  {\bibfnamefont {A.}~\bibnamefont {Rosch}}, \bibinfo {author} {\bibfnamefont
  {A.}~\bibnamefont {Neubauer}}, \bibinfo {author} {\bibfnamefont
  {R.}~\bibnamefont {Georgii}},\ and\ \bibinfo {author} {\bibfnamefont
  {P.}~\bibnamefont {Bo\"oni}},\ }\bibfield  {title} {\bibinfo {title}
  {Skyrmion lattice in a chiral magnet},\ }\href
  {https://www.science.org/doi/10.1126/science.1166767} {\bibfield  {journal}
  {\bibinfo  {journal} {Science}\ }\textbf {\bibinfo {volume} {323}},\ \bibinfo
  {pages} {915} (\bibinfo {year} {2009})}\BibitemShut {NoStop}%
\bibitem [{\citenamefont {Yu}\ \emph {et~al.}(2010)\citenamefont {Yu},
  \citenamefont {Onose}, \citenamefont {Kanazawa}, \citenamefont {Park},
  \citenamefont {Han}, \citenamefont {Matsui}, \citenamefont {Nagaosa},\ and\
  \citenamefont {Tokura}}]{yu2010real}%
  \BibitemOpen
  \bibfield  {author} {\bibinfo {author} {\bibfnamefont {X.}~\bibnamefont
  {Yu}}, \bibinfo {author} {\bibfnamefont {Y.}~\bibnamefont {Onose}}, \bibinfo
  {author} {\bibfnamefont {N.}~\bibnamefont {Kanazawa}}, \bibinfo {author}
  {\bibfnamefont {J.~H.}\ \bibnamefont {Park}}, \bibinfo {author}
  {\bibfnamefont {J.}~\bibnamefont {Han}}, \bibinfo {author} {\bibfnamefont
  {Y.}~\bibnamefont {Matsui}}, \bibinfo {author} {\bibfnamefont
  {N.}~\bibnamefont {Nagaosa}},\ and\ \bibinfo {author} {\bibfnamefont
  {Y.}~\bibnamefont {Tokura}},\ }\bibfield  {title} {\bibinfo {title}
  {Real-space observation of a two-dimensional skyrmion crystal},\ }\href
  {https://www.nature.com/articles/nature09124} {\bibfield  {journal} {\bibinfo
   {journal} {Nature}\ }\textbf {\bibinfo {volume} {465}},\ \bibinfo {pages}
  {901} (\bibinfo {year} {2010})}\BibitemShut {NoStop}%
\bibitem [{\citenamefont {Katukuri}\ \emph {et~al.}(2014)\citenamefont
  {Katukuri}, \citenamefont {Nishimoto}, \citenamefont {Yushankhai},
  \citenamefont {Stoyanova}, \citenamefont {Kandpal}, \citenamefont {Choi},
  \citenamefont {Coldea}, \citenamefont {Rousochatzakis}, \citenamefont
  {Hozoi},\ and\ \citenamefont {Van Den~Brink}}]{Kiatev2014}%
  \BibitemOpen
  \bibfield  {author} {\bibinfo {author} {\bibfnamefont {V.~M.}\ \bibnamefont
  {Katukuri}}, \bibinfo {author} {\bibfnamefont {S.}~\bibnamefont {Nishimoto}},
  \bibinfo {author} {\bibfnamefont {V.}~\bibnamefont {Yushankhai}}, \bibinfo
  {author} {\bibfnamefont {A.}~\bibnamefont {Stoyanova}}, \bibinfo {author}
  {\bibfnamefont {H.}~\bibnamefont {Kandpal}}, \bibinfo {author} {\bibfnamefont
  {S.}~\bibnamefont {Choi}}, \bibinfo {author} {\bibfnamefont {R.}~\bibnamefont
  {Coldea}}, \bibinfo {author} {\bibfnamefont {I.}~\bibnamefont
  {Rousochatzakis}}, \bibinfo {author} {\bibfnamefont {L.}~\bibnamefont
  {Hozoi}},\ and\ \bibinfo {author} {\bibfnamefont {J.}~\bibnamefont {Van
  Den~Brink}},\ }\bibfield  {title} {\bibinfo {title} {Kitaev interactions
  between $j= 1/2$ moments in honeycomb
  $\mathrm{Na}_{2}\mathrm{Ir}\mathrm{O}_{3}$ are large and ferromagnetic:
  insights from \emph{ab} \emph{initio} quantum chemistry calculations},\
  }\href {https://iopscience.iop.org/article/10.1088/1367-2630/16/1/013056}
  {\bibfield  {journal} {\bibinfo  {journal} {New Journal of Physics}\ }\textbf
  {\bibinfo {volume} {16}},\ \bibinfo {pages} {013056} (\bibinfo {year}
  {2014})}\BibitemShut {NoStop}%
\bibitem [{\citenamefont {Rau}\ \emph {et~al.}(2014)\citenamefont {Rau},
  \citenamefont {Lee},\ and\ \citenamefont {Kee}}]{kiatev2014b}%
  \BibitemOpen
  \bibfield  {author} {\bibinfo {author} {\bibfnamefont {J.~G.}\ \bibnamefont
  {Rau}}, \bibinfo {author} {\bibfnamefont {E.~K.-H.}\ \bibnamefont {Lee}},\
  and\ \bibinfo {author} {\bibfnamefont {H.-Y.}\ \bibnamefont {Kee}},\
  }\bibfield  {title} {\bibinfo {title} {Generic spin model for the honeycomb
  iridates beyond the kitaev limit},\ }\href
  {https://doi.org/10.1103/PhysRevLett.112.077204} {\bibfield  {journal}
  {\bibinfo  {journal} {Phys. Rev. Lett.}\ }\textbf {\bibinfo {volume} {112}},\
  \bibinfo {pages} {077204} (\bibinfo {year} {2014})}\BibitemShut {NoStop}%
\bibitem [{\citenamefont {Reuther}\ \emph {et~al.}(2011)\citenamefont
  {Reuther}, \citenamefont {Thomale},\ and\ \citenamefont {Trebst}}]{HK1}%
  \BibitemOpen
  \bibfield  {author} {\bibinfo {author} {\bibfnamefont {J.}~\bibnamefont
  {Reuther}}, \bibinfo {author} {\bibfnamefont {R.}~\bibnamefont {Thomale}},\
  and\ \bibinfo {author} {\bibfnamefont {S.}~\bibnamefont {Trebst}},\
  }\bibfield  {title} {\bibinfo {title} {Finite-temperature phase diagram of
  the {H}eisenberg-{K}itaev model},\ }\href
  {https://doi.org/10.1103/PhysRevB.84.100406} {\bibfield  {journal} {\bibinfo
  {journal} {Phys. Rev. B}\ }\textbf {\bibinfo {volume} {84}},\ \bibinfo
  {pages} {100406} (\bibinfo {year} {2011})}\BibitemShut {NoStop}%
\bibitem [{\citenamefont {Trousselet}\ \emph {et~al.}(2011)\citenamefont
  {Trousselet}, \citenamefont {Khaliullin},\ and\ \citenamefont
  {Horsch}}]{HK2}%
  \BibitemOpen
  \bibfield  {author} {\bibinfo {author} {\bibfnamefont {F.}~\bibnamefont
  {Trousselet}}, \bibinfo {author} {\bibfnamefont {G.}~\bibnamefont
  {Khaliullin}},\ and\ \bibinfo {author} {\bibfnamefont {P.}~\bibnamefont
  {Horsch}},\ }\bibfield  {title} {\bibinfo {title} {Effects of spin vacancies
  on magnetic properties of the {K}itaev-{H}eisenberg model},\ }\href
  {https://doi.org/10.1103/PhysRevB.84.054409} {\bibfield  {journal} {\bibinfo
  {journal} {Phys. Rev. B}\ }\textbf {\bibinfo {volume} {84}},\ \bibinfo
  {pages} {054409} (\bibinfo {year} {2011})}\BibitemShut {NoStop}%
\bibitem [{\citenamefont {Ducatman}\ \emph {et~al.}(2018)\citenamefont
  {Ducatman}, \citenamefont {Rousochatzakis},\ and\ \citenamefont
  {Perkins}}]{material2018}%
  \BibitemOpen
  \bibfield  {author} {\bibinfo {author} {\bibfnamefont {S.}~\bibnamefont
  {Ducatman}}, \bibinfo {author} {\bibfnamefont {I.}~\bibnamefont
  {Rousochatzakis}},\ and\ \bibinfo {author} {\bibfnamefont {N.~B.}\
  \bibnamefont {Perkins}},\ }\bibfield  {title} {\bibinfo {title} {Magnetic
  structure and excitation spectrum of the hyperhoneycomb {K}itaev magnet
  $\ensuremath{\beta}\text{\ensuremath{-}}\mathrm{{L}i}_{2}\mathrm{{I}rO}_{3}$},\
  }\href {https://doi.org/10.1103/PhysRevB.97.125125} {\bibfield  {journal}
  {\bibinfo  {journal} {Phys. Rev. B}\ }\textbf {\bibinfo {volume} {97}},\
  \bibinfo {pages} {125125} (\bibinfo {year} {2018})}\BibitemShut {NoStop}%
\bibitem [{\citenamefont {Smit}\ \emph {et~al.}(2020)\citenamefont {Smit},
  \citenamefont {Keupert}, \citenamefont {Tsyplyatyev}, \citenamefont
  {Maksimov}, \citenamefont {Chernyshev},\ and\ \citenamefont
  {Kopietz}}]{material2020}%
  \BibitemOpen
  \bibfield  {author} {\bibinfo {author} {\bibfnamefont {R.~L.}\ \bibnamefont
  {Smit}}, \bibinfo {author} {\bibfnamefont {S.}~\bibnamefont {Keupert}},
  \bibinfo {author} {\bibfnamefont {O.}~\bibnamefont {Tsyplyatyev}}, \bibinfo
  {author} {\bibfnamefont {P.~A.}\ \bibnamefont {Maksimov}}, \bibinfo {author}
  {\bibfnamefont {A.~L.}\ \bibnamefont {Chernyshev}},\ and\ \bibinfo {author}
  {\bibfnamefont {P.}~\bibnamefont {Kopietz}},\ }\bibfield  {title} {\bibinfo
  {title} {Magnon damping in the zigzag phase of the
  {K}itaev-{H}eisenberg-$\mathrm{\ensuremath{\Gamma}}$ model on a honeycomb
  lattice},\ }\href {https://doi.org/10.1103/PhysRevB.101.054424} {\bibfield
  {journal} {\bibinfo  {journal} {Phys. Rev. B}\ }\textbf {\bibinfo {volume}
  {101}},\ \bibinfo {pages} {054424} (\bibinfo {year} {2020})}\BibitemShut
  {NoStop}%
\bibitem [{\citenamefont {Takikawa}\ and\ \citenamefont
  {Fujimoto}(2019)}]{Ru2019}%
  \BibitemOpen
  \bibfield  {author} {\bibinfo {author} {\bibfnamefont {D.}~\bibnamefont
  {Takikawa}}\ and\ \bibinfo {author} {\bibfnamefont {S.}~\bibnamefont
  {Fujimoto}},\ }\bibfield  {title} {\bibinfo {title} {Impact of off-diagonal
  exchange interactions on the {K}itaev spin-liquid state of
  $\ensuremath{\alpha}\text{\ensuremath{-}}\mathrm{RuCl}_{3}$},\ }\href
  {https://doi.org/10.1103/PhysRevB.99.224409} {\bibfield  {journal} {\bibinfo
  {journal} {Phys. Rev. B}\ }\textbf {\bibinfo {volume} {99}},\ \bibinfo
  {pages} {224409} (\bibinfo {year} {2019})}\BibitemShut {NoStop}%
\bibitem [{\citenamefont {Xu}\ \emph {et~al.}(2020)\citenamefont {Xu},
  \citenamefont {Feng}, \citenamefont {Kawamura}, \citenamefont {Yamaji},
  \citenamefont {Nahas}, \citenamefont {Prokhorenko}, \citenamefont {Qi},
  \citenamefont {Xiang},\ and\ \citenamefont {Bellaiche}}]{xu20possible}%
  \BibitemOpen
  \bibfield  {author} {\bibinfo {author} {\bibfnamefont {C.}~\bibnamefont
  {Xu}}, \bibinfo {author} {\bibfnamefont {J.}~\bibnamefont {Feng}}, \bibinfo
  {author} {\bibfnamefont {M.}~\bibnamefont {Kawamura}}, \bibinfo {author}
  {\bibfnamefont {Y.}~\bibnamefont {Yamaji}}, \bibinfo {author} {\bibfnamefont
  {Y.}~\bibnamefont {Nahas}}, \bibinfo {author} {\bibfnamefont
  {S.}~\bibnamefont {Prokhorenko}}, \bibinfo {author} {\bibfnamefont
  {Y.}~\bibnamefont {Qi}}, \bibinfo {author} {\bibfnamefont {H.}~\bibnamefont
  {Xiang}},\ and\ \bibinfo {author} {\bibfnamefont {L.}~\bibnamefont
  {Bellaiche}},\ }\bibfield  {title} {\bibinfo {title} {Possible kitaev quantum
  spin liquid state in 2d materials with {S}=$3/2$},\ }\href
  {https://doi.org/10.1103/PhysRevLett.124.087205} {\bibfield  {journal}
  {\bibinfo  {journal} {Phys. Rev. Lett.}\ }\textbf {\bibinfo {volume} {124}},\
  \bibinfo {pages} {087205} (\bibinfo {year} {2020})}\BibitemShut {NoStop}%
\bibitem [{\citenamefont {Feynman}(1982)}]{feynman1982computer}%
  \BibitemOpen
  \bibfield  {author} {\bibinfo {author} {\bibfnamefont {R.}~\bibnamefont
  {Feynman}},\ }\bibfield  {title} {\bibinfo {title} {Simulating physics with
  computers},\ }\href
  {http://mmcp2009.jinr.ru/pdf/Fritzsche_b09.dubna-mmcp-jul.pdf} {\bibfield
  {journal} {\bibinfo  {journal} {International Journal of Theoretical
  Physics}\ }\textbf {\bibinfo {volume} {21}},\ \bibinfo {pages} {467}
  (\bibinfo {year} {1982})}\BibitemShut {NoStop}%
\bibitem [{\citenamefont {Trotzky}\ \emph {et~al.}(2008)\citenamefont
  {Trotzky}, \citenamefont {Cheinet}, \citenamefont {Folling}, \citenamefont
  {Feld}, \citenamefont {Schnorrberger}, \citenamefont {Rey}, \citenamefont
  {Polkovnikov}, \citenamefont {Demler}, \citenamefont {Lukin},\ and\
  \citenamefont {Bloch}}]{trotzky2008time}%
  \BibitemOpen
  \bibfield  {author} {\bibinfo {author} {\bibfnamefont {S.}~\bibnamefont
  {Trotzky}}, \bibinfo {author} {\bibfnamefont {P.}~\bibnamefont {Cheinet}},
  \bibinfo {author} {\bibfnamefont {S.}~\bibnamefont {Folling}}, \bibinfo
  {author} {\bibfnamefont {M.}~\bibnamefont {Feld}}, \bibinfo {author}
  {\bibfnamefont {U.}~\bibnamefont {Schnorrberger}}, \bibinfo {author}
  {\bibfnamefont {A.~M.}\ \bibnamefont {Rey}}, \bibinfo {author} {\bibfnamefont
  {A.}~\bibnamefont {Polkovnikov}}, \bibinfo {author} {\bibfnamefont {E.~A.}\
  \bibnamefont {Demler}}, \bibinfo {author} {\bibfnamefont {M.~D.}\
  \bibnamefont {Lukin}},\ and\ \bibinfo {author} {\bibfnamefont
  {I.}~\bibnamefont {Bloch}},\ }\bibfield  {title} {\bibinfo {title}
  {Time-resolved observation and control of superexchange interactions with
  ultracold atoms in optical lattices},\ }\href
  {https://www.science.org/doi/10.1126/science.1150841} {\bibfield  {journal}
  {\bibinfo  {journal} {Science}\ }\textbf {\bibinfo {volume} {319}},\ \bibinfo
  {pages} {295} (\bibinfo {year} {2008})}\BibitemShut {NoStop}%
\bibitem [{\citenamefont {Yan}\ \emph {et~al.}(2013)\citenamefont {Yan},
  \citenamefont {Moses}, \citenamefont {Gadway}, \citenamefont {Covey},
  \citenamefont {Hazzard}, \citenamefont {Rey}, \citenamefont {Jin},\ and\
  \citenamefont {Ye}}]{yanbo13nature}%
  \BibitemOpen
  \bibfield  {author} {\bibinfo {author} {\bibfnamefont {B.}~\bibnamefont
  {Yan}}, \bibinfo {author} {\bibfnamefont {S.~A.}\ \bibnamefont {Moses}},
  \bibinfo {author} {\bibfnamefont {B.}~\bibnamefont {Gadway}}, \bibinfo
  {author} {\bibfnamefont {J.~P.}\ \bibnamefont {Covey}}, \bibinfo {author}
  {\bibfnamefont {K.~R.~A.}\ \bibnamefont {Hazzard}}, \bibinfo {author}
  {\bibfnamefont {A.~M.}\ \bibnamefont {Rey}}, \bibinfo {author} {\bibfnamefont
  {D.~S.}\ \bibnamefont {Jin}},\ and\ \bibinfo {author} {\bibfnamefont
  {J.}~\bibnamefont {Ye}},\ }\bibfield  {title} {\bibinfo {title} {Observation
  of dipolar spin-exchange interactions with lattice-confined polar
  molecules},\ }\href {https://doi.org/10.1038/nature12483} {\bibfield
  {journal} {\bibinfo  {journal} {Nature}\ }\textbf {\bibinfo {volume} {501}},\
  \bibinfo {pages} {521} (\bibinfo {year} {2013})}\BibitemShut {NoStop}%
\bibitem [{\citenamefont {Jurcevic}\ \emph {et~al.}(2014)\citenamefont
  {Jurcevic}, \citenamefont {Lanyon}, \citenamefont {Hauke}, \citenamefont
  {Hempel}, \citenamefont {Zoller}, \citenamefont {Blatt},\ and\ \citenamefont
  {Roos}}]{roos14nature}%
  \BibitemOpen
  \bibfield  {author} {\bibinfo {author} {\bibfnamefont {P.}~\bibnamefont
  {Jurcevic}}, \bibinfo {author} {\bibfnamefont {B.~P.}\ \bibnamefont
  {Lanyon}}, \bibinfo {author} {\bibfnamefont {P.}~\bibnamefont {Hauke}},
  \bibinfo {author} {\bibfnamefont {C.}~\bibnamefont {Hempel}}, \bibinfo
  {author} {\bibfnamefont {P.}~\bibnamefont {Zoller}}, \bibinfo {author}
  {\bibfnamefont {R.}~\bibnamefont {Blatt}},\ and\ \bibinfo {author}
  {\bibfnamefont {C.~F.}\ \bibnamefont {Roos}},\ }\bibfield  {title} {\bibinfo
  {title} {Quasiparticle engineering and entanglement propagation in a quantum
  many-body system},\ }\href {https://doi.org/10.1038/nature13461} {\bibfield
  {journal} {\bibinfo  {journal} {Nature}\ }\textbf {\bibinfo {volume} {511}},\
  \bibinfo {pages} {202} (\bibinfo {year} {2014})}\BibitemShut {NoStop}%
\bibitem [{\citenamefont {Richerme}\ \emph {et~al.}(2014)\citenamefont
  {Richerme}, \citenamefont {Gong}, \citenamefont {Lee}, \citenamefont {Senko},
  \citenamefont {Smith}, \citenamefont {Foss-Feig}, \citenamefont {Michalakis},
  \citenamefont {Gorshkov},\ and\ \citenamefont {Monroe}}]{monroe14nature}%
  \BibitemOpen
  \bibfield  {author} {\bibinfo {author} {\bibfnamefont {P.}~\bibnamefont
  {Richerme}}, \bibinfo {author} {\bibfnamefont {Z.-X.}\ \bibnamefont {Gong}},
  \bibinfo {author} {\bibfnamefont {A.}~\bibnamefont {Lee}}, \bibinfo {author}
  {\bibfnamefont {C.}~\bibnamefont {Senko}}, \bibinfo {author} {\bibfnamefont
  {J.}~\bibnamefont {Smith}}, \bibinfo {author} {\bibfnamefont
  {M.}~\bibnamefont {Foss-Feig}}, \bibinfo {author} {\bibfnamefont
  {S.}~\bibnamefont {Michalakis}}, \bibinfo {author} {\bibfnamefont {A.~V.}\
  \bibnamefont {Gorshkov}},\ and\ \bibinfo {author} {\bibfnamefont
  {C.}~\bibnamefont {Monroe}},\ }\bibfield  {title} {\bibinfo {title}
  {Non-local propagation of correlations in quantum systems with long-range
  interactions},\ }\href {https://doi.org/10.1038/nature13450} {\bibfield
  {journal} {\bibinfo  {journal} {Nature}\ }\textbf {\bibinfo {volume} {511}},\
  \bibinfo {pages} {198} (\bibinfo {year} {2014})}\BibitemShut {NoStop}%
\bibitem [{\citenamefont {Hung}\ \emph {et~al.}(2016)\citenamefont {Hung},
  \citenamefont {González-Tudela}, \citenamefont {Cirac},\ and\ \citenamefont
  {Kimble}}]{hung2016quantum}%
  \BibitemOpen
  \bibfield  {author} {\bibinfo {author} {\bibfnamefont {C.-L.}\ \bibnamefont
  {Hung}}, \bibinfo {author} {\bibfnamefont {A.}~\bibnamefont
  {González-Tudela}}, \bibinfo {author} {\bibfnamefont {J.~I.}\ \bibnamefont
  {Cirac}},\ and\ \bibinfo {author} {\bibfnamefont {H.~J.}\ \bibnamefont
  {Kimble}},\ }\bibfield  {title} {\bibinfo {title} {Quantum spin dynamics with
  pairwise-tunable, long-range interactions},\ }\href
  {https://doi.org/10.1073/pnas.1603777113} {\bibfield  {journal} {\bibinfo
  {journal} {Proc. Natl. Acad. Sci.}\ }\textbf {\bibinfo {volume} {113}},\
  \bibinfo {pages} {E4946} (\bibinfo {year} {2016})}\BibitemShut {NoStop}%
\bibitem [{\citenamefont {Douglas}\ \emph {et~al.}(2015)\citenamefont
  {Douglas}, \citenamefont {Habibian}, \citenamefont {Hung}, \citenamefont
  {Gorshkov}, \citenamefont {Kimble},\ and\ \citenamefont
  {Chang}}]{Douglas15quantum}%
  \BibitemOpen
  \bibfield  {author} {\bibinfo {author} {\bibfnamefont {J.~S.}\ \bibnamefont
  {Douglas}}, \bibinfo {author} {\bibfnamefont {H.}~\bibnamefont {Habibian}},
  \bibinfo {author} {\bibfnamefont {C.~L.}\ \bibnamefont {Hung}}, \bibinfo
  {author} {\bibfnamefont {A.~V.}\ \bibnamefont {Gorshkov}}, \bibinfo {author}
  {\bibfnamefont {H.~J.}\ \bibnamefont {Kimble}},\ and\ \bibinfo {author}
  {\bibfnamefont {D.~E.}\ \bibnamefont {Chang}},\ }\bibfield  {title} {\bibinfo
  {title} {Quantum many-body models with cold atoms coupled to photonic
  crystals},\ }\href {https://doi.org/10.1038/nphoton.2015.57} {\bibfield
  {journal} {\bibinfo  {journal} {Nature Photonics}\ }\textbf {\bibinfo
  {volume} {9}},\ \bibinfo {pages} {326} (\bibinfo {year} {2015})}\BibitemShut
  {NoStop}%
\bibitem [{\citenamefont {Gonz{\'a}lez-Tudela}\ \emph
  {et~al.}(2015)\citenamefont {Gonz{\'a}lez-Tudela}, \citenamefont {Hung},
  \citenamefont {Chang}, \citenamefont {Cirac},\ and\ \citenamefont
  {Kimble}}]{Gonz15quantum}%
  \BibitemOpen
  \bibfield  {author} {\bibinfo {author} {\bibfnamefont {A.}~\bibnamefont
  {Gonz{\'a}lez-Tudela}}, \bibinfo {author} {\bibfnamefont {C.~L.}\
  \bibnamefont {Hung}}, \bibinfo {author} {\bibfnamefont {D.~E.}\ \bibnamefont
  {Chang}}, \bibinfo {author} {\bibfnamefont {J.~I.}\ \bibnamefont {Cirac}},\
  and\ \bibinfo {author} {\bibfnamefont {H.~J.}\ \bibnamefont {Kimble}},\
  }\bibfield  {title} {\bibinfo {title} {Subwavelength vacuum lattices and
  atom--atom interactions in two-dimensional photonic crystals},\ }\href
  {https://doi.org/10.1038/nphoton.2015.54} {\bibfield  {journal} {\bibinfo
  {journal} {Nature Photonics}\ }\textbf {\bibinfo {volume} {9}},\ \bibinfo
  {pages} {320} (\bibinfo {year} {2015})}\BibitemShut {NoStop}%
\bibitem [{\citenamefont {Browaeys}\ and\ \citenamefont
  {Lahaye}(2020)}]{ABrowaeys20np}%
  \BibitemOpen
  \bibfield  {author} {\bibinfo {author} {\bibfnamefont {A.}~\bibnamefont
  {Browaeys}}\ and\ \bibinfo {author} {\bibfnamefont {T.}~\bibnamefont
  {Lahaye}},\ }\bibfield  {title} {\bibinfo {title} {Many-body physics with
  individually controlled {R}ydberg atoms},\ }\href
  {https://doi.org/10.1038/s41567-019-0733-z} {\bibfield  {journal} {\bibinfo
  {journal} {Nature Physics}\ }\textbf {\bibinfo {volume} {16}},\ \bibinfo
  {pages} {132} (\bibinfo {year} {2020})}\BibitemShut {NoStop}%
\bibitem [{\citenamefont {Ebadi}\ \emph {et~al.}(2021)\citenamefont {Ebadi},
  \citenamefont {Wang}, \citenamefont {Levine}, \citenamefont {Keesling},
  \citenamefont {Semeghini}, \citenamefont {Omran}, \citenamefont {Bluvstein},
  \citenamefont {Samajdar}, \citenamefont {Pichler}, \citenamefont {Ho} \emph
  {et~al.}}]{lukin21nature256}%
  \BibitemOpen
  \bibfield  {author} {\bibinfo {author} {\bibfnamefont {S.}~\bibnamefont
  {Ebadi}}, \bibinfo {author} {\bibfnamefont {T.~T.}\ \bibnamefont {Wang}},
  \bibinfo {author} {\bibfnamefont {H.}~\bibnamefont {Levine}}, \bibinfo
  {author} {\bibfnamefont {A.}~\bibnamefont {Keesling}}, \bibinfo {author}
  {\bibfnamefont {G.}~\bibnamefont {Semeghini}}, \bibinfo {author}
  {\bibfnamefont {A.}~\bibnamefont {Omran}}, \bibinfo {author} {\bibfnamefont
  {D.}~\bibnamefont {Bluvstein}}, \bibinfo {author} {\bibfnamefont
  {R.}~\bibnamefont {Samajdar}}, \bibinfo {author} {\bibfnamefont
  {H.}~\bibnamefont {Pichler}}, \bibinfo {author} {\bibfnamefont {W.~W.}\
  \bibnamefont {Ho}}, \emph {et~al.},\ }\bibfield  {title} {\bibinfo {title}
  {Quantum phases of matter on a 256-atom programmable quantum simulator},\
  }\href {https://doi.org/10.1038/s41586-021-03582-4} {\bibfield  {journal}
  {\bibinfo  {journal} {Nature}\ }\textbf {\bibinfo {volume} {595}},\ \bibinfo
  {pages} {227} (\bibinfo {year} {2021})}\BibitemShut {NoStop}%
\bibitem [{\citenamefont {Davis}\ \emph {et~al.}(2020)\citenamefont {Davis},
  \citenamefont {Periwal}, \citenamefont {Cooper}, \citenamefont {Bentsen},
  \citenamefont {Evered}, \citenamefont {Van~Kirk},\ and\ \citenamefont
  {Schleier-Smith}}]{monika20heisenberg}%
  \BibitemOpen
  \bibfield  {author} {\bibinfo {author} {\bibfnamefont {E.~J.}\ \bibnamefont
  {Davis}}, \bibinfo {author} {\bibfnamefont {A.}~\bibnamefont {Periwal}},
  \bibinfo {author} {\bibfnamefont {E.~S.}\ \bibnamefont {Cooper}}, \bibinfo
  {author} {\bibfnamefont {G.}~\bibnamefont {Bentsen}}, \bibinfo {author}
  {\bibfnamefont {S.~J.}\ \bibnamefont {Evered}}, \bibinfo {author}
  {\bibfnamefont {K.}~\bibnamefont {Van~Kirk}},\ and\ \bibinfo {author}
  {\bibfnamefont {M.~H.}\ \bibnamefont {Schleier-Smith}},\ }\bibfield  {title}
  {\bibinfo {title} {Protecting spin coherence in a tunable heisenberg model},\
  }\href {https://doi.org/10.1103/PhysRevLett.125.060402} {\bibfield  {journal}
  {\bibinfo  {journal} {Phys. Rev. Lett.}\ }\textbf {\bibinfo {volume} {125}},\
  \bibinfo {pages} {060402} (\bibinfo {year} {2020})}\BibitemShut {NoStop}%
\bibitem [{\citenamefont {Chiocchetta}\ \emph {et~al.}(2021)\citenamefont
  {Chiocchetta}, \citenamefont {Kiese}, \citenamefont {Zelle}, \citenamefont
  {Piazza},\ and\ \citenamefont {Diehl}}]{chiocchetta2021cavity}%
  \BibitemOpen
  \bibfield  {author} {\bibinfo {author} {\bibfnamefont {A.}~\bibnamefont
  {Chiocchetta}}, \bibinfo {author} {\bibfnamefont {D.}~\bibnamefont {Kiese}},
  \bibinfo {author} {\bibfnamefont {C.~P.}\ \bibnamefont {Zelle}}, \bibinfo
  {author} {\bibfnamefont {F.}~\bibnamefont {Piazza}},\ and\ \bibinfo {author}
  {\bibfnamefont {S.}~\bibnamefont {Diehl}},\ }\bibfield  {title} {\bibinfo
  {title} {Cavity-induced quantum spin liquids},\ }\href
  {https://www.nature.com/articles/s41467-021-26076-3} {\bibfield  {journal}
  {\bibinfo  {journal} {Nature Communications}\ }\textbf {\bibinfo {volume}
  {12}},\ \bibinfo {pages} {5901} (\bibinfo {year} {2021})}\BibitemShut
  {NoStop}%
\bibitem [{\citenamefont {Glaetzle}\ \emph {et~al.}(2015)\citenamefont
  {Glaetzle}, \citenamefont {Dalmonte}, \citenamefont {Nath}, \citenamefont
  {Gross}, \citenamefont {Bloch},\ and\ \citenamefont
  {Zoller}}]{zoller15designing}%
  \BibitemOpen
  \bibfield  {author} {\bibinfo {author} {\bibfnamefont {A.~W.}\ \bibnamefont
  {Glaetzle}}, \bibinfo {author} {\bibfnamefont {M.}~\bibnamefont {Dalmonte}},
  \bibinfo {author} {\bibfnamefont {R.}~\bibnamefont {Nath}}, \bibinfo {author}
  {\bibfnamefont {C.}~\bibnamefont {Gross}}, \bibinfo {author} {\bibfnamefont
  {I.}~\bibnamefont {Bloch}},\ and\ \bibinfo {author} {\bibfnamefont
  {P.}~\bibnamefont {Zoller}},\ }\bibfield  {title} {\bibinfo {title}
  {Designing frustrated quantum magnets with laser-dressed rydberg atoms},\
  }\href {https://doi.org/10.1103/PhysRevLett.114.173002} {\bibfield  {journal}
  {\bibinfo  {journal} {Phys. Rev. Lett.}\ }\textbf {\bibinfo {volume} {114}},\
  \bibinfo {pages} {173002} (\bibinfo {year} {2015})}\BibitemShut {NoStop}%
\bibitem [{\citenamefont {Mivehvar}\ \emph {et~al.}(2019)\citenamefont
  {Mivehvar}, \citenamefont {Ritsch},\ and\ \citenamefont
  {Piazza}}]{Mivehvar19toolbox}%
  \BibitemOpen
  \bibfield  {author} {\bibinfo {author} {\bibfnamefont {F.}~\bibnamefont
  {Mivehvar}}, \bibinfo {author} {\bibfnamefont {H.}~\bibnamefont {Ritsch}},\
  and\ \bibinfo {author} {\bibfnamefont {F.}~\bibnamefont {Piazza}},\
  }\bibfield  {title} {\bibinfo {title} {Cavity-quantum-electrodynamical
  toolbox for quantum magnetism},\ }\href
  {https://doi.org/10.1103/PhysRevLett.122.113603} {\bibfield  {journal}
  {\bibinfo  {journal} {Phys. Rev. Lett.}\ }\textbf {\bibinfo {volume} {122}},\
  \bibinfo {pages} {113603} (\bibinfo {year} {2019})}\BibitemShut {NoStop}%
\bibitem [{\citenamefont {Mivehvar}\ \emph {et~al.}(2021)\citenamefont
  {Mivehvar}, \citenamefont {Piazza}, \citenamefont {Donner},\ and\
  \citenamefont {Ritsch}}]{Mivehvar21review}%
  \BibitemOpen
  \bibfield  {author} {\bibinfo {author} {\bibfnamefont {F.}~\bibnamefont
  {Mivehvar}}, \bibinfo {author} {\bibfnamefont {F.}~\bibnamefont {Piazza}},
  \bibinfo {author} {\bibfnamefont {T.}~\bibnamefont {Donner}},\ and\ \bibinfo
  {author} {\bibfnamefont {H.}~\bibnamefont {Ritsch}},\ }\bibfield  {title}
  {\bibinfo {title} {Cavity {QED} with quantum gases: new paradigms in
  many-body physics},\ }\href {https://doi.org/10.1080/00018732.2021.1969727}
  {\bibfield  {journal} {\bibinfo  {journal} {Advances in Physics}\ }\textbf
  {\bibinfo {volume} {70}},\ \bibinfo {pages} {1} (\bibinfo {year}
  {2021})}\BibitemShut {NoStop}%
\bibitem [{\citenamefont {Gopalakrishnan}\ \emph {et~al.}(2011)\citenamefont
  {Gopalakrishnan}, \citenamefont {Lev},\ and\ \citenamefont
  {Goldbart}}]{lev11frustration}%
  \BibitemOpen
  \bibfield  {author} {\bibinfo {author} {\bibfnamefont {S.}~\bibnamefont
  {Gopalakrishnan}}, \bibinfo {author} {\bibfnamefont {B.~L.}\ \bibnamefont
  {Lev}},\ and\ \bibinfo {author} {\bibfnamefont {P.~M.}\ \bibnamefont
  {Goldbart}},\ }\bibfield  {title} {\bibinfo {title} {Frustration and
  glassiness in spin models with cavity-mediated interactions},\ }\href
  {https://doi.org/10.1103/PhysRevLett.107.277201} {\bibfield  {journal}
  {\bibinfo  {journal} {Phys. Rev. Lett.}\ }\textbf {\bibinfo {volume} {107}},\
  \bibinfo {pages} {277201} (\bibinfo {year} {2011})}\BibitemShut {NoStop}%
\bibitem [{\citenamefont {Mivehvar}\ \emph {et~al.}(2017)\citenamefont
  {Mivehvar}, \citenamefont {Piazza},\ and\ \citenamefont
  {Ritsch}}]{Mivehvar17disorder}%
  \BibitemOpen
  \bibfield  {author} {\bibinfo {author} {\bibfnamefont {F.}~\bibnamefont
  {Mivehvar}}, \bibinfo {author} {\bibfnamefont {F.}~\bibnamefont {Piazza}},\
  and\ \bibinfo {author} {\bibfnamefont {H.}~\bibnamefont {Ritsch}},\
  }\bibfield  {title} {\bibinfo {title} {Disorder-driven density and spin
  self-ordering of a bose-einstein condensate in a cavity},\ }\href
  {https://doi.org/10.1103/PhysRevLett.119.063602} {\bibfield  {journal}
  {\bibinfo  {journal} {Phys. Rev. Lett.}\ }\textbf {\bibinfo {volume} {119}},\
  \bibinfo {pages} {063602} (\bibinfo {year} {2017})}\BibitemShut {NoStop}%
\bibitem [{\citenamefont {Landini}\ \emph {et~al.}(2018)\citenamefont
  {Landini}, \citenamefont {Dogra}, \citenamefont {Kroeger}, \citenamefont
  {Hruby}, \citenamefont {Donner},\ and\ \citenamefont
  {Esslinger}}]{Esslinger18formation}%
  \BibitemOpen
  \bibfield  {author} {\bibinfo {author} {\bibfnamefont {M.}~\bibnamefont
  {Landini}}, \bibinfo {author} {\bibfnamefont {N.}~\bibnamefont {Dogra}},
  \bibinfo {author} {\bibfnamefont {K.}~\bibnamefont {Kroeger}}, \bibinfo
  {author} {\bibfnamefont {L.}~\bibnamefont {Hruby}}, \bibinfo {author}
  {\bibfnamefont {T.}~\bibnamefont {Donner}},\ and\ \bibinfo {author}
  {\bibfnamefont {T.}~\bibnamefont {Esslinger}},\ }\bibfield  {title} {\bibinfo
  {title} {Formation of a spin texture in a quantum gas coupled to a cavity},\
  }\href {https://doi.org/10.1103/PhysRevLett.120.223602} {\bibfield  {journal}
  {\bibinfo  {journal} {Phys. Rev. Lett.}\ }\textbf {\bibinfo {volume} {120}},\
  \bibinfo {pages} {223602} (\bibinfo {year} {2018})}\BibitemShut {NoStop}%
\bibitem [{\citenamefont {Ren}\ \emph {et~al.}(2022)\citenamefont {Ren},
  \citenamefont {Wang}, \citenamefont {Chen},\ and\ \citenamefont
  {You}}]{ren22longrange}%
  \BibitemOpen
  \bibfield  {author} {\bibinfo {author} {\bibfnamefont {J.}~\bibnamefont
  {Ren}}, \bibinfo {author} {\bibfnamefont {Z.}~\bibnamefont {Wang}}, \bibinfo
  {author} {\bibfnamefont {W.-X.}\ \bibnamefont {Chen}},\ and\ \bibinfo
  {author} {\bibfnamefont {W.-L.}\ \bibnamefont {You}},\ }\bibfield  {title}
  {\bibinfo {title} {Long-range order and quantum criticality in
  antiferromagnetic chains with long-range staggered interactions},\ }\href
  {https://doi.org/10.1103/PhysRevE.105.034128} {\bibfield  {journal} {\bibinfo
   {journal} {Phys. Rev. E}\ }\textbf {\bibinfo {volume} {105}},\ \bibinfo
  {pages} {034128} (\bibinfo {year} {2022})}\BibitemShut {NoStop}%
\bibitem [{\citenamefont {You}(2009)}]{you2009long}%
  \BibitemOpen
  \bibfield  {author} {\bibinfo {author} {\bibfnamefont {W.-L.}\ \bibnamefont
  {You}},\ }\bibfield  {title} {\bibinfo {title} {Long-range order in
  two-dimensional {XXZ} model},\ }\href
  {https://doi.org/10.1142/S0217979209052145} {\bibfield  {journal} {\bibinfo
  {journal} {International Journal of Modern Physics B}\ }\textbf {\bibinfo
  {volume} {23}},\ \bibinfo {pages} {2195} (\bibinfo {year}
  {2009})}\BibitemShut {NoStop}%
\bibitem [{\citenamefont {Tang}\ and\ \citenamefont
  {Hirsch}(1989)}]{tang1989long}%
  \BibitemOpen
  \bibfield  {author} {\bibinfo {author} {\bibfnamefont {S.}~\bibnamefont
  {Tang}}\ and\ \bibinfo {author} {\bibfnamefont {J.~E.}\ \bibnamefont
  {Hirsch}},\ }\bibfield  {title} {\bibinfo {title} {Long-range order without
  broken symmetry: Two-dimensional {H}eisenberg antiferromagnet at zero
  temperature},\ }\href {https://doi.org/10.1103/PhysRevB.39.4548} {\bibfield
  {journal} {\bibinfo  {journal} {Phys. Rev. B}\ }\textbf {\bibinfo {volume}
  {39}},\ \bibinfo {pages} {4548} (\bibinfo {year} {1989})}\BibitemShut
  {NoStop}%
\bibitem [{\citenamefont {Lin}\ and\ \citenamefont
  {Campbell}(1992)}]{lin92long}%
  \BibitemOpen
  \bibfield  {author} {\bibinfo {author} {\bibfnamefont {H.~Q.}\ \bibnamefont
  {Lin}}\ and\ \bibinfo {author} {\bibfnamefont {D.~K.}\ \bibnamefont
  {Campbell}},\ }\bibfield  {title} {\bibinfo {title} {Long-range order in the
  2{D} antiferromagnetic {H}eisenberg model: A renormalization perspective},\
  }\href {https://doi.org/10.1103/PhysRevLett.69.2415} {\bibfield  {journal}
  {\bibinfo  {journal} {Phys. Rev. Lett.}\ }\textbf {\bibinfo {volume} {69}},\
  \bibinfo {pages} {2415} (\bibinfo {year} {1992})}\BibitemShut {NoStop}%
\bibitem [{\citenamefont {Li}\ \emph {et~al.}(2021)\citenamefont {Li},
  \citenamefont {Choudhury},\ and\ \citenamefont {Liu}}]{li21long}%
  \BibitemOpen
  \bibfield  {author} {\bibinfo {author} {\bibfnamefont {Z.}~\bibnamefont
  {Li}}, \bibinfo {author} {\bibfnamefont {S.}~\bibnamefont {Choudhury}},\ and\
  \bibinfo {author} {\bibfnamefont {W.~V.}\ \bibnamefont {Liu}},\ }\bibfield
  {title} {\bibinfo {title} {Long-range-ordered phase in a quantum {H}eisenberg
  chain with interactions beyond nearest neighbors},\ }\href
  {https://doi.org/10.1103/PhysRevA.104.013303} {\bibfield  {journal} {\bibinfo
   {journal} {Phys. Rev. A}\ }\textbf {\bibinfo {volume} {104}},\ \bibinfo
  {pages} {013303} (\bibinfo {year} {2021})}\BibitemShut {NoStop}%
\bibitem [{\citenamefont
  {Dzyaloshinsky}(1958)}]{dzyaloshinsky1958thermodynamic}%
  \BibitemOpen
  \bibfield  {author} {\bibinfo {author} {\bibfnamefont {I.}~\bibnamefont
  {Dzyaloshinsky}},\ }\bibfield  {title} {\bibinfo {title} {A thermodynamic
  theory of “weak” ferromagnetism of antiferromagnetics},\ }\href
  {https://doi.org/10.1016/0022-3697(58)90076-3} {\bibfield  {journal}
  {\bibinfo  {journal} {Journal of physics and chemistry of solids}\ }\textbf
  {\bibinfo {volume} {4}},\ \bibinfo {pages} {241} (\bibinfo {year}
  {1958})}\BibitemShut {NoStop}%
\bibitem [{\citenamefont {Anderson}(1959)}]{DM1959}%
  \BibitemOpen
  \bibfield  {author} {\bibinfo {author} {\bibfnamefont {P.~W.}\ \bibnamefont
  {Anderson}},\ }\bibfield  {title} {\bibinfo {title} {New approach to the
  theory of superexchange interactions},\ }\href
  {https://doi.org/10.1103/PhysRev.115.2} {\bibfield  {journal} {\bibinfo
  {journal} {Phys. Rev.}\ }\textbf {\bibinfo {volume} {115}},\ \bibinfo {pages}
  {2} (\bibinfo {year} {1959})}\BibitemShut {NoStop}%
\bibitem [{\citenamefont {Moriya}(1960{\natexlab{a}})}]{XYDM1960}%
  \BibitemOpen
  \bibfield  {author} {\bibinfo {author} {\bibfnamefont {T.}~\bibnamefont
  {Moriya}},\ }\bibfield  {title} {\bibinfo {title} {Anisotropic superexchange
  interaction and weak ferromagnetism},\ }\href
  {https://doi.org/10.1103/PhysRev.120.91} {\bibfield  {journal} {\bibinfo
  {journal} {Phys. Rev.}\ }\textbf {\bibinfo {volume} {120}},\ \bibinfo {pages}
  {91} (\bibinfo {year} {1960}{\natexlab{a}})}\BibitemShut {NoStop}%
\bibitem [{\citenamefont {Moriya}(1960{\natexlab{b}})}]{DM1960b}%
  \BibitemOpen
  \bibfield  {author} {\bibinfo {author} {\bibfnamefont {T.}~\bibnamefont
  {Moriya}},\ }\bibfield  {title} {\bibinfo {title} {New mechanism of
  anisotropic superexchange interaction},\ }\href
  {https://doi.org/10.1103/PhysRevLett.4.228} {\bibfield  {journal} {\bibinfo
  {journal} {Phys. Rev. Lett.}\ }\textbf {\bibinfo {volume} {4}},\ \bibinfo
  {pages} {228} (\bibinfo {year} {1960}{\natexlab{b}})}\BibitemShut {NoStop}%
\bibitem [{\citenamefont {Luo}\ \emph {et~al.}(2021{\natexlab{a}})\citenamefont
  {Luo}, \citenamefont {Zhao}, \citenamefont {Wang},\ and\ \citenamefont
  {Kee}}]{Luo2021unveiling}%
  \BibitemOpen
  \bibfield  {author} {\bibinfo {author} {\bibfnamefont {Q.}~\bibnamefont
  {Luo}}, \bibinfo {author} {\bibfnamefont {J.}~\bibnamefont {Zhao}}, \bibinfo
  {author} {\bibfnamefont {X.}~\bibnamefont {Wang}},\ and\ \bibinfo {author}
  {\bibfnamefont {H.-Y.}\ \bibnamefont {Kee}},\ }\bibfield  {title} {\bibinfo
  {title} {Unveiling the phase diagram of a bond-alternating spin-$\frac{1}{2}$
  {K}$\text{\ensuremath{-}}\mathrm{\ensuremath{\Gamma}}$ chain},\ }\href
  {https://doi.org/10.1103/PhysRevB.103.144423} {\bibfield  {journal} {\bibinfo
   {journal} {Phys. Rev. B}\ }\textbf {\bibinfo {volume} {103}},\ \bibinfo
  {pages} {144423} (\bibinfo {year} {2021}{\natexlab{a}})}\BibitemShut
  {NoStop}%
\bibitem [{\citenamefont {Luo}\ \emph {et~al.}(2021{\natexlab{b}})\citenamefont
  {Luo}, \citenamefont {Hu},\ and\ \citenamefont {Kee}}]{Luo2021unusual}%
  \BibitemOpen
  \bibfield  {author} {\bibinfo {author} {\bibfnamefont {Q.}~\bibnamefont
  {Luo}}, \bibinfo {author} {\bibfnamefont {S.}~\bibnamefont {Hu}},\ and\
  \bibinfo {author} {\bibfnamefont {H.-Y.}\ \bibnamefont {Kee}},\ }\bibfield
  {title} {\bibinfo {title} {Unusual excitations and double-peak specific heat
  in a bond-alternating spin-1 {K}$\ensuremath{-}\mathrm{\ensuremath{\Gamma}}$
  chain},\ }\href {https://doi.org/10.1103/PhysRevResearch.3.033048} {\bibfield
   {journal} {\bibinfo  {journal} {Phys. Rev. Research}\ }\textbf {\bibinfo
  {volume} {3}},\ \bibinfo {pages} {033048} (\bibinfo {year}
  {2021}{\natexlab{b}})}\BibitemShut {NoStop}%
\bibitem [{\citenamefont {Liu}\ \emph {et~al.}(2020)\citenamefont {Liu},
  \citenamefont {Yi}, \citenamefont {Sun}, \citenamefont {Dong},\ and\
  \citenamefont {You}}]{you2020Lifshitz}%
  \BibitemOpen
  \bibfield  {author} {\bibinfo {author} {\bibfnamefont {Z.-A.}\ \bibnamefont
  {Liu}}, \bibinfo {author} {\bibfnamefont {T.-C.}\ \bibnamefont {Yi}},
  \bibinfo {author} {\bibfnamefont {J.-H.}\ \bibnamefont {Sun}}, \bibinfo
  {author} {\bibfnamefont {Y.-L.}\ \bibnamefont {Dong}},\ and\ \bibinfo
  {author} {\bibfnamefont {W.-L.}\ \bibnamefont {You}},\ }\bibfield  {title}
  {\bibinfo {title} {Lifshitz phase transitions in a one-dimensional gamma
  model},\ }\href {https://doi.org/10.1103/PhysRevE.102.032127} {\bibfield
  {journal} {\bibinfo  {journal} {Phys. Rev. E}\ }\textbf {\bibinfo {volume}
  {102}},\ \bibinfo {pages} {032127} (\bibinfo {year} {2020})}\BibitemShut
  {NoStop}%
\bibitem [{\citenamefont {Yang}\ \emph
  {et~al.}(2020{\natexlab{a}})\citenamefont {Yang}, \citenamefont {Nocera},
  \citenamefont {Tummuru}, \citenamefont {Kee},\ and\ \citenamefont
  {Affleck}}]{Alberto2020phase}%
  \BibitemOpen
  \bibfield  {author} {\bibinfo {author} {\bibfnamefont {W.}~\bibnamefont
  {Yang}}, \bibinfo {author} {\bibfnamefont {A.}~\bibnamefont {Nocera}},
  \bibinfo {author} {\bibfnamefont {T.}~\bibnamefont {Tummuru}}, \bibinfo
  {author} {\bibfnamefont {H.-Y.}\ \bibnamefont {Kee}},\ and\ \bibinfo {author}
  {\bibfnamefont {I.}~\bibnamefont {Affleck}},\ }\bibfield  {title} {\bibinfo
  {title} {Phase diagram of the spin-$1/2$ kitaev-gamma chain and emergent
  {SU}(2) symmetry},\ }\href {https://doi.org/10.1103/PhysRevLett.124.147205}
  {\bibfield  {journal} {\bibinfo  {journal} {Phys. Rev. Lett.}\ }\textbf
  {\bibinfo {volume} {124}},\ \bibinfo {pages} {147205} (\bibinfo {year}
  {2020}{\natexlab{a}})}\BibitemShut {NoStop}%
\bibitem [{\citenamefont {S\o{}rensen}\ \emph {et~al.}(2021)\citenamefont
  {S\o{}rensen}, \citenamefont {Catuneanu}, \citenamefont {Gordon},\ and\
  \citenamefont {Kee}}]{Erik2021entanglement}%
  \BibitemOpen
  \bibfield  {author} {\bibinfo {author} {\bibfnamefont {E.~S.}\ \bibnamefont
  {S\o{}rensen}}, \bibinfo {author} {\bibfnamefont {A.}~\bibnamefont
  {Catuneanu}}, \bibinfo {author} {\bibfnamefont {J.~S.}\ \bibnamefont
  {Gordon}},\ and\ \bibinfo {author} {\bibfnamefont {H.-Y.}\ \bibnamefont
  {Kee}},\ }\bibfield  {title} {\bibinfo {title} {Heart of entanglement:
  Chiral, nematic, and incommensurate phases in the {K}itaev-{G}amma ladder in
  a field},\ }\href {https://doi.org/10.1103/PhysRevX.11.011013} {\bibfield
  {journal} {\bibinfo  {journal} {Phys. Rev. X}\ }\textbf {\bibinfo {volume}
  {11}},\ \bibinfo {pages} {011013} (\bibinfo {year} {2021})}\BibitemShut
  {NoStop}%
\bibitem [{\citenamefont {Yang}\ \emph
  {et~al.}(2020{\natexlab{b}})\citenamefont {Yang}, \citenamefont {Nocera},\
  and\ \citenamefont {Affleck}}]{Yang2020comprehensive}%
  \BibitemOpen
  \bibfield  {author} {\bibinfo {author} {\bibfnamefont {W.}~\bibnamefont
  {Yang}}, \bibinfo {author} {\bibfnamefont {A.}~\bibnamefont {Nocera}},\ and\
  \bibinfo {author} {\bibfnamefont {I.}~\bibnamefont {Affleck}},\ }\bibfield
  {title} {\bibinfo {title} {Comprehensive study of the phase diagram of the
  spin-$\frac{1}{2}$ {K}itaev-{H}eisenberg-{G}amma chain},\ }\href
  {https://doi.org/10.1103/PhysRevResearch.2.033268} {\bibfield  {journal}
  {\bibinfo  {journal} {Phys. Rev. Research}\ }\textbf {\bibinfo {volume}
  {2}},\ \bibinfo {pages} {033268} (\bibinfo {year}
  {2020}{\natexlab{b}})}\BibitemShut {NoStop}%
\bibitem [{\citenamefont {Vaidya}\ \emph {et~al.}(2018)\citenamefont {Vaidya},
  \citenamefont {Guo}, \citenamefont {Kroeze}, \citenamefont {Ballantine},
  \citenamefont {Koll\'ar}, \citenamefont {Keeling},\ and\ \citenamefont
  {Lev}}]{Lev18tunable}%
  \BibitemOpen
  \bibfield  {author} {\bibinfo {author} {\bibfnamefont {V.~D.}\ \bibnamefont
  {Vaidya}}, \bibinfo {author} {\bibfnamefont {Y.}~\bibnamefont {Guo}},
  \bibinfo {author} {\bibfnamefont {R.~M.}\ \bibnamefont {Kroeze}}, \bibinfo
  {author} {\bibfnamefont {K.~E.}\ \bibnamefont {Ballantine}}, \bibinfo
  {author} {\bibfnamefont {A.~J.}\ \bibnamefont {Koll\'ar}}, \bibinfo {author}
  {\bibfnamefont {J.}~\bibnamefont {Keeling}},\ and\ \bibinfo {author}
  {\bibfnamefont {B.~L.}\ \bibnamefont {Lev}},\ }\bibfield  {title} {\bibinfo
  {title} {Tunable-range, photon-mediated atomic interactions in multimode
  cavity qed},\ }\href {https://doi.org/10.1103/PhysRevX.8.011002} {\bibfield
  {journal} {\bibinfo  {journal} {Phys. Rev. X}\ }\textbf {\bibinfo {volume}
  {8}},\ \bibinfo {pages} {011002} (\bibinfo {year} {2018})}\BibitemShut
  {NoStop}%
\bibitem [{\citenamefont {Guo}\ \emph {et~al.}(2021)\citenamefont {Guo},
  \citenamefont {Kroeze}, \citenamefont {Marsh}, \citenamefont
  {Gopalakrishnan}, \citenamefont {Keeling},\ and\ \citenamefont
  {Lev}}]{Guo21nature}%
  \BibitemOpen
  \bibfield  {author} {\bibinfo {author} {\bibfnamefont {Y.}~\bibnamefont
  {Guo}}, \bibinfo {author} {\bibfnamefont {R.~M.}\ \bibnamefont {Kroeze}},
  \bibinfo {author} {\bibfnamefont {B.~P.}\ \bibnamefont {Marsh}}, \bibinfo
  {author} {\bibfnamefont {S.}~\bibnamefont {Gopalakrishnan}}, \bibinfo
  {author} {\bibfnamefont {J.}~\bibnamefont {Keeling}},\ and\ \bibinfo {author}
  {\bibfnamefont {B.~L.}\ \bibnamefont {Lev}},\ }\bibfield  {title} {\bibinfo
  {title} {An optical lattice with sound},\ }\href
  {https://doi.org/10.1038/s41586-021-03945-x} {\bibfield  {journal} {\bibinfo
  {journal} {Nature}\ }\textbf {\bibinfo {volume} {599}},\ \bibinfo {pages}
  {211} (\bibinfo {year} {2021})}\BibitemShut {NoStop}%
\bibitem [{\citenamefont {Periwal}\ \emph {et~al.}(2021)\citenamefont
  {Periwal}, \citenamefont {Cooper}, \citenamefont {Kunkel}, \citenamefont
  {Wienand}, \citenamefont {Davis},\ and\ \citenamefont
  {Schleier-Smith}}]{periwal2021programmable}%
  \BibitemOpen
  \bibfield  {author} {\bibinfo {author} {\bibfnamefont {A.}~\bibnamefont
  {Periwal}}, \bibinfo {author} {\bibfnamefont {E.~S.}\ \bibnamefont {Cooper}},
  \bibinfo {author} {\bibfnamefont {P.}~\bibnamefont {Kunkel}}, \bibinfo
  {author} {\bibfnamefont {J.~F.}\ \bibnamefont {Wienand}}, \bibinfo {author}
  {\bibfnamefont {E.~J.}\ \bibnamefont {Davis}},\ and\ \bibinfo {author}
  {\bibfnamefont {M.}~\bibnamefont {Schleier-Smith}},\ }\bibfield  {title}
  {\bibinfo {title} {Programmable interactions and emergent geometry in an
  array of atom clouds},\ }\href
  {https://www.nature.com/articles/s41586-021-04156-0} {\bibfield  {journal}
  {\bibinfo  {journal} {Nature}\ }\textbf {\bibinfo {volume} {600}},\ \bibinfo
  {pages} {630} (\bibinfo {year} {2021})}\BibitemShut {NoStop}%
\bibitem [{\citenamefont {Bitko}\ \emph {et~al.}(1996)\citenamefont {Bitko},
  \citenamefont {Rosenbaum},\ and\ \citenamefont {Aeppli}}]{bitko1996quantum}%
  \BibitemOpen
  \bibfield  {author} {\bibinfo {author} {\bibfnamefont {D.}~\bibnamefont
  {Bitko}}, \bibinfo {author} {\bibfnamefont {T.~F.}\ \bibnamefont
  {Rosenbaum}},\ and\ \bibinfo {author} {\bibfnamefont {G.}~\bibnamefont
  {Aeppli}},\ }\bibfield  {title} {\bibinfo {title} {Quantum critical behavior
  for a model magnet},\ }\href {https://doi.org/10.1103/PhysRevLett.77.940}
  {\bibfield  {journal} {\bibinfo  {journal} {Phys. Rev. Lett.}\ }\textbf
  {\bibinfo {volume} {77}},\ \bibinfo {pages} {940} (\bibinfo {year}
  {1996})}\BibitemShut {NoStop}%
\bibitem [{\citenamefont {Cui}\ \emph {et~al.}(2019)\citenamefont {Cui},
  \citenamefont {Zou}, \citenamefont {Xi}, \citenamefont {He}, \citenamefont
  {Yang}, \citenamefont {Shu}, \citenamefont {Zhang}, \citenamefont {Hu},
  \citenamefont {Chen}, \citenamefont {Yu} \emph {et~al.}}]{cui2019quantum}%
  \BibitemOpen
  \bibfield  {author} {\bibinfo {author} {\bibfnamefont {Y.}~\bibnamefont
  {Cui}}, \bibinfo {author} {\bibfnamefont {H.}~\bibnamefont {Zou}}, \bibinfo
  {author} {\bibfnamefont {N.}~\bibnamefont {Xi}}, \bibinfo {author}
  {\bibfnamefont {Z.}~\bibnamefont {He}}, \bibinfo {author} {\bibfnamefont
  {Y.~X.}\ \bibnamefont {Yang}}, \bibinfo {author} {\bibfnamefont
  {L.}~\bibnamefont {Shu}}, \bibinfo {author} {\bibfnamefont {G.~H.}\
  \bibnamefont {Zhang}}, \bibinfo {author} {\bibfnamefont {Z.}~\bibnamefont
  {Hu}}, \bibinfo {author} {\bibfnamefont {T.}~\bibnamefont {Chen}}, \bibinfo
  {author} {\bibfnamefont {R.}~\bibnamefont {Yu}}, \emph {et~al.},\ }\bibfield
  {title} {\bibinfo {title} {Quantum criticality of the ising-like screw chain
  antiferromagnet {${\mathrm{SrCo}}_{2}{\mathrm{V}}_{2}{\mathrm{O}}_{8}$} in a
  transverse magnetic field},\ }\href
  {https://doi.org/10.1103/PhysRevLett.123.067203} {\bibfield  {journal}
  {\bibinfo  {journal} {Phys. Rev. Lett.}\ }\textbf {\bibinfo {volume} {123}},\
  \bibinfo {pages} {067203} (\bibinfo {year} {2019})}\BibitemShut {NoStop}%
\bibitem [{\citenamefont {Kenzelmann}\ \emph {et~al.}(2002)\citenamefont
  {Kenzelmann}, \citenamefont {Coldea}, \citenamefont {Tennant}, \citenamefont
  {Visser}, \citenamefont {Hofmann}, \citenamefont {Smeibidl},\ and\
  \citenamefont {Tylczynski}}]{kenzelmann2002order}%
  \BibitemOpen
  \bibfield  {author} {\bibinfo {author} {\bibfnamefont {M.}~\bibnamefont
  {Kenzelmann}}, \bibinfo {author} {\bibfnamefont {R.}~\bibnamefont {Coldea}},
  \bibinfo {author} {\bibfnamefont {D.~A.}\ \bibnamefont {Tennant}}, \bibinfo
  {author} {\bibfnamefont {D.}~\bibnamefont {Visser}}, \bibinfo {author}
  {\bibfnamefont {M.}~\bibnamefont {Hofmann}}, \bibinfo {author} {\bibfnamefont
  {P.}~\bibnamefont {Smeibidl}},\ and\ \bibinfo {author} {\bibfnamefont
  {Z.}~\bibnamefont {Tylczynski}},\ }\bibfield  {title} {\bibinfo {title}
  {Order-to-disorder transition in the $\mathrm{XY}$-like quantum magnet
  $\mathrm{Cs}_{2}\mathrm{CoCl}_{4}$ induced by noncommuting applied fields},\
  }\href {https://doi.org/10.1103/PhysRevB.65.144432} {\bibfield  {journal}
  {\bibinfo  {journal} {Phys. Rev. B}\ }\textbf {\bibinfo {volume} {65}},\
  \bibinfo {pages} {144432} (\bibinfo {year} {2002})}\BibitemShut {NoStop}%
\bibitem [{\citenamefont {Coldea}\ \emph {et~al.}(2010)\citenamefont {Coldea},
  \citenamefont {Tennant}, \citenamefont {Wheeler}, \citenamefont {Wawrzynska},
  \citenamefont {Prabhakaran}, \citenamefont {Telling}, \citenamefont
  {Habicht}, \citenamefont {Smeibidl},\ and\ \citenamefont
  {Kiefer}}]{coldea2010quantum}%
  \BibitemOpen
  \bibfield  {author} {\bibinfo {author} {\bibfnamefont {R.}~\bibnamefont
  {Coldea}}, \bibinfo {author} {\bibfnamefont {D.}~\bibnamefont {Tennant}},
  \bibinfo {author} {\bibfnamefont {E.}~\bibnamefont {Wheeler}}, \bibinfo
  {author} {\bibfnamefont {E.}~\bibnamefont {Wawrzynska}}, \bibinfo {author}
  {\bibfnamefont {D.}~\bibnamefont {Prabhakaran}}, \bibinfo {author}
  {\bibfnamefont {M.}~\bibnamefont {Telling}}, \bibinfo {author} {\bibfnamefont
  {K.}~\bibnamefont {Habicht}}, \bibinfo {author} {\bibfnamefont
  {P.}~\bibnamefont {Smeibidl}},\ and\ \bibinfo {author} {\bibfnamefont
  {K.}~\bibnamefont {Kiefer}},\ }\bibfield  {title} {\bibinfo {title} {Quantum
  criticality in an {I}sing chain: experimental evidence for emergent
  $\mathrm{E}_{8}$ symmetry},\ }\href
  {https://www.science.org/doi/10.1126/science.1180085} {\bibfield  {journal}
  {\bibinfo  {journal} {Science}\ }\textbf {\bibinfo {volume} {327}},\ \bibinfo
  {pages} {177} (\bibinfo {year} {2010})}\BibitemShut {NoStop}%
\bibitem [{\citenamefont {Faure}\ \emph {et~al.}(2019)\citenamefont {Faure},
  \citenamefont {Takayoshi}, \citenamefont {Simonet}, \citenamefont {Grenier},
  \citenamefont {M\aa{}nsson}, \citenamefont {White}, \citenamefont {Tucker},
  \citenamefont {R\"uegg}, \citenamefont {Lejay}, \citenamefont {Giamarchi},\
  and\ \citenamefont {Petit}}]{PhysRevLett.123.027204}%
  \BibitemOpen
  \bibfield  {author} {\bibinfo {author} {\bibfnamefont {Q.}~\bibnamefont
  {Faure}}, \bibinfo {author} {\bibfnamefont {S.}~\bibnamefont {Takayoshi}},
  \bibinfo {author} {\bibfnamefont {V.}~\bibnamefont {Simonet}}, \bibinfo
  {author} {\bibfnamefont {B.}~\bibnamefont {Grenier}}, \bibinfo {author}
  {\bibfnamefont {M.}~\bibnamefont {M\aa{}nsson}}, \bibinfo {author}
  {\bibfnamefont {J.~S.}\ \bibnamefont {White}}, \bibinfo {author}
  {\bibfnamefont {G.~S.}\ \bibnamefont {Tucker}}, \bibinfo {author}
  {\bibfnamefont {C.}~\bibnamefont {R\"uegg}}, \bibinfo {author} {\bibfnamefont
  {P.}~\bibnamefont {Lejay}}, \bibinfo {author} {\bibfnamefont
  {T.}~\bibnamefont {Giamarchi}},\ and\ \bibinfo {author} {\bibfnamefont
  {S.}~\bibnamefont {Petit}},\ }\bibfield  {title} {\bibinfo {title}
  {Tomonaga-luttinger liquid spin dynamics in the quasi-one-dimensional
  {I}sing-like antiferromagnet
  $\mathrm{BaCo}_{2}\mathrm{V}_{2}\mathrm{O}_{8}$},\ }\href
  {https://doi.org/10.1103/PhysRevLett.123.027204} {\bibfield  {journal}
  {\bibinfo  {journal} {Phys. Rev. Lett.}\ }\textbf {\bibinfo {volume} {123}},\
  \bibinfo {pages} {027204} (\bibinfo {year} {2019})}\BibitemShut {NoStop}%
\bibitem [{\citenamefont {Chauhan}\ \emph {et~al.}(2020)\citenamefont
  {Chauhan}, \citenamefont {Mahmood}, \citenamefont {Changlani}, \citenamefont
  {Koohpayeh},\ and\ \citenamefont {Armitage}}]{chauhan20tunable}%
  \BibitemOpen
  \bibfield  {author} {\bibinfo {author} {\bibfnamefont {P.}~\bibnamefont
  {Chauhan}}, \bibinfo {author} {\bibfnamefont {F.}~\bibnamefont {Mahmood}},
  \bibinfo {author} {\bibfnamefont {H.~J.}\ \bibnamefont {Changlani}}, \bibinfo
  {author} {\bibfnamefont {S.~M.}\ \bibnamefont {Koohpayeh}},\ and\ \bibinfo
  {author} {\bibfnamefont {N.~P.}\ \bibnamefont {Armitage}},\ }\bibfield
  {title} {\bibinfo {title} {Tunable magnon interactions in a ferromagnetic
  spin-1 chain},\ }\href {https://doi.org/10.1103/PhysRevLett.124.037203}
  {\bibfield  {journal} {\bibinfo  {journal} {Phys. Rev. Lett.}\ }\textbf
  {\bibinfo {volume} {124}},\ \bibinfo {pages} {037203} (\bibinfo {year}
  {2020})}\BibitemShut {NoStop}%
\bibitem [{\citenamefont {Yi}\ \emph {et~al.}(2018)\citenamefont {Yi},
  \citenamefont {Ding}, \citenamefont {Ren}, \citenamefont {Wang},\ and\
  \citenamefont {You}}]{XYDMYTC}%
  \BibitemOpen
  \bibfield  {author} {\bibinfo {author} {\bibfnamefont {T.-C.}\ \bibnamefont
  {Yi}}, \bibinfo {author} {\bibfnamefont {Y.-R.}\ \bibnamefont {Ding}},
  \bibinfo {author} {\bibfnamefont {J.}~\bibnamefont {Ren}}, \bibinfo {author}
  {\bibfnamefont {Y.-M.}\ \bibnamefont {Wang}},\ and\ \bibinfo {author}
  {\bibfnamefont {W.-L.}\ \bibnamefont {You}},\ }\bibfield  {title} {\bibinfo
  {title} {Quantum coherence of {XY} model with {D}zyaloshinskii-{M}oriya
  interaction},\ }\href
  {https://wulixb.iphy.ac.cn/en/article/doi/10.7498/aps.67.20172755} {\bibfield
   {journal} {\bibinfo  {journal} {Acta Physica Sinica}\ }\textbf {\bibinfo
  {volume} {67}},\ \bibinfo {pages} {140303} (\bibinfo {year}
  {2018})}\BibitemShut {NoStop}%
\bibitem [{\citenamefont {Rufo}\ \emph {et~al.}(2019)\citenamefont {Rufo},
  \citenamefont {Lopes}, \citenamefont {Continentino},\ and\ \citenamefont
  {Griffith}}]{Rufo2019Multicritical}%
  \BibitemOpen
  \bibfield  {author} {\bibinfo {author} {\bibfnamefont {S.}~\bibnamefont
  {Rufo}}, \bibinfo {author} {\bibfnamefont {N.}~\bibnamefont {Lopes}},
  \bibinfo {author} {\bibfnamefont {M.~A.}\ \bibnamefont {Continentino}},\ and\
  \bibinfo {author} {\bibfnamefont {M.~A.~R.}\ \bibnamefont {Griffith}},\
  }\bibfield  {title} {\bibinfo {title} {Multicritical behavior in topological
  phase transitions},\ }\href {https://doi.org/10.1103/PhysRevB.100.195432}
  {\bibfield  {journal} {\bibinfo  {journal} {Phys. Rev. B}\ }\textbf {\bibinfo
  {volume} {100}},\ \bibinfo {pages} {195432} (\bibinfo {year}
  {2019})}\BibitemShut {NoStop}%
\bibitem [{\citenamefont {Su}\ \emph {et~al.}(2006)\citenamefont {Su},
  \citenamefont {Song},\ and\ \citenamefont {Gu}}]{scalling2006}%
  \BibitemOpen
  \bibfield  {author} {\bibinfo {author} {\bibfnamefont {S.-Q.}\ \bibnamefont
  {Su}}, \bibinfo {author} {\bibfnamefont {J.-L.}\ \bibnamefont {Song}},\ and\
  \bibinfo {author} {\bibfnamefont {S.-J.}\ \bibnamefont {Gu}},\ }\bibfield
  {title} {\bibinfo {title} {Local entanglement and quantum phase transition in
  a one-dimensional transverse field {I}sing model},\ }\href
  {https://doi.org/10.1103/PhysRevA.74.032308} {\bibfield  {journal} {\bibinfo
  {journal} {Phys. Rev. A}\ }\textbf {\bibinfo {volume} {74}},\ \bibinfo
  {pages} {032308} (\bibinfo {year} {2006})}\BibitemShut {NoStop}%
\bibitem [{\citenamefont {Liu}\ \emph {et~al.}(2021{\natexlab{b}})\citenamefont
  {Liu}, \citenamefont {Dong}, \citenamefont {Wu}, \citenamefont {Wang},\ and\
  \citenamefont {You}}]{GammaLZA}%
  \BibitemOpen
  \bibfield  {author} {\bibinfo {author} {\bibfnamefont {Z.-A.}\ \bibnamefont
  {Liu}}, \bibinfo {author} {\bibfnamefont {Y.-L.}\ \bibnamefont {Dong}},
  \bibinfo {author} {\bibfnamefont {N.}~\bibnamefont {Wu}}, \bibinfo {author}
  {\bibfnamefont {Y.}~\bibnamefont {Wang}},\ and\ \bibinfo {author}
  {\bibfnamefont {W.-L.}\ \bibnamefont {You}},\ }\bibfield  {title} {\bibinfo
  {title} {Quantum criticality and correlations in the {I}sing-{G}amma chain},\
  }\href
  {https://www.sciencedirect.com/science/article/abs/pii/S0378437121003952}
  {\bibfield  {journal} {\bibinfo  {journal} {Physica A: Statistical Mechanics
  and its Applications}\ }\textbf {\bibinfo {volume} {579}},\ \bibinfo {pages}
  {126122} (\bibinfo {year} {2021}{\natexlab{b}})}\BibitemShut {NoStop}%
\bibitem [{\citenamefont {Barouch}\ and\ \citenamefont
  {McCoy}(1971)}]{Pfaffian}%
  \BibitemOpen
  \bibfield  {author} {\bibinfo {author} {\bibfnamefont {E.}~\bibnamefont
  {Barouch}}\ and\ \bibinfo {author} {\bibfnamefont {B.~M.}\ \bibnamefont
  {McCoy}},\ }\bibfield  {title} {\bibinfo {title} {Statistical mechanics of
  the ${XY}$ model. {II}. spin-correlation functions},\ }\href
  {https://doi.org/10.1103/PhysRevA.3.786} {\bibfield  {journal} {\bibinfo
  {journal} {Phys. Rev. A}\ }\textbf {\bibinfo {volume} {3}},\ \bibinfo {pages}
  {786} (\bibinfo {year} {1971})}\BibitemShut {NoStop}%
\bibitem [{\citenamefont {McCulloch}\ \emph {et~al.}(2008)\citenamefont
  {McCulloch}, \citenamefont {Kube}, \citenamefont {Kurz}, \citenamefont
  {Kleine}, \citenamefont {Schollw\"ock},\ and\ \citenamefont
  {Kolezhuk}}]{mcculloch08vector}%
  \BibitemOpen
  \bibfield  {author} {\bibinfo {author} {\bibfnamefont {I.~P.}\ \bibnamefont
  {McCulloch}}, \bibinfo {author} {\bibfnamefont {R.}~\bibnamefont {Kube}},
  \bibinfo {author} {\bibfnamefont {M.}~\bibnamefont {Kurz}}, \bibinfo {author}
  {\bibfnamefont {A.}~\bibnamefont {Kleine}}, \bibinfo {author} {\bibfnamefont
  {U.}~\bibnamefont {Schollw\"ock}},\ and\ \bibinfo {author} {\bibfnamefont
  {A.~K.}\ \bibnamefont {Kolezhuk}},\ }\bibfield  {title} {\bibinfo {title}
  {Vector chiral order in frustrated spin chains},\ }\href
  {https://doi.org/10.1103/PhysRevB.77.094404} {\bibfield  {journal} {\bibinfo
  {journal} {Phys. Rev. B}\ }\textbf {\bibinfo {volume} {77}},\ \bibinfo
  {pages} {094404} (\bibinfo {year} {2008})}\BibitemShut {NoStop}%
\bibitem [{\citenamefont {Ueda}\ and\ \citenamefont
  {Onoda}(2014)}]{ueda2014vector}%
  \BibitemOpen
  \bibfield  {author} {\bibinfo {author} {\bibfnamefont {H.}~\bibnamefont
  {Ueda}}\ and\ \bibinfo {author} {\bibfnamefont {S.}~\bibnamefont {Onoda}},\
  }\bibfield  {title} {\bibinfo {title} {Vector-spin-chirality order in a
  dimerized frustrated spin-$\frac{1}{2}$ chain},\ }\href
  {https://doi.org/10.1103/PhysRevB.89.024407} {\bibfield  {journal} {\bibinfo
  {journal} {Phys. Rev. B}\ }\textbf {\bibinfo {volume} {89}},\ \bibinfo
  {pages} {024407} (\bibinfo {year} {2014})}\BibitemShut {NoStop}%
\bibitem [{\citenamefont {Vidal}\ \emph {et~al.}(2003)\citenamefont {Vidal},
  \citenamefont {Latorre}, \citenamefont {Rico},\ and\ \citenamefont
  {Kitaev}}]{vidal2003entanglement}%
  \BibitemOpen
  \bibfield  {author} {\bibinfo {author} {\bibfnamefont {G.}~\bibnamefont
  {Vidal}}, \bibinfo {author} {\bibfnamefont {J.~I.}\ \bibnamefont {Latorre}},
  \bibinfo {author} {\bibfnamefont {E.}~\bibnamefont {Rico}},\ and\ \bibinfo
  {author} {\bibfnamefont {A.}~\bibnamefont {Kitaev}},\ }\bibfield  {title}
  {\bibinfo {title} {Entanglement in quantum critical phenomena},\ }\href
  {https://doi.org/10.1103/PhysRevLett.90.227902} {\bibfield  {journal}
  {\bibinfo  {journal} {Phys. Rev. Lett.}\ }\textbf {\bibinfo {volume} {90}},\
  \bibinfo {pages} {227902} (\bibinfo {year} {2003})}\BibitemShut {NoStop}%
\bibitem [{\citenamefont {Osterloh}\ \emph {et~al.}(2002)\citenamefont
  {Osterloh}, \citenamefont {Amico}, \citenamefont {Falci},\ and\ \citenamefont
  {Fazio}}]{osterloh2002scaling}%
  \BibitemOpen
  \bibfield  {author} {\bibinfo {author} {\bibfnamefont {A.}~\bibnamefont
  {Osterloh}}, \bibinfo {author} {\bibfnamefont {L.}~\bibnamefont {Amico}},
  \bibinfo {author} {\bibfnamefont {G.}~\bibnamefont {Falci}},\ and\ \bibinfo
  {author} {\bibfnamefont {R.}~\bibnamefont {Fazio}},\ }\bibfield  {title}
  {\bibinfo {title} {Scaling of entanglement close to a quantum phase
  transition},\ }\href {https://www.nature.com/articles/416608a} {\bibfield
  {journal} {\bibinfo  {journal} {Nature}\ }\textbf {\bibinfo {volume} {416}},\
  \bibinfo {pages} {608} (\bibinfo {year} {2002})}\BibitemShut {NoStop}%
\bibitem [{\citenamefont {Gu}\ \emph {et~al.}(2003)\citenamefont {Gu},
  \citenamefont {Lin},\ and\ \citenamefont {Li}}]{gu2003entanglement}%
  \BibitemOpen
  \bibfield  {author} {\bibinfo {author} {\bibfnamefont {S.-J.}\ \bibnamefont
  {Gu}}, \bibinfo {author} {\bibfnamefont {H.-Q.}\ \bibnamefont {Lin}},\ and\
  \bibinfo {author} {\bibfnamefont {Y.-Q.}\ \bibnamefont {Li}},\ }\bibfield
  {title} {\bibinfo {title} {Entanglement, quantum phase transition, and
  scaling in the $\mathrm{XXZ}$ chain},\ }\href
  {https://doi.org/10.1103/PhysRevA.68.042330} {\bibfield  {journal} {\bibinfo
  {journal} {Phys. Rev. A}\ }\textbf {\bibinfo {volume} {68}},\ \bibinfo
  {pages} {042330} (\bibinfo {year} {2003})}\BibitemShut {NoStop}%
\bibitem [{\citenamefont {Ollivier}\ and\ \citenamefont
  {Zurek}(2001)}]{ollivier2001quantum}%
  \BibitemOpen
  \bibfield  {author} {\bibinfo {author} {\bibfnamefont {H.}~\bibnamefont
  {Ollivier}}\ and\ \bibinfo {author} {\bibfnamefont {W.~H.}\ \bibnamefont
  {Zurek}},\ }\bibfield  {title} {\bibinfo {title} {Quantum discord: A measure
  of the quantumness of correlations},\ }\href
  {https://doi.org/10.1103/PhysRevLett.88.017901} {\bibfield  {journal}
  {\bibinfo  {journal} {Phys. Rev. Lett.}\ }\textbf {\bibinfo {volume} {88}},\
  \bibinfo {pages} {017901} (\bibinfo {year} {2001})}\BibitemShut {NoStop}%
\bibitem [{\citenamefont {Modi}\ \emph {et~al.}(2012)\citenamefont {Modi},
  \citenamefont {Brodutch}, \citenamefont {Cable}, \citenamefont {Paterek},\
  and\ \citenamefont {Vedral}}]{modi2012classical}%
  \BibitemOpen
  \bibfield  {author} {\bibinfo {author} {\bibfnamefont {K.}~\bibnamefont
  {Modi}}, \bibinfo {author} {\bibfnamefont {A.}~\bibnamefont {Brodutch}},
  \bibinfo {author} {\bibfnamefont {H.}~\bibnamefont {Cable}}, \bibinfo
  {author} {\bibfnamefont {T.}~\bibnamefont {Paterek}},\ and\ \bibinfo {author}
  {\bibfnamefont {V.}~\bibnamefont {Vedral}},\ }\bibfield  {title} {\bibinfo
  {title} {The classical-quantum boundary for correlations: Discord and related
  measures},\ }\href {https://doi.org/10.1103/RevModPhys.84.1655} {\bibfield
  {journal} {\bibinfo  {journal} {Rev. Mod. Phys.}\ }\textbf {\bibinfo {volume}
  {84}},\ \bibinfo {pages} {1655} (\bibinfo {year} {2012})}\BibitemShut
  {NoStop}%
\bibitem [{\citenamefont {Chen}\ \emph {et~al.}(2016)\citenamefont {Chen},
  \citenamefont {Cui}, \citenamefont {Zhang},\ and\ \citenamefont
  {Fan}}]{chen2016coherence}%
  \BibitemOpen
  \bibfield  {author} {\bibinfo {author} {\bibfnamefont {J.-J.}\ \bibnamefont
  {Chen}}, \bibinfo {author} {\bibfnamefont {J.}~\bibnamefont {Cui}}, \bibinfo
  {author} {\bibfnamefont {Y.-R.}\ \bibnamefont {Zhang}},\ and\ \bibinfo
  {author} {\bibfnamefont {H.}~\bibnamefont {Fan}},\ }\bibfield  {title}
  {\bibinfo {title} {Coherence susceptibility as a probe of quantum phase
  transitions},\ }\href {https://doi.org/10.1103/PhysRevA.94.022112} {\bibfield
   {journal} {\bibinfo  {journal} {Phys. Rev. A}\ }\textbf {\bibinfo {volume}
  {94}},\ \bibinfo {pages} {022112} (\bibinfo {year} {2016})}\BibitemShut
  {NoStop}%
\bibitem [{\citenamefont {You}\ \emph {et~al.}(2007)\citenamefont {You},
  \citenamefont {Li},\ and\ \citenamefont {Gu}}]{you2007fidelity}%
  \BibitemOpen
  \bibfield  {author} {\bibinfo {author} {\bibfnamefont {W.-L.}\ \bibnamefont
  {You}}, \bibinfo {author} {\bibfnamefont {Y.-W.}\ \bibnamefont {Li}},\ and\
  \bibinfo {author} {\bibfnamefont {S.-J.}\ \bibnamefont {Gu}},\ }\bibfield
  {title} {\bibinfo {title} {Fidelity, dynamic structure factor, and
  susceptibility in critical phenomena},\ }\href
  {https://doi.org/10.1103/PhysRevE.76.022101} {\bibfield  {journal} {\bibinfo
  {journal} {Phys. Rev. E}\ }\textbf {\bibinfo {volume} {76}},\ \bibinfo
  {pages} {022101} (\bibinfo {year} {2007})}\BibitemShut {NoStop}%
\bibitem [{\citenamefont {Bloch}(2008)}]{Bloch2008quantum}%
  \BibitemOpen
  \bibfield  {author} {\bibinfo {author} {\bibfnamefont {I.}~\bibnamefont
  {Bloch}},\ }\bibfield  {title} {\bibinfo {title} {Quantum coherence and
  entanglement with ultracold atoms in optical lattices},\ }\href
  {https://doi.org/10.1038/nature07126} {\bibfield  {journal} {\bibinfo
  {journal} {Nature}\ }\textbf {\bibinfo {volume} {453}},\ \bibinfo {pages}
  {1016} (\bibinfo {year} {2008})}\BibitemShut {NoStop}%
\bibitem [{\citenamefont {Baumgratz}\ \emph {et~al.}(2014)\citenamefont
  {Baumgratz}, \citenamefont {Cramer},\ and\ \citenamefont
  {Plenio}}]{Baumgratz14quantify}%
  \BibitemOpen
  \bibfield  {author} {\bibinfo {author} {\bibfnamefont {T.}~\bibnamefont
  {Baumgratz}}, \bibinfo {author} {\bibfnamefont {M.}~\bibnamefont {Cramer}},\
  and\ \bibinfo {author} {\bibfnamefont {M.~B.}\ \bibnamefont {Plenio}},\
  }\bibfield  {title} {\bibinfo {title} {Quantifying coherence},\ }\href
  {https://doi.org/10.1103/PhysRevLett.113.140401} {\bibfield  {journal}
  {\bibinfo  {journal} {Phys. Rev. Lett.}\ }\textbf {\bibinfo {volume} {113}},\
  \bibinfo {pages} {140401} (\bibinfo {year} {2014})}\BibitemShut {NoStop}%
\bibitem [{\citenamefont {You}\ \emph {et~al.}(2017)\citenamefont {You},
  \citenamefont {Zhang}, \citenamefont {Ni}, \citenamefont {Gong},\ and\
  \citenamefont {Ole\ifmmode~\acute{s}\else \'{s}\fi{}}}]{you17emergent}%
  \BibitemOpen
  \bibfield  {author} {\bibinfo {author} {\bibfnamefont {W.-L.}\ \bibnamefont
  {You}}, \bibinfo {author} {\bibfnamefont {C.-J.}\ \bibnamefont {Zhang}},
  \bibinfo {author} {\bibfnamefont {W.}~\bibnamefont {Ni}}, \bibinfo {author}
  {\bibfnamefont {M.}~\bibnamefont {Gong}},\ and\ \bibinfo {author}
  {\bibfnamefont {A.~M.}\ \bibnamefont {Ole\ifmmode~\acute{s}\else
  \'{s}\fi{}}},\ }\bibfield  {title} {\bibinfo {title} {Emergent phases in a
  compass chain with multisite interactions},\ }\href
  {https://doi.org/10.1103/PhysRevB.95.224404} {\bibfield  {journal} {\bibinfo
  {journal} {Phys. Rev. B}\ }\textbf {\bibinfo {volume} {95}},\ \bibinfo
  {pages} {224404} (\bibinfo {year} {2017})}\BibitemShut {NoStop}%
\bibitem [{\citenamefont {You}\ \emph {et~al.}(2018)\citenamefont {You},
  \citenamefont {Wang}, \citenamefont {Yi}, \citenamefont {Zhang},\ and\
  \citenamefont {Ole\ifmmode~\acute{s}\else \'{s}\fi{}}}]{you18quantum}%
  \BibitemOpen
  \bibfield  {author} {\bibinfo {author} {\bibfnamefont {W.-L.}\ \bibnamefont
  {You}}, \bibinfo {author} {\bibfnamefont {Y.}~\bibnamefont {Wang}}, \bibinfo
  {author} {\bibfnamefont {T.-C.}\ \bibnamefont {Yi}}, \bibinfo {author}
  {\bibfnamefont {C.}~\bibnamefont {Zhang}},\ and\ \bibinfo {author}
  {\bibfnamefont {A.~M.}\ \bibnamefont {Ole\ifmmode~\acute{s}\else
  \'{s}\fi{}}},\ }\bibfield  {title} {\bibinfo {title} {Quantum coherence in a
  compass chain under an alternating magnetic field},\ }\href
  {https://doi.org/10.1103/PhysRevB.97.224420} {\bibfield  {journal} {\bibinfo
  {journal} {Phys. Rev. B}\ }\textbf {\bibinfo {volume} {97}},\ \bibinfo
  {pages} {224420} (\bibinfo {year} {2018})}\BibitemShut {NoStop}%
\bibitem [{\citenamefont {Yi}\ \emph {et~al.}(2019)\citenamefont {Yi},
  \citenamefont {You}, \citenamefont {Wu},\ and\ \citenamefont
  {Ole\ifmmode~\acute{s}\else \'{s}\fi{}}}]{yi19criticality}%
  \BibitemOpen
  \bibfield  {author} {\bibinfo {author} {\bibfnamefont {T.-C.}\ \bibnamefont
  {Yi}}, \bibinfo {author} {\bibfnamefont {W.-L.}\ \bibnamefont {You}},
  \bibinfo {author} {\bibfnamefont {N.}~\bibnamefont {Wu}},\ and\ \bibinfo
  {author} {\bibfnamefont {A.~M.}\ \bibnamefont {Ole\ifmmode~\acute{s}\else
  \'{s}\fi{}}},\ }\bibfield  {title} {\bibinfo {title} {Criticality and
  factorization in the {H}eisenberg chain with {D}zyaloshinskii-{M}oriya
  interaction},\ }\href {https://doi.org/10.1103/PhysRevB.100.024423}
  {\bibfield  {journal} {\bibinfo  {journal} {Phys. Rev. B}\ }\textbf {\bibinfo
  {volume} {100}},\ \bibinfo {pages} {024423} (\bibinfo {year}
  {2019})}\BibitemShut {NoStop}%
\bibitem [{\citenamefont {Hu}\ and\ \citenamefont {Fan}(2018)}]{M.L.Hu2018}%
  \BibitemOpen
  \bibfield  {author} {\bibinfo {author} {\bibfnamefont {M.-L.}\ \bibnamefont
  {Hu}}\ and\ \bibinfo {author} {\bibfnamefont {H.}~\bibnamefont {Fan}},\
  }\bibfield  {title} {\bibinfo {title} {Nonlocal advantage of quantum
  coherence in high-dimensional states},\ }\href
  {https://doi.org/10.1103/PhysRevA.98.022312} {\bibfield  {journal} {\bibinfo
  {journal} {Phys. Rev. A}\ }\textbf {\bibinfo {volume} {98}},\ \bibinfo
  {pages} {022312} (\bibinfo {year} {2018})}\BibitemShut {NoStop}%
\bibitem [{\citenamefont {Hu}\ \emph {et~al.}(2020)\citenamefont {Hu},
  \citenamefont {Gao},\ and\ \citenamefont {Fan}}]{M.L.Hu2020}%
  \BibitemOpen
  \bibfield  {author} {\bibinfo {author} {\bibfnamefont {M.-L.}\ \bibnamefont
  {Hu}}, \bibinfo {author} {\bibfnamefont {Y.-Y.}\ \bibnamefont {Gao}},\ and\
  \bibinfo {author} {\bibfnamefont {H.}~\bibnamefont {Fan}},\ }\bibfield
  {title} {\bibinfo {title} {Steered quantum coherence as a signature of
  quantum phase transitions in spin chains},\ }\href
  {https://doi.org/10.1103/PhysRevA.101.032305} {\bibfield  {journal} {\bibinfo
   {journal} {Phys. Rev. A}\ }\textbf {\bibinfo {volume} {101}},\ \bibinfo
  {pages} {032305} (\bibinfo {year} {2020})}\BibitemShut {NoStop}%
\bibitem [{\citenamefont {Hu}\ \emph {et~al.}(2021)\citenamefont {Hu},
  \citenamefont {Fang},\ and\ \citenamefont {Fan}}]{M.L.Hu2021}%
  \BibitemOpen
  \bibfield  {author} {\bibinfo {author} {\bibfnamefont {M.-L.}\ \bibnamefont
  {Hu}}, \bibinfo {author} {\bibfnamefont {F.}~\bibnamefont {Fang}},\ and\
  \bibinfo {author} {\bibfnamefont {H.}~\bibnamefont {Fan}},\ }\bibfield
  {title} {\bibinfo {title} {Finite-size scaling of coherence and steered
  coherence in the lipkin-meshkov-glick model},\ }\href
  {https://doi.org/10.1103/PhysRevA.104.062416} {\bibfield  {journal} {\bibinfo
   {journal} {Phys. Rev. A}\ }\textbf {\bibinfo {volume} {104}},\ \bibinfo
  {pages} {062416} (\bibinfo {year} {2021})}\BibitemShut {NoStop}%
\bibitem [{\citenamefont {Sinha}\ \emph {et~al.}(2019)\citenamefont {Sinha},
  \citenamefont {Rams},\ and\ \citenamefont {Dziarmaga}}]{PhysRevB.99.094203}%
  \BibitemOpen
  \bibfield  {author} {\bibinfo {author} {\bibfnamefont {A.}~\bibnamefont
  {Sinha}}, \bibinfo {author} {\bibfnamefont {M.~M.}\ \bibnamefont {Rams}},\
  and\ \bibinfo {author} {\bibfnamefont {J.}~\bibnamefont {Dziarmaga}},\
  }\bibfield  {title} {\bibinfo {title} {Kibble-{Z}urek mechanism with a single
  particle: Dynamics of the localization-delocalization transition in the
  {A}ubry-{A}ndr\'e model},\ }\href
  {https://doi.org/10.1103/PhysRevB.99.094203} {\bibfield  {journal} {\bibinfo
  {journal} {Phys. Rev. B}\ }\textbf {\bibinfo {volume} {99}},\ \bibinfo
  {pages} {094203} (\bibinfo {year} {2019})}\BibitemShut {NoStop}%
\bibitem [{\citenamefont {Chepiga}\ and\ \citenamefont
  {Mila}(2021)}]{chepiga2021kibble}%
  \BibitemOpen
  \bibfield  {author} {\bibinfo {author} {\bibfnamefont {N.}~\bibnamefont
  {Chepiga}}\ and\ \bibinfo {author} {\bibfnamefont {F.}~\bibnamefont {Mila}},\
  }\bibfield  {title} {\bibinfo {title} {Kibble-{Z}urek exponent and chiral
  transition of the period-4 phase of {R}ydberg chains},\ }\href
  {https://www.nature.com/articles/s41467-020-20641-y} {\bibfield  {journal}
  {\bibinfo  {journal} {Nature Communications}\ }\textbf {\bibinfo {volume}
  {12}},\ \bibinfo {pages} {414} (\bibinfo {year} {2021})}\BibitemShut
  {NoStop}%
\end{thebibliography}%

\appendix
\begin{appendix}
\twocolumngrid

\section{Effective Hamiltonian}\label{DerivationOfEffectiveHamiltonian}

The time-independent Hamiltonian in the rotating frame is $\tilde{H}= UHU^\dagger+i(\partial_t U)U^\dagger$,
and therefore becomes Eq.\,(\ref{eq:Hrot}).
In the second quantization form of field operators $(\hat{\psi}_s, \hat{\psi}_g, \hat{\psi}_1,\hat{\psi}_2)$,
\begin{eqnarray}
  \tilde{H}&=&-\sum_{\mathbf{k}} \Delta_\mathbf{k} \hat{a}^\dagger_\mathbf{k}\hat{a}_\mathbf{k}
  +\int d\mathbf{r} \left[\delta \hat{\psi}_g^\dagger\hat{\psi}_g  -\Delta_1\hat{\psi}_1^\dagger\hat{\psi}_1-\Delta_2\hat{\psi}_2^\dagger\hat{\psi}_2\right]\nonumber\\
  &&+\int d\mathbf{r}\left[{\Omega_1(\mathbf{r})}\hat{\psi}^\dagger_1\hat{\psi}_g
  +{\Omega_2(\mathbf{r})}\hat{\psi}^\dagger_2\hat{\psi}_s+{\rm H.c.}\right]\nonumber\\
  &&+\int d\mathbf{r}\sum_{\mathbf{k}}
  \left[G_{\mathbf{k}}(\mathbf{r})\hat{a}_{\mathbf{k}}(\hat{\psi}^\dagger_1\hat{\psi}_s+\hat{\psi}^\dagger_2\hat{\psi}_g)+{\rm H.c.}\right],
  \end{eqnarray}
in which we use the relation $\hat\psi_\alpha^\dagger\hat{\psi}_\beta = \sum_m |\alpha\rangle_m\langle \beta|$.

\emph{Heisenberg picture.--} Considering the Heisenberg equation for field operators, we have
\begin{eqnarray}
  i\partial_t\hat{\psi}_1&=&[\hat{\psi}_1, \tilde{H}]
  =-\Delta_1\hat{\psi}_1 +\Omega_1\hat{\psi}_g +
  \sum_{\mathbf{k}}G_\mathbf{k} \hat{a}_\mathbf{k} \hat{\psi}_s,\\
  i\partial_t\hat{\psi}_2&=& [\hat{\psi}_2, \tilde{H}]=
   -\Delta_2\hat{\psi}_2 +\Omega_2\hat{\psi}_s + \sum_{\mathbf{k}}G_\mathbf{k} \hat{a}_\mathbf{k} \hat{\psi}_g,\\
   i\partial_t\hat{a}_\mathbf{k}&=&[\hat{a}_\mathbf{k}, \tilde{H}] =
   -(\Delta_\mathbf{k}+i\kappa)\hat{a}_\mathbf{k}
   +\int d\mathbf{r}\,G^\ast_\mathbf{k}(\hat{\psi}^\dagger_s\hat{\psi}_1+\hat{\psi}^\dagger_g\hat{\psi}_2), \nonumber \\ \label{eq:a_eom}
\end{eqnarray}
wherein we have included the cavity dissipation $\kappa$ phenomenologically. To this end, the steady-state values of $\hat{\psi}_{1,2}$ are
\begin{eqnarray}
  \hat{\psi}_{1,\rm{ss}}=\frac{1}{\Delta_1}(\Omega_1\hat{\psi}_g+
  \sum_{\mathbf{k}}{G_\mathbf{k}} \hat{a}_\mathbf{k} \hat{\psi}_s),\\
  \hat{\psi}_{2,\rm{ss}}=\frac{1}{\Delta_2}(\Omega_2\hat{\psi}_s+
  \sum_{\mathbf{k}}{G_\mathbf{k}} \hat{a}_\mathbf{k} \hat{\psi}_g).
\end{eqnarray}
Then we can obtain the Heisenberg equation of motion for $\hat{\psi}_{s(g)}$,
\begin{eqnarray}
  i\partial_t\hat{\psi}_g
  &=&(\delta+V_g)\hat{\psi}_g+\hat{\Omega}\hat{\psi}_s,\\
  i\partial_t\hat{\psi}_s
  &=&V_s\hat{\psi}_s+\hat{\Omega}^\dagger\hat{\psi}_g,
\end{eqnarray}
and the Hamiltonian in the ground-sate manifold is
\begin{eqnarray}
  \hat{H}_{gs}&=&\int d\mathbf{r} \left[(\delta+V_g(\mathbf{r}))\hat{\psi}_g^\dagger(\mathbf{r})\hat{\psi}_g(\mathbf{r})+V_s(\mathbf{r})\hat{\psi}^\dagger_s(\mathbf{r})\hat{\psi}_s(\mathbf{r})\right]\nonumber\\
  &&
  +\int d\mathbf{r}\left[\hat{\psi}^\dagger_g(\mathbf{r})\hat{\Omega}\hat{\psi}_s(\mathbf{r})
  +\hat{\psi}_s^\dagger(\mathbf{r})\hat{\Omega}^\dagger\hat{\psi}_g(\mathbf{r})\right].\label{eq:Hgs_b}
\end{eqnarray}
Here the light-induced potential $V_{g(s)}$ and Raman coupling operator $\hat{\Omega}$ are defined as follows:
\begin{eqnarray}
  V_g(\mathbf{r})&=&\frac{|\Omega_1|^2}{\Delta_1}+\frac{\sum_{kk^\prime}G^\ast_\mathbf{k}G_{\mathbf{k}^\prime}\hat{a}^\dagger_{\mathbf{k}}\hat{a}_{\mathbf{k}^\prime}}{\Delta_2},\\
  V_s(\mathbf{r})&=&\frac{|\Omega_2|^2}{\Delta_2}+\frac{\sum_{kk^\prime}G^\ast_\mathbf{k}G_{\mathbf{k}^\prime}\hat{a}^\dagger_{\mathbf{k}}\hat{a}_{\mathbf{k}^\prime}}{\Delta_2},\\
  \hat{\Omega}(\mathbf{r})&=&\frac{\Omega_1^\ast}{\Delta_1}\sum_\mathbf{k}G_\mathbf{k}(\mathbf{r})\hat{a}_{\mathbf{k}}
  +\frac{\Omega_2}{\Delta_2}\sum_\mathbf{k}G^\ast_\mathbf{k}(\mathbf{r})\hat{a}^\dagger_{\mathbf{k}}.
  \end{eqnarray}
Substituting $\hat{\psi}_{1,\rm{ss}}$,  $\hat{\psi}_{2,\rm{ss}}$ into Eq.~(\ref{eq:a_eom}),  yields
  \begin{eqnarray}
  i\partial_t\hat{a}_\mathbf{k}=
     -(\Delta_\mathbf{k}+i\kappa)\hat{a}_\mathbf{k}
     + \int d\mathbf{r}\,G^\ast_\mathbf{k}(\hat{\psi}^\dagger_s\hat{\psi}_1+\hat{\psi}^\dagger_g\hat{\psi}_2).\quad \quad
  \end{eqnarray}
We neglect the light shift from different bosonic modes, which can be negligible small or even zero if we take orthogonal spatial modes for different bosonic modes.
In the adiabatic limit due to large detuning or large dissipation $\kappa$
of modes $\hat{a}_\mathbf{k}$,
the photon field can be approximated by its steady-state value,
\begin{eqnarray}
\hat{a}_{\mathbf{k}}^{\rm{ss}}&\approx&
  ({1}/{\tilde{\Delta}_\mathbf{k}})
\int d\mathbf{r}\, G^\ast_{\mathbf{k}}
  (\frac{\Omega_1}{\Delta_1}\hat{\psi}^\dagger_s\hat{\psi}_g+\frac{\Omega_2}{\Delta_2}\hat{\psi}^\dagger_g\hat{\psi}_s),
\end{eqnarray}
wherein we have defined $\tilde{\Delta}_\mathbf{k}=\Delta_\mathbf{k}+i\kappa-
\int d\mathbf{r}\,|G_\mathbf{k}|^2[({1}/{\Delta_1})\hat{\psi}^\dagger_s\hat{\psi}_s+
({1}/{\Delta_2})\hat{\psi}^\dagger_g\hat{\psi}_g]$.
In this case, $\hat{\Omega}(\mathbf{r})\approx\int d\mathbf{r}^\prime \Lambda_{0}(\mathbf{r},\mathbf{r}^\prime)\hat{\psi}^\dagger_s(\mathbf{r}^\prime)\hat{\psi}_g(\mathbf{r}^\prime)+\int d\mathbf{r}^\prime\Lambda_{1}(\mathbf{r},\mathbf{r}^\prime)
\hat{\psi}^\dagger_g(\mathbf{r}^\prime)\hat{\psi}_s(\mathbf{r}^\prime),$
where
\begin{eqnarray}
  \Lambda_0(\mathbf{r},\mathbf{r}^\prime)&=&\eta_{11}(\mathbf{r},\mathbf{r}^\prime)+\eta_{22}(\mathbf{r},\mathbf{r}^\prime),\\
  \Lambda_1(\mathbf{r},\mathbf{r}^\prime)&=& \eta_{12}(\mathbf{r},\mathbf{r}^\prime)+\eta_{21}(\mathbf{r},\mathbf{r}^\prime),
\end{eqnarray}
with coefficients
\begin{eqnarray}
  \eta_{11}(\mathbf{r},\mathbf{r}^\prime)&=& \sum_\mathbf{k}
  \frac{\Omega_1^\ast(\mathbf{r})\Omega_1(\mathbf{r}^\prime)}{\Delta_1^2\tilde{\Delta}_\mathbf{k}}
  G_\mathbf{k}(\mathbf{r})G^\ast_{\mathbf{k}}(\mathbf{r}^\prime), \\
  \eta_{12}(\mathbf{r},\mathbf{r}^\prime)&=&\sum_\mathbf{k}
  \frac{\Omega_1^\ast(\mathbf{r})\Omega_2(\mathbf{r}^\prime)}{\Delta_1\Delta_2\tilde{\Delta}_\mathbf{k}}
  G_\mathbf{k}(\mathbf{r})G^\ast_{\mathbf{k}}(\mathbf{r}^\prime),\\
  \eta_{21}(\mathbf{r},\mathbf{r}^\prime)&=&\sum_\mathbf{k}
  \frac{\Omega_2(\mathbf{r})\Omega_1^\ast(\mathbf{r}^\prime)}{\Delta_1\Delta_2\tilde{\Delta}^\ast_\mathbf{k}}
  G_\mathbf{k}^\ast(\mathbf{r})G_{\mathbf{k}}(\mathbf{r}^\prime),\\
  \eta_{22}(\mathbf{r},\mathbf{r}^\prime)&=&\sum_\mathbf{k}
  \frac{\Omega_2(\mathbf{r})\Omega_2^\ast(\mathbf{r}^\prime)}{\Delta^2_2\tilde{\Delta}^\ast_\mathbf{k}}
  G_\mathbf{k}^\ast(\mathbf{r})G_{\mathbf{k}}(\mathbf{r}^\prime).
\end{eqnarray}

Therefore, $\hat{H}_{gs}$ in Eq.~(\ref{eq:Hgs_b}) becomes
\begin{eqnarray}
  \hat{H}_{gs}&=&\hat{H}_{\rm 0}
  +\hat{H}_{\rm int},
  \end{eqnarray}
  where
  \begin{eqnarray}
  \hat{H}_{\rm 0}\!&=&\!\int\!\!d\mathbf{r} [(\delta+V_g(\mathbf{r}))\!\hat{\psi}_g^\dagger(\mathbf{r})\hat{\psi}_g(\mathbf{r})\!+\!
  V_s(\mathbf{r})\!\hat{\psi}^\dagger_s(\mathbf{r})\hat{\psi}_s(\mathbf{r})], \quad \quad \quad   \\
  \hat{H}_{\rm int}&=&\iint d\mathbf{r}^\prime d\mathbf{r}\left[\Lambda_{0}(\mathbf{r},\mathbf{r}^\prime)
  \hat{\mathbf{s}}_+(\mathbf{r})  \hat{\mathbf{s}}_-(\mathbf{r}^\prime)\right.
  \nonumber \\&& \quad  \quad +\left.\Lambda_1(\mathbf{r},\mathbf{r}^\prime) \hat{\mathbf{s}}_+(\mathbf{r})\hat{\mathbf{s}}_+(\mathbf{r}^\prime)
  +{\rm H.c.}\right].\label{eq:Hgs_c}
\end{eqnarray}
We use site index to label the trapped atoms, and now the
interaction Hamiltonian can be recast as
\begin{eqnarray}
  \hat{H}_{\rm int}&=&\sum_{mn}\left[\Lambda_{0}(\mathbf{r}_m,\mathbf{r}_n)
  \hat{\sigma}_m^+ \hat{\sigma}_n^-
  +\Lambda_1(\mathbf{r}_m,\mathbf{r}_n) \hat{\sigma}_m^+ \hat{\sigma}_n^+
  +{\rm H.c.}\right]\label{eq:Hgs_d}\nonumber \\
  &=&\sum_{mn}\,2(\Re{[\Lambda_0]}+\Re{[\Lambda_1]})  \hat{\sigma}_m^x \hat{\sigma}_n^x\nonumber\\
  &&+\sum_{mn}\,2(\Re{[\Lambda_0]}-\Re{[\Lambda_1]}) \hat{\sigma}_m^y \hat{\sigma}_n^y\nonumber\\
  &&+\sum_{mn}\, 2\Im[\Lambda_0]\left[\hat{\sigma}_m^x \hat{\sigma}_n^y
  -\hat{\sigma}_m^y \hat{\sigma}_n^x\right]\nonumber\\
  &&-\sum_{mn} 2\Im[\Lambda_1] \left[\hat{\sigma}_m^y \hat{\sigma}_n^x
  + \hat{\sigma}_m^x \hat{\sigma}_n^y\right].
\end{eqnarray}

\section{Exact Solution of  \emph{XY}-Gamma Model and Correlations}
\label{Diagonalization}

\subsection{Energy spectrum and finite-size scaling}
\label{energyspect1}
\emph{Energy spectrum.--}The Jordan-Winger transformation,  provides an efficient nonlocal mapping between  spin operators and spinless fermion operators through the following relation:
\begin{eqnarray}
\sigma _{j}^{x} &=&-\prod\limits_{l<j}\left( 1-2c_{l}^{\dagger }c_{l}\right)
\left( c_{j}+c_{j}^+\right) ,  \label{JW2}\\ \sigma _{j}^{z} &=&1-2c_{j}^{\dagger }c_{j}\text{, }\sigma _{j}^{y}=\mathrm{i%
}\sigma _{j}^{x}\sigma _{j}^{z},  \label{JW1}
\end{eqnarray}%
in which $c_{j}$ ($c_{j}^{\dagger}$) is the annihilation (creation)
operator of spinless fermion at site $j$ obeying the standard
anticommutation relations, $\{c_{i},c_{j}\}$=$\{c_{i}^{\dagger},c_{j}^{\dagger}\}=0$ and
$\{c_{i}^{\dagger},c_{j}\}=\delta_{ij}$.
Then, the Hamiltonian (\ref{eq:Hamiltonian1}) can be cast into a quadratic form of spinless fermions:
\begin{equation}
\hat{H}=\hat{ H}_{\rm{b}}+\hat{H}_{\rm{e}},
\label{fer}
\end{equation}%
with
\begin{eqnarray}
\hat{H}_{\rm b}=&&\sum_{j=1}^{N-1}[ (-c_{j}c_{j+1}^\dagger-\gamma c_{j}c_{j+1}+\gamma c_{j}^\dagger c_{j+1}^\dagger+c_{j}^\dagger c_{j+1})\nonumber\\
&& + i\Gamma(-c_{j}c_{j+1}^\dagger+c_{j}c_{j+1}+c_{j}^\dagger c_{j+1}^\dagger-c_{j}^\dagger c_{j+1})\nonumber\\
&& + i\Gamma\alpha(c_{j}c_{j+1}^\dagger+c_{j}c_{j+1}+c_{j}^\dagger c_{j+1}^\dagger+c_{j}^\dagger c_{j+1})]\nonumber\\
&& + h\sum_{j=1}^{N}(1-2c_{j}^\dagger c_{j})
\end{eqnarray}
and
\begin{eqnarray}
\hat{H}_{\rm e}&=& s[(-c_{N}c_{1}^\dagger-\gamma c_{N}c_{ 1}+\gamma c_{N}^\dagger c_{1}^\dagger+c_{N}^\dagger c_{1})\nonumber\\
&+&i\Gamma(-c_{N}c_{1}^\dagger+c_{N}c_{1}+c_{N}^\dagger c_{1}^\dagger-c_{N}^\dagger c_{1})\nonumber\\
&+&i\Gamma\alpha(c_{N}c_{1}^\dagger+c_{N}c_{1}+c_{N}^\dagger c_{1}^\dagger+c_{N}^\dagger c_{1})].
\label{ferb}
\end{eqnarray}
$\hat{ H}_{\mathrm{b}}$ and $\hat{H}_{\mathrm{e}}$  represent the contribution from the bulk and the edges, respectively. One can find that an extra phase factor $s=(-1)^{N_p+1}$ in Eq.~(\ref{ferb}) with total fermion number $N_p$=$\sum_{j=1}^N c_{j}^\dagger c_{j}$ makes the Hilbert space decompose into odd and even parity subspaces, leading to either a periodic boundary condition ($c_{N+1}$= $c_1$) or antiperiodic boundary condition ($c_{N+1}= -c_1$) for the spinless fermionic chain. In the thermodynamic limit, the $1/N$ correction due to the subtle boundary term becomes negligible. In this regard, the Hamiltonian (\ref{fer}) can be further diagonalized in terms of Fourier transformations:
\begin{eqnarray}
c_{j}=\frac{1}{\sqrt{N}}\sum_{k}e^{-ik j}c_{k},  \quad c_{j}^\dagger=\frac{1}{\sqrt{N}}\sum_{k}e^{ ik j}c_{k}^\dagger,
\end{eqnarray}
where the "half-integer" momenta in the antiperiodic boundary condition channel are employed, i.e., $k= n\pi/ N $, $n= -(N-1), -(N-3), \ldots, (N-1)$. The bilinear Hamiltonian can thereby be rewritten as
\begin{eqnarray}
  \hat{H}&=&\sum_{k} \left[(2\cos{k}+ 2\Gamma(\alpha-1)\sin{k} -2h)\right] c_{k}^\dagger c_{k}+Nh\nonumber\\
  &&+\sum_{k}\left\{\left[\Gamma(\alpha+1)+i\gamma\right]\sin{k}~c_{-k}c_{k}+{\rm H.c.}\right\}.
  \label{calH1}
\end{eqnarray}
Equation~(\ref{calH1}) is an extended  mean-field model for a 1D triplet superconductor,
which can then be arranged straightforwardly in the Bogoliubov-de Gennes (BdG) representation:
\begin{equation}
  \hat{H}=\sum_k
\begin{pmatrix}
c^\dagger_k & c_{-k}
\end{pmatrix}
\mathcal{H}_k
\begin{pmatrix}
c_k\\
c^\dagger_{-k}
\end{pmatrix},
\end{equation}
where
\begin{equation}
\mathcal{H}_k=\begin{pmatrix}
A_k & B_{k}\\
B_k^\ast & -A_{-k}
\end{pmatrix}~, \label{Mk}
\end{equation}
with $A_k=\cos{k}+\Gamma(\alpha-1)\sin{k}-h$,
$B_k=-i\gamma \sin{k}+(\alpha+1)\sin{k}$. Next, it can be diagonalized by using the Bogoliubov transformation
\begin{eqnarray}
b_k&=&u_k c_k+v_k e^{i\varphi_k}c_{-k}^\dagger, b_{-k}=u_k c_{-k}-v_k e^{i\varphi_k}c_{k}^\dagger, \nonumber \quad \\
b_{k}^\dagger&=&u_k c_{k}^\dagger+v_k e^{-i\varphi_k}c_{-k}, b_{-k}^\dagger=u_k c_{-k}^\dagger-v_k e^{-i\varphi_k}c_{k},\quad \quad \quad
\label{BogoliubovT}
\end{eqnarray}
where $u_k=u_{-k}$, $v_{-k}=-v_{k}$, $\varphi_{k}=\varphi_{-k}$ are real numbers.
Finally, the Hamiltonian in the diagonal form is given by
\begin{eqnarray}
\hat{H}&=& \sum_{k}{\varepsilon}_{k}({b}_{k}^{+}{b}_{k}-\frac{1}{2}),
\end{eqnarray}
with the energy spectrum being
\begin{eqnarray}
{\varepsilon}_{k}&=&
2\sqrt{\Gamma^2(\alpha+1)^2\sin^2{k}+(\cos{k}-h)^2+\gamma^2\sin^2{k}} \nonumber\\ &&-2\Gamma(1-\alpha)\sin{k} .
\label{varepsilonx}
\end{eqnarray}
In the thermodynamic limit ($N \to \infty$), the ground state of the system corresponds to the configuration where all the states with ${\varepsilon}_{k} < 0$ are filled and ${\varepsilon}_{k} \ge 0$ are vacant. The ground state $\vert GS\rangle$ is defined by
\begin{eqnarray}
b_k  \vert GS\rangle =0 \quad {\rm if} \quad {\varepsilon}_{k} \ge 0, \nonumber \\ b_k^\dagger  \vert GS\rangle =0 \quad  {\rm if}  \quad  {\varepsilon}_{k} <0.
\end{eqnarray}
The ground-state energy is given by
\begin{eqnarray}
E_0=-\frac{1}{2}\sum_k \vert {\varepsilon}_{k}\vert.
\end{eqnarray}

\emph{Critical lines.--}  In terms of Eq.~(\ref{varepsilonx}), the critical points
can be identified by the fact that the gap is vanishing, i.e., ${\varepsilon}_{k}=0$. The critical lines 
are given by
(1) CP-1: $4\alpha\Gamma^2+\gamma^2>0$, the critical mode $k_c=0$,   and the critical field $h_{c,1}=1$;
(2) CP-2: $4\alpha\Gamma^2+\gamma^2=0$ and $h\le1$, the critical mode $k_c=\arccos h$,   and the critical line $\alpha_{c,1}=-\frac{\gamma^2}{4\Gamma^2}$;
and (3) CP-3: $ 4\alpha\Gamma^2+\gamma^2<0$, $\alpha_{c,2}=\frac{1-h^2-\gamma^2}{4\Gamma^2}$, or equivalently, $h_{c,2}=\sqrt{1-\gamma^2-4\Gamma^2\alpha}$ with the critical mode $k_c =\arccos h_{c,2}^{-1}$.
In the gapless phase, which is encompassed by CP-2 and CP-3,
the excitation spectrum $\varepsilon_k$ consist of two fermion points $k_L$, $k_R$, given by
\begin{eqnarray}
  k_{L,R} = \arccos \frac{h \pm \sqrt{(h^2-1)X+X^2}}{1-X}, X=4\alpha\Gamma^2+\gamma^2. \nonumber \\
\end{eqnarray}
When $h$ approaches $h_{c,2}$, $k_L$, and $k_R$ merge together into $\arccos h_{c,2}^{-1}$.

\emph{Critical exponents.--}
Now we show how to extract the critical exponent $z$ and $\nu$ through the ansatz $\Delta \sim(\lambda-\lambda_c)^{z\nu}$ and $\Delta (\lambda_c)$$\sim(k-k_c)^z$. First, we consider the dispersion on CP-3,
where $h=h_{c,2}$, $\cos k_c=h_{c,2}^{-1}$, ${\varepsilon} (k_c)=0$.
In this case, we expand ${\varepsilon}_{k} $ around $k_c$ to the second order of $\delta_k=k-k_c$,
\begin{eqnarray}
  \varepsilon(k_c+\delta k) &=&
   \left[\frac{\Gamma(1-\alpha)\cos^2k_c}{2\sin k_c}+\frac{( {h_{c,2}}^{-1}-h_{c,2})^2}{2\Gamma\sin k_c(1-\alpha)}\right]\delta^2_k, \nonumber\\
\end{eqnarray}
which implies $z=2$. Similarly, we expand $\Delta$ around $h_c$ with $\delta_h=h-h_{c,2}$,
\begin{eqnarray}
  \Delta &\approx&\varepsilon(k_c)
  =\frac{2(h_{c,2}-h_{c,2}^{-1})}{\Gamma(1-\alpha)\sin k_c}(h-h_{c,2}),
\end{eqnarray}
which suggests $\nu z=1$, showing $\nu=1/2$, $z=2$.
Then we focus on $\varepsilon_k$ around $h_{c,1}=1$ with $k_c=0$,
\begin{eqnarray}
\varepsilon_k
&\approx&2\sqrt{[\Gamma^2(1+\alpha)^2+\gamma^2]}|k|-2\Gamma(1-\alpha)k,
\end{eqnarray}
suggesting $z=1$. Using
\begin{eqnarray}
\Delta &=& \varepsilon(k_c=0) =2|h-1|=2(h-h_{c,1})^{\nu z}.
\end{eqnarray}
In this respect, the critical exponents $\nu=1$ and $z=1$.
Finally we expand $\varepsilon_k$ around $\alpha_{c,1}=- \gamma^2/(4\Gamma^2)$ with respect to $\delta_k=k-k_c$ with $k_c=\arccos h$,
\begin{eqnarray}
  \varepsilon_k   &=&
  \frac{\sqrt{1-h^2}}{\Gamma(1-\alpha_{c,1})}(k-k_c)^2.
\end{eqnarray}
Thus we can get $z=2$.
We also expand the gap around $\alpha_{c,1}$ with $\delta_\alpha= \alpha-\alpha_{c,1}$ as
\begin{eqnarray}
  \Delta=\frac{4\Gamma\sin k_c}{1-\alpha_{c,1}}(\alpha-\alpha_{c,1}).
\end{eqnarray}
In this regard, $\nu z=1$.

\subsection{Correlation function}
\label{correlation function1}
\begin{figure}[ht]
\includegraphics[width=0.9\columnwidth]{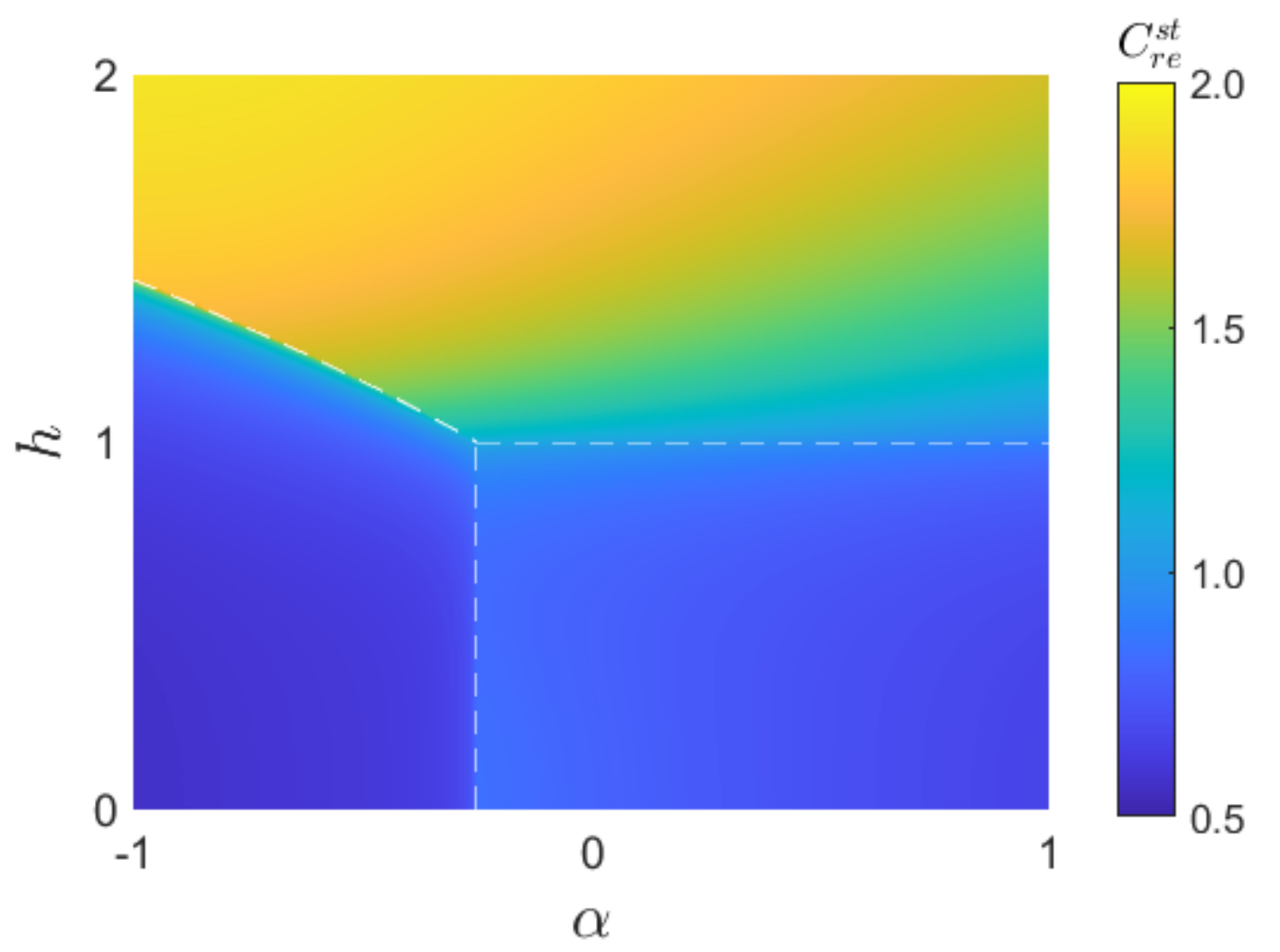}
\caption{The contour map of steered quantum coherence $C_{re}^{st}$ with respect to $\alpha$ and $h$ with $N=2000$, $J=1.00$, $\Gamma=0.60, \gamma=0.60$. The white dashed lines correspond to the critical lines.
}
\label{SQCcontour}
\end{figure}

To calculate the two-qubit correlation, we define
\begin{eqnarray}
A_i=c_i^\dagger + c_i, B_i=c_i^\dagger-c_i,
\end{eqnarray}
and it can be easily verified that the following relationships hold:
\begin{eqnarray}
  \{{A_i,A_j}\} =2\delta_{ij}, \quad\{ {B_i,B_j}\} =-2\delta_{ij},\quad\{ {A_i,B_j}\}=0. \notag
\end{eqnarray}
In this case, the Pauli matrices can be rewritten as
\begin{eqnarray}
\sigma_i^{x}=A_i \prod_{j=1}^{i-1} A_j B_j,
\sigma_i^{y}=i B_i \prod_{j=1}^{i-1} A_j B_j,
\sigma_i^{z}=A_i B_i.\notag
\end{eqnarray}
Accordingly, the two-qubit correlation of  the $x$ component can be written into fermion form using the Jordan-Wigner transformation:
\begin{eqnarray}
  G_{i,j}^{xx}  &=& \langle\sigma_i^x\sigma_j^x\rangle =\langle(\sigma_i^++\sigma_i^-)(\sigma_j^++\sigma_j^-)\rangle \notag \\
               &=&\langle(e^{i\pi\sum_{n=1}^{i-1}c_n^+c_n}c_i+ e^{-i\pi\sum_{n=1}^{i-1}c_n^+c_n}c_i^+) \notag  \\
               &\quad&(e^{i\pi\sum_{n=1}^{j-1}c_n^+c_n}c_j+ e^{-i\pi\sum_{n=1}^{j-1}c_n^+c_n}c_j^+)\rangle \notag  \\
               &=&\langle B_i(\prod_{n=i+1}^{j-1}A_nB_n)A_j\rangle\notag\\
               &=&\langle B_iA_{i+1}B_{i+1}A_{i+2}B_{i+2}\cdots A_{j-1}B_{j-1}A_j\rangle. \quad
\end{eqnarray}
Similarly, the $y$- and $z$-component correlations
\begin{eqnarray}
G_{i,j}^{yy}&=&(-1)^{j-i}\langle A_{i}B_{i+1}A_{i+1} \cdots B_{j-1}A_{j-1}B_{j}\rangle, \quad \\
G_{i,j}^{zz}&=&\langle A_{i}B_{i}A_{j}B_{j}\rangle. \quad
\end{eqnarray}
In addition, the cross-correlations $G_{i,j}^{xy}$ can also be obtained through
\begin{eqnarray}
G_{i,j}^{xy}&=& i\langle B_{i}A_{i+1}B_{i+1}\cdots A_{j-1}B_{j-1}B_{j}\rangle, \\
G_{i,j}^{yx}&=& i\langle A_{i}A_{i+1}B_{i+1}\cdots A_{j-1}B_{j-1}A_{j}\rangle.
\end{eqnarray}
Using Wick's theorem~\cite{Pfaffian}, these correlations
can be expanded by the
contractions $\langle A_k A_l \rangle$, $\langle B_k B_l \rangle$, and $\langle B_k A_l \rangle$.
To this end, their expansion formulas can be expressed as a Pfaffian,
which can be cast into a $2r\times 2r$ ($r\equiv \vert j-i \vert$) antisymmetric determinant.
In the case of preserving reflection symmetry  with $\Gamma=0$ in Hamiltonian (\ref{eq:Hamiltonian1}),
it is easy to verify $\langle A_k A_l \rangle$
=$\delta_{kl}$, $\langle B_k B_l \rangle$=-$\delta_{kl}$, which are vanishing for $k\neq l$.
Therefore, the Pfaffian can be simplified as a $ r\times r$ Toeplitz determinant.
Be aware that the reflection symmetry is broken owing to the introduction of off-diagonal
exchange $\Gamma$ interactions and the excitation spectrum in Eq.~(\ref{varepsilon}) is not always positive.
In this case, $\langle A_k A_l \rangle$ and $\langle B_k B_l \rangle$
are otherwise finite for $k\neq l$ in gapless phase, implying that $\langle\sigma_i^x\sigma_j^y\rangle $=$\langle\sigma_i^y\sigma_j^x\rangle $ are not necessarily vanishing. Simultaneously, we can rewrite the $z$-component correlation as
\begin{eqnarray}
  G_{i,j}^{zz}&=&\langle B_{i} A_{i}\rangle\langle B_{j}A_{j}\rangle - \langle B_{j} A_{i} \rangle\langle B_{i} A_{j}\rangle-\langle A_{i}A_{j}\rangle\!\langle B_{i}B_{j}\rangle.\nonumber \\ \label{eq:Gzz}
\end{eqnarray}
The last term in Eq.~(\ref{eq:Gzz}) is usually wrongly discarded in literatures.
\begin{figure}[ht]
\includegraphics[width=\columnwidth]{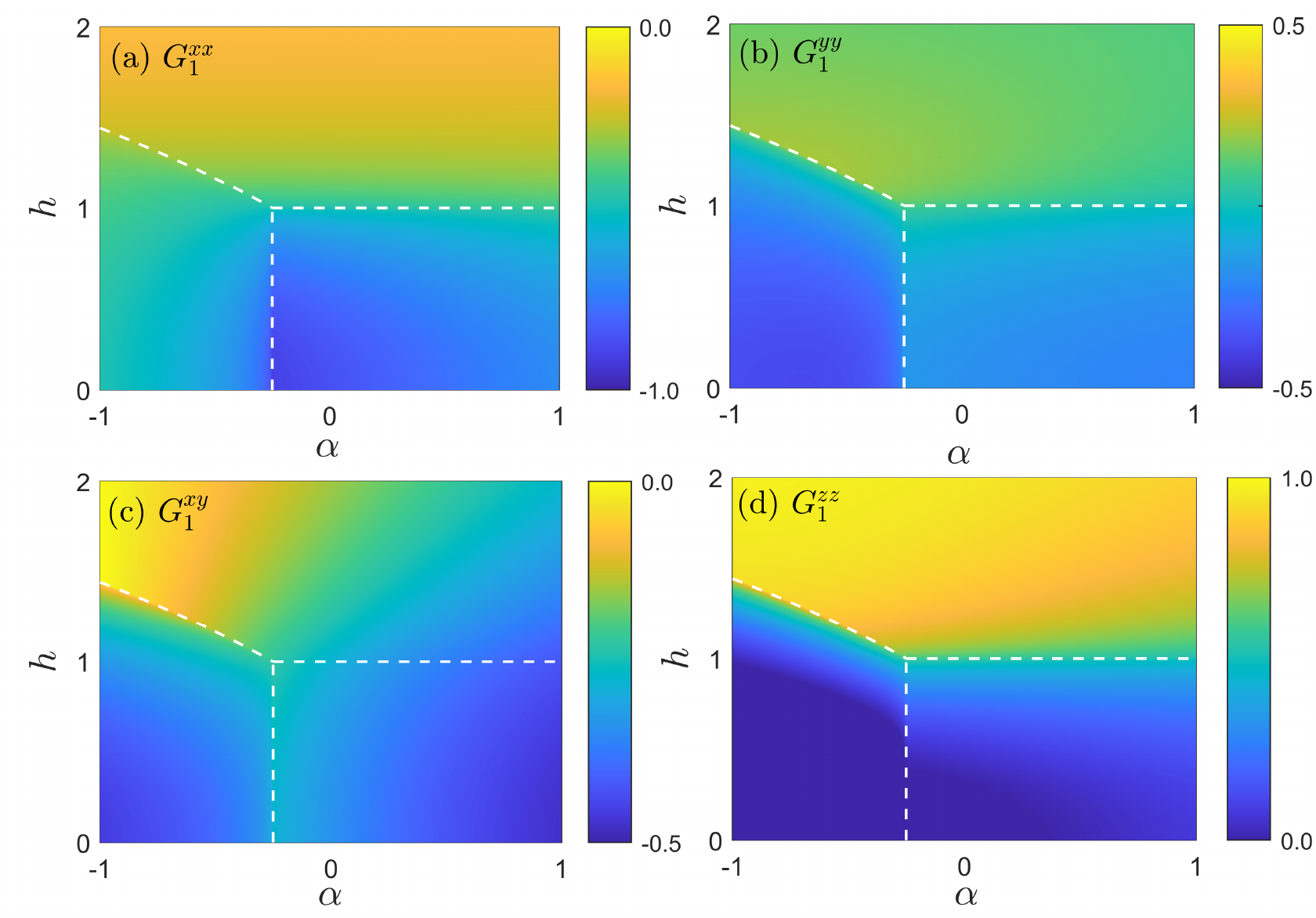}
\caption{The contour map of correlation functions (a) $G_{1}^{xx}$, (b) $G_{1}^{yy}$, (c) $G_{1}^{xy}$ and (d) $G_{1}^{zz}$ with respect to $\alpha$ and $h$ with $N=2000$, $J=1.00$, $\Gamma=0.60, \gamma=0.60$. The white dashed lines correspond to the critical lines.
}
\label{Gcontour}
\end{figure}
To be concrete, it is easy to calculate the nearest-neighbor correlations (i.e., $r=1$); we have
\begin{eqnarray}
G_{i,i+1}^{xx}&=&\langle B_iA_{i+1}\rangle, \quad
G_{i,i+1}^{yy}=-\langle A_{i}B_{i+1} \rangle,  \nonumber\\
G_{i,i+1}^{xy}&=&i\langle B_i B_{i+1}\rangle, \quad
G_{i,i+1}^{yx}= i\langle A_iA_{i+1}\rangle, \nonumber \\
G_{i,i+1}^{zz}&=&\langle A_{i}B_{i}A_{i+1}B_{i+1}\rangle \nonumber \\
&=&\langle\sigma_{i}^{z}\rangle
\langle\sigma_{i+1}^{z}\rangle\!-\!\langle\sigma_{i}^{x}\sigma_{i+1}^{x}\rangle
\langle\sigma_{i}^{y}\sigma_{i+1}^{y}\rangle
\!+\!\langle\sigma_{i}^{x}\sigma_{i+1}^{y}\rangle
\langle\sigma_{i}^{y}\sigma_{i+1}^{x}\rangle.\nonumber \\
\end{eqnarray}
The contour plots of nearest-neighbor correlation functions are shown in Fig. \ref{Gcontour} and provide a full scope of Figs. \ref{G1} and \ref{Gxx}. Similarly, the contour plot of the steered quantum coherence in Fig.\ref{SQCcontour} complements the slice plot in  Fig.\ref{SQC1}.

The four-qubit correlation is described by the $z$ component vector chiral order parameter~\cite{mcculloch08vector,ueda2014vector}
\begin{eqnarray}
\kappa_i=\left(\sigma_i \times \sigma_{i+1}\right)^z.
\end{eqnarray}
As for the consecutive four qubits, it yields \begin{eqnarray}
\langle\kappa_i\kappa_{i+1} \rangle=\langle B_iB_{i+2}\rangle.
\end{eqnarray}
When the dimers are farther than the nearest neighbor, we have for $r\equiv j-i>1$
\begin{eqnarray}
\langle\kappa_i \kappa_j \rangle&=&\langle B_i B_j\rangle\langle B_{i+1}B_{j+1}\rangle-\langle B_iB_{i+1}\rangle \langle B_j B_{j+1}\rangle \nonumber\\
&-&\langle B_iB_{j+1}\rangle\langle B_{i+1}B_j\rangle.
\end{eqnarray}
It is recognized that the nonvanishing cross-correlations arouse nontrivial effect in reflection-symmetry-broken systems and lead to the gapless phase with quasi-long-range order.

\end{appendix}

\end{document}